\documentclass[11pt, a4paper]{article}
\pdfoutput=1

\usepackage{jheppub}
\usepackage[dvipsnames]{xcolor}
\usepackage[utf8]{inputenc}
\usepackage{lipsum}
\usepackage{physics}
\usepackage{mathtools}
\usepackage{multirow}
\usepackage{diagbox}
\usepackage{booktabs}
\usepackage{ytableau}
\usepackage{cancel}
\usepackage{tikz}
\usepackage{subfig}
\usepackage[export]{adjustbox}
\usepackage{amsmath,amssymb,amsfonts,amsthm}
\usepackage{algorithm}
\usepackage{algpseudocode}

\newcommand{\numberset}[1]{\mathbb{#1}}
\newcommand{\N}{\numberset{N}}
\newcommand{\Z}{\numberset{Z}}

\newcommand{\C}{\numberset{C}}

\newcommand{\groupset}[1]{\mathrm{#1}}
\newcommand{\GL}[1]{\groupset{GL}(#1)}
\newcommand{\U}[1]{\groupset{U}(#1)}
\newcommand{\SU}[1]{\groupset{SU}(#1)}

\newcommand{\Aut}[1]{\groupset{Aut}(#1)}
\newcommand{\End}[1]{\groupset{End}(#1)}
\newcommand{\sym}[1]{\mathfrak{S}_{#1}}

\newcommand{\liealgebra}[2]{\mathfrak{#1}(#2)}

\newcommand{\ttr}[2]{\tr_{#1}\!\left(#2\right)}

\newcommand{\id}{\mathrm{id}}

\newcommand{\bigO}[1]{O\mathopen{}\left(#1\right)}

\newcommand{\enstq}[2]{\left\{#1\mathrel{} : \mathrel{}#2\right\}}
\newcommand{\pt}{\, .}
\newcommand{\comma}{\, ,}

\title{An exact algorithm for $\U{N}$ matrix models in the gauge-invariant singlet sector}

\author{Enrico Brehm}
\author{and Jean Cazalis}

\affiliation{Deutsches Elektronen-Synchrotron DESY, \\
    Platanenallee 6, 15738 Zeuthen, Germany}

\emailAdd{enrico.brehm@desy.de}
\emailAdd{jean.cazalis@desy.de}

\abstract{Matrix models appear as fundamental descriptions of M-theory and D-brane dynamics, and via the gauge/gravity duality their gauge-invariant, or singlet, sector describes the purely gravitational degrees of freedom in the holographic dual. We present a new exact algorithm for computing observables of bosonic $\U{N}$ matrix models in the gauge-invariant singlet sector. This sector is spanned by an orthogonal basis of Schur polynomials (for a single matrix) and restricted Schur polynomials (for multiple matrices), which diagonalizes the free Hamiltonian and provides a natural truncation of the Hilbert space by excitation number. Matrix elements of the interaction Hamiltonian, or any gauge-invariant observable, are evaluated through a group-theoretic reduction to cosets and double cosets of suitable subgroups of the symmetric group, together with character sums on the symmetric group. The resulting entries are closed-form polynomials in the gauge-group rank~$N$, assembled from group-theoretic data that are precomputed once and can be reused for any $N$ and any coupling constants. We validate the one-matrix implementation against the exact mapping to $N$ non-interacting fermions, demonstrating rapid convergence of the low-lying spectrum with the cutoff. The multi-matrix extension is outlined; its main bottleneck is the computation of restricted characters of the symmetric group, for which no algorithm comparable to the Murnaghan--Nakayama rule is currently known. The framework gives direct access to finite-$N$, finite-coupling dynamics of gauge-invariant states and opens a new computational window on the non-planar regime of holographic matrix models.}


\arxivnumber{2607.13725}
\preprint{Published in JHEP \textbf{10} (2026) 048, \href{https://doi.org/10.1007/JHEP10(2026)048}{doi:10.1007/JHEP10(2026)048}}

\begin{document} 
\maketitle
\flushbottom

\section{General introduction}

Matrix models are quantum mechanical systems whose degrees of freedom are organized as matrices. Despite having no explicit spatial dependence, they provide a powerful framework for studying theories in higher dimensions, particularly supersymmetric gauge theories and aspects of quantum gravity. A central challenge in the analysis of these systems is the efficient treatment of their gauge symmetry and the explicit construction of the gauge-invariant Hilbert space, together with the computation of Hamiltonian matrix elements within it. This work addresses these challenges through a new exact algorithm.

\subsection{Matrix models in high-energy physics}

The study of matrix models has a rich history dating back to the early developments in string theory and quantum gravity research. Through the gauge/gravity duality~\cite{Maldacena:1997re}, matrix models establish deep connections between strongly-coupled quantum field theories and weakly-curved gravitational theories in higher dimensions. This duality has motivated extensive investigation of matrix models as tools for understanding quantum gravitational phenomena, including features of black holes (see e.g.~\cite{Maldacena:2023acv}).

Notable examples include the BFSS matrix model~\cite{Banks:1996vh,Seiberg:1997ad}, which describes M-theory in the infinite momentum frame and arises as the effective description of interacting D0-branes, and the BMN matrix model~\cite{Berenstein:2002jq,Dasgupta:2002hx}, obtained through mass deformation of the BFSS model. These constructions have advanced our understanding of string theory and quantum gravitational theories, enabling computations of graviton scattering amplitudes and properties of black hole systems~\cite{Becker:1997wh,Becker:1997xw,Helling:1999js,Herderschee:2023pza,Banks:1997hz,Banks:1997tn,Klebanov:1997kv,Halyo:1997wj,Horowitz:1997fr,Kabat:1997im,Hyakutake:2018bht,Du:2018dmi,Miller:2022fvc,Tropper:2023fjr,Herderschee:2023bnc,Berenstein:2002jq,Shin:2005iy,Asplund:2011qj,Brady:2013opa,Costa:2014wya,Pramodh:2014jha,Asano:2018nol,Asano:2020yry,Axenides:2017nwp,Gray:2020zov}.

A key advantage of matrix models is that they are free from the need to discretize spacetime, potentially offering a more natural path to simulate aspects of higher-dimensional quantum field theories and string theory compared to traditional lattice formulations~\cite{Gharibyan:2020bab}.

\subsection{\texorpdfstring{$\U{N}$}{U(N)} symmetry of the matrix model}

Matrix models typically possess rich symmetry structures that can be exploited to reduce computational complexity. For a matrix model with degrees of freedom organized in the Hermitian matrix $X$ transforming in the adjoint representation of $\U{N}$, the Hamiltonian is invariant under (gauge) transformations of the form $X \to UXU^\dagger$. This $\U{N}$ symmetry allows us to organize the Hilbert space into sectors labeled by irreducible representations.

In the quantum mechanical setting, we can treat this symmetry in two ways: either as a genuine gauge redundancy requiring physical states to be gauge singlets, or as a global symmetry allowing for charged sectors. The singlet constraint is implemented through the Gauß law, which requires physical states to be annihilated by the generators of $\U{N}$.

In this work, we focus on constructing the invariant or singlet sector of matrix models. Physical states in this sector are those that remain unchanged under all $\U{N}$ gauge transformations, a requirement that has implications both for computational efficiency and physical interpretation.

From the physical perspective, the singlet sector carries special significance through the lens of holography. Via the gauge/gravity correspondence, the restriction to gauge singlets is conjectured to correspond to considering only local degrees of freedom in the dual gravitational description. This excludes, for instance, Wilson line insertions extending to some boundary. Thus, constructing and analyzing the singlet sector allows us to study the quantum gravitational theory in its most fundamental setting, focusing on the closed-string or purely gravitational degrees of freedom without additional probe charges~\cite{Maldacena:2018vsr}.

From the computational perspective, restricting to gauge singlets significantly reduces the dimension of the Hilbert space that must be explored. Rather than working with the full space, we only need to consider gauge-invariant combinations of the matrix degrees of freedom. By the first fundamental theorem of invariant theory~\cite{goodman2009Symmetry}, any gauge-invariant polynomial can be expressed in terms of traces. More systematically, exploiting Schur-Weyl duality between the symmetric group $\sym{n}$ and the unitary group $\U{N}$, the gauge-invariant sector admits an explicit orthogonal basis of Schur polynomials (for a single matrix) and restricted Schur polynomials (for multiple matrices) that in particular diagonalizes the free Hamiltonian. In this work, we present an exact algorithm to compute all matrix elements of the interaction Hamiltonian in this basis. The key output is a Hamiltonian matrix whose entries are explicit polynomials in $N$, assembled from group-theoretical data that are precomputed once and are independent of the physical couplings. The matrix can then be evaluated and diagonalized for any value of $N$ and any coupling constants without repeating the algebraic computation, making numerical simulations tractable at larger matrix sizes while giving direct analytical access to the $N$-dependence of the spectrum.

We note that the gauge group considered here is $\U{N}$ rather than $\SU{N}$, in contrast to the BFSS and BMN models. While the Schur polynomial basis provides an orthogonal and complete basis for $\U{N}$ singlets, extending this construction to $\SU{N}$ is technically nontrivial: the tracelessness of $\SU{N}$ generators introduces additional constraints that render the set of Schur polynomial states overcomplete, requiring additional steps to extract a proper basis. We discuss this issue and potential strategies in the concluding section. We also do not include the cubic interaction term $\epsilon_{IJK}\tr(X_I[X_J,X_K])$ present in the BMN model; this term is handled by the same algorithm, and its omission simplifies the presentation without affecting the generality of the method. 

\subsection{Novelty of this work}

The (restricted) Schur polynomial literature is largely concerned with the free theory and with one-loop effects. On the one hand, the Schur and restricted Schur bases were introduced to diagonalize the free two-point function exactly at finite $N$~\cite{corley2002Exact,bhattacharyya2008Exact,bhattacharyya2008Exacta}, and the associated algebraic structure was subsequently developed in~\cite{mattioli2016Permutation,padellaro2025Eigenvalue,ramgoolam2026Finitedimensional}. On the other hand, the one-loop dilatation operator was diagonalized in special large-$N$, non-planar limits -- the distant-corners and double-coset constructions of~\cite{decomarmond2011Surprisingly,demellokoch2011Giant,demellokoch2012Double}. In all these works, however, the full interacting Hamiltonian matrix in the singlet sector is never assembled: the results are either free-field correlators, one-loop spectra in restricted limits, or algebraic constructions of orthogonal bases.

The numerical approaches to matrix models, by contrast, are approximate. Hamiltonian truncation methods (TCSA and its conformal/lightcone variants) truncate an infinite Hilbert space and require renormalization~\cite{yurov1990TCSA,hogervorst2015TCSA,katz2016LCT}, while the bootstrap and tensor-network methods give bounds or variational approximations rather than exact matrix elements~\cite{Han:2020bkb,brehm2025Simulating}. None of these is built on a group-theoretically complete, orthogonal singlet basis.

The new contribution of the present work is precisely the systematic coset/double-coset and symmetric-group character-sum reduction that yields exact, closed-form polynomial-in-$N$ matrix elements for the full interaction Hamiltonian in the singlet sector, valid at finite $N$ and finite coupling and reusable for any $N$ and any couplings. This extends the finite-$N$ free-theory diagonalization of the restricted-Schur basis into the interacting regime, and complements the large-$N$ non-planar one-loop results with an exact finite-$N$, finite-coupling algorithm.

The paper is organized as follows. Section~\ref{sec:technical_introduction} introduces the Hamiltonian formulation of the $\U{N}$ matrix models, constructs the Schur and restricted Schur polynomial bases of the gauge-invariant Hilbert space, and establishes their orthogonality. Section~\ref{sec:computation_matrix_elements} develops the algorithm for computing the Hamiltonian matrix elements in this basis, working through the single-matrix case in complete detail and outlining the multi-matrix extension together with a complexity analysis and numerical validation. Section~\ref{sec:conclusion} summarizes the results and discusses directions for future work. The appendices collect the necessary background from representation theory, Schur-Weyl duality, and the bosonic Fock space formalism, together with the fermion mapping used for numerical validation.

\section{Technical introduction}
\label{sec:technical_introduction} 

In this section, we describe the bosonic $\U{N}$ matrix models considered in this work. They are toy models whose physical states shall remain invariant under the adjoint action of the gauge group -- that is, they should be $\U{N}$ singlets. We therefore introduce an orthogonal basis for the singlet sector of these models. We first detail its construction for one-matrix models, and then extend it to systems with several matrices. Some of the notation and conventions are adapted from~\cite{brehm2025Simulating}. A reference for the construction of a singlet basis is~\cite{koch2024Pedagogical}.

\subsection{Hamiltonian formulation of matrix models}

We consider $D$ copies of ladder operators $A_I$ and $A_I^\dagger$, $I=1,\dots,D$, each a complex $N\times N$ matrix jointly transforming in the adjoint representation of $\U{N}$, that is,
\begin{align}
    A_I \longrightarrow U A_I U^\dagger \qq{and} A_I^\dagger \longrightarrow U A_I^\dagger U^\dagger \comma \quad \forall I \pt
\end{align}
To leverage the representation theory of the symmetric group, it will be convenient to decompose each of them in the \emph{matrix-entry} basis, as exposed in~\cite{koch2024Pedagogical} or in Appendix~\ref{app:bosonic_fock_space}. This amounts to writing
\begin{align}
    A_I = (A_I)_k^\ell E^k_\ell \qq{and} A_I^\dagger = (A_I^\dagger)_\ell^k E_k^\ell \comma
\end{align}
where $(A_I^\dagger)_\ell^k = ((A_I)^\ell_k)^\dagger$, and $E^k_\ell$ denotes the $N\times N$ matrix whose $(\ell,k)$-entry is~$1$ and all other entries vanish. In this setup, the canonical commutation relations read
\begin{align}
    \comm{(A_I)^\ell_k}{(A_J^\dagger)^m_n} = \delta_{IJ} \delta_{k}^m \delta^{\ell}_n \pt
\end{align}
The bosonic matrices and their conjugate momenta, or quadrature operators, are constructed from these ladder operators as
\begin{align}
    \label{eq:quadrature_operators}
    X_I = \frac{1}{\sqrt{2m}}(A_I+A_I^\dagger) \qq{and} P_I = i \sqrt{\frac{m}{2}}(A_I^\dagger - A_I) \comma
\end{align}
where $m>0$ is the mass of a single bosonic mode.

A multi-matrix $\U{N}$ model can be described by a Hamiltonian of the form
\begin{align}
    \label{eq:hamiltonian_form_2}
    H_{\mathcal{O}} = \frac{1}{2} \sum_{I=1}^D \tr( P_I^2 + m^2 X_I^2) + \mathcal{O} = \sum_{I=1}^D \sum_{\ell,k=1}^N m((N_I)^k_\ell + \tfrac{1}{2}) + \mathcal{O}\comma
\end{align}
where $(N_I)^k_\ell \coloneqq (A_I^\dagger)^\ell_k (A_I)^k_\ell$ is the number operator in the mode $(k,\ell)$ for species $I$, and $\mathcal{O}$ is any gauge-invariant observable such that $H_{\mathcal{O}}$ is bounded from below. As explained in Appendix~\ref{app:gauge_invariant_observables}, any gauge-invariant observable can be expressed as a linear combination of products of traces of words drawn from the alphabet $\{A_I,A_I^\dagger\}_{I=1}^D$, or equivalently $\{X_I,P_I\}_{I=1}^D$.

For $D=1$, this reduces to the single-matrix model. A simple interaction leading to a Hamiltonian bounded from below with a unique ground state is the quartic self-interaction $\tr(X^4)$. The single-matrix Hamiltonian reads
\begin{align}
    \label{eq:hamiltonian_form_1}
    H = \frac{1}{2} \tr( P^2 + m^2 X^2) + \frac{g^2}{2} \tr(X^4)
      = \sum_{\ell,k=1}^N m(N^k_\ell + \tfrac{1}{2}) + \frac{g^2}{2} \tr(X^4) \comma
\end{align}
where $g>0$ describes the interaction strength. This model's mean-field regime and its connection to a system of $N$ non-interacting fermions are discussed in~\cite{brezin1978Planar}. More recently, it was also analyzed using bootstrap methods~\cite{Han:2020bkb} and tensor networks~\cite{brehm2025Simulating}.

This fermion mapping is in fact more general: it applies to any one-matrix $\U{N}$ Hamiltonian of the form~\eqref{eq:hamiltonian_form_2} with $D=1$, even though the fermions may interact nontrivially when the interaction terms contain product of traces. A concise review is given in Appendix~\ref{app:fermion_mapping}. The mapping enables efficient numerical approximation of $\U{N}$ one-matrix models up to large cutoffs. It therefore provides a perfect toy model on which to develop and evaluate the algorithm presented in this work.

For $D>1$, the natural gauge-invariant interaction generalizing the quartic term is the commutator-squared interaction,
\begin{align}
    \label{eq:hamiltonian_multi}
    H = \frac{1}{2} \sum_{I=1}^D \tr( P_I^2 + m^2 X_I^2) - \frac{g^2}{2} \sum_{I,J=1}^D \tr([X_I,X_J]^2) \comma
\end{align}
which is the even bosonic part of the BFSS and BMN matrix models~\cite{Banks:1996vh,Berenstein:2002jq}.  The multi-matrix case cannot be reduced to a system of fermions, since the $X_I$ do not commute and cannot be simultaneously diagonalized.

\subsection{Singlet states in one-matrix models}

In this section, we review a systematic procedure for constructing gauge-invariant states in the Fock space. This approach was pioneered in~\cite{corley2002Exact} in the context of $\mathcal{N}=4$ SYM theory and is also described in detail in the pedagogical introduction~\cite{koch2024Pedagogical}, from which several parts of this section are adapted. We first describe the general form of a singlet state, then introduce the Young projectors, which correspond to an orthogonal basis of the center of the group algebra $\C[\sym{n}]$ of the symmetric group. From these we construct a basis of singlet states and prove its orthogonality. Finally, we reformulate the result in a graphical notation that will be used throughout the paper.

\paragraph{Gauge-invariant states.}

Consider a general state with at most $\Lambda$ excitations, which may be expressed in the number basis as
\begin{align}
    \ket{\psi} = \sum_{\{n_\ell^k\}} c_{\{n_\ell^k\}}\ket{\{n_\ell^k\}}  \comma
\end{align}
where $c_{\{n_\ell^k\}} \in \C$ and $n_\ell^k \in \N$. The sum extends over all occupation number configurations $\{n_\ell^k\}_{\ell,k}$ satisfying the excitation bound $\sum_{\ell,k} n_\ell^k \leq \Lambda$. 

Equivalently, this state admits a polynomial representation
\begin{align}
    \ket{\psi} = \ket{P} \coloneqq P((A^{\dagger})_1^1,\dots,(A^{\dagger})_N^N) \ket{\Omega_{\mathrm{free}}}  \comma
\end{align}
where $\ket{\Omega_{\mathrm{free}}}$ is the vacuum of the non-interacting theory, $P \in \C[Y] = \C[Y_1^1,\dots,Y_N^N]$ is a multivariate polynomial in $N^2$ variables with total degree at most $\Lambda$, and here $Y= A^\dagger$.

Under a transformation $Y\mapsto UYU^\dagger$ with $Y$ a complex $N\times N$ matrix and $U\in\U{N}$, the state $\ket{P}$ transforms according to $\ket{P} \mapsto \ket{P_U}$, where $P_U(Y) = P(UYU^\dagger)$. This transformation preserves both the total degree bound $\Lambda$ and homogeneity: if $P$ is homogeneous of degree $k$, then so is $P_U$. Our objective is therefore to identify the subspace of gauge-invariant polynomials, i.e. those satisfying
\begin{align}
    P = P_U ,~\forall U \in \U{N}
    \Longleftrightarrow P(Y) = P(UYU^\dagger) ,~\forall U \in \U{N}  ,~ \forall Y \pt
\end{align}
Such polynomials generate states that are singlets under the adjoint action of $\U{N}$ and span the physical Hilbert space of the gauge theory. 

\paragraph{Young projectors.}

At fixed homogeneous degree $n$, it is convenient to package the coefficients into a $\U{N}$-invariant operator $Q \in \End{(\C^N)^{\otimes n}}$ and write
\begin{align}
    \ket{Q} \coloneqq Q^K_L (A^{\dagger\otimes n})_K^L \ket{\Omega_{\mathrm{free}}}  \comma
\end{align}
where $K = (k_1,\dots,k_n)$ and $L = (\ell_1,\dots,\ell_n)$ are multi-indices. The Schur-Weyl duality~\cite[Chapter 9]{goodman2009Symmetry} implies that the $\U{N}$-invariant tensors are generated by permutations of tensor slots, i.e. the entries in the multi-indices. This means for gauge singlets one has to consider $Q \in \C[\sym{n}]$, the group algebra of $\sym{n}$, see Appendix~\ref{app:group_algebra} for more details. In addition, all tensor factors are the same creation operator $A^\dagger$. Consequently, $Q$ must be chosen up to a reordering of its indices, that is as an element of the center
\begin{align}
    Z(\C[\sym{n}]) = \enstq{v \in \C[\sym{n}]}{\sigma v = v \sigma, ~\forall \sigma \in \sym{n}}  \pt
\end{align}
An orthogonal basis for $Z(\C[\sym{n}])$ can be constructed from the projections onto the isotypic components in the Schur-Weyl decomposition
\begin{align}
    \label{eq:schur_weyl_decomposition}
    (\C^N)^{\otimes n} = \bigoplus_{\substack{R\vdash n \\ \ell(R)\leq N}} V^R_{\U{N}}\otimes V^R_{\sym{n}}  \comma
\end{align}
where the sum runs over all Young diagrams $R$ with $n$ boxes (denoted $R\vdash n$) and at most $N$ rows (denoted $\ell(R)\leq N$). Here, $V^R_{\U{N}}$ carries the irreducible representation $R$ of $\U{N}$, and $V^R_{\sym{n}}$ carries the irreducible representation $R$ of $\sym{n}$. We refer to Appendix~\ref{app:young_diagrams} and Appendix~\ref{app:schur-weyl-duality} for a brief review of, respectively, the Young diagrams and the Schur-Weyl decomposition.

Let $\chi_R(\sigma)$ denote the character of the irreducible representation $V_{\sym{n}}^R$ of $\sym{n}$. In this context, the \emph{Young projector} $P_R$ is the $\U{N}$-invariant projector onto the $R$-isotypic component $V_{\U{N}}^R\otimes V_{\sym{n}}^R$ in the Schur-Weyl decomposition~\eqref{eq:schur_weyl_decomposition}. More explicitly, we have
\begin{align}
    \label{eq:young_projector}
    (P_R)^K_L = (P_R)^{k_1 \dots k_n}_{\ell_1 \dots \ell_n} \coloneqq \frac{\dim_R}{n!} \sum_{\sigma \in \sym{n}} \chi_R(\sigma) \sigma^K_L   \comma
\end{align}
where $\dim_R$ is the dimension of the irrep $V_{\sym{n}}^R$, and where $\sigma^K_L = \sigma^{k_1\cdots k_n}_{\ell_1\cdots \ell_n} = \delta_{\ell_1}^{k_{\sigma_1}}\cdots\delta_{\ell_n}^{k_{\sigma_n}}$. 

Since $P_R$ is $\U{N}$-invariant, the Schur-Weyl duality identifies it with an element of the group algebra of the symmetric group, $P_R \in \C[\sym{n}]$. In the sequel, we will freely identify $\U{N}$-invariant tensors in $\End{(\C^N)^{\otimes n}}$ with the corresponding elements of $\C[\sym{n}]$; our index conventions guarantee that multiplications in the two algebras agree, as detailed in Appendix~\ref{app:notation}.

For fixed $n$, the family $\{P_R\}_{R\vdash n}$ is precisely the Young-diagram-indexed basis of the center $Z(\C[\sym{n}])$ discussed in Appendix~\ref{app:center_group_algebra}. Since the trace of $P_R$ is the dimension of the space it projects onto, we have
\begin{align}
    \label{eq:young_projectors_trace}
    \tr(P_R) \coloneqq (P_R)_K^K = (P_R)_{k_1 \dots k_n}^{k_1 \dots k_n} = \dim_R \mathrm{Dim}_R  \comma
\end{align}
where $\mathrm{Dim}_R$ is the dimension of the representation space $V_{\U{N}}^R$. Both $\dim_R$ and $\mathrm{Dim}_R$ admit a closed formula, see Appendices~\ref{app:young_diagrams} and~\ref{app:schur-weyl-duality}.

Two fundamental properties of the Young projectors $P_R$ are particularly important: first, their orthogonality relation,
\begin{align}
    \label{eq:young_projectors_orthonormality}
    P_R \cdot P_S = \delta_{RS}  P_R  \comma
\end{align}
and second, that they belong to the center of the group algebra 
\begin{align}
    \label{eq:young_projectors_center}
    P_R \rho = \rho P_R  , \quad \forall \rho \in \sym{n} \pt
\end{align}

\paragraph{An orthogonal basis of singlet states.}

The same data also define the associated Schur polynomial~\cite{james1984Representation,sagan2010Symmetric} by contraction with $Y^{\otimes n}$:
\begin{align}
    \label{eq:schur_polynomials}
    \chi_R(Y)
    \coloneqq \frac{1}{\dim_R}(P_R)^K_L (Y^{\otimes n})_K^L  \comma
\end{align}
where $Y$ is an $N\times N$ matrix of indeterminates, and $R$ ranges over all Young diagrams with $n$ boxes and at most $N$ rows. The Schur polynomial $\chi_R$ is homogeneous of degree $n$.

To construct gauge-invariant states, we evaluate Schur polynomials at the creation operator matrix $A^\dagger$ and apply the result to the Fock vacuum:
\begin{align}\label{eq:Schurr_singlets}
    \ket{R} \coloneqq \chi_R(A^\dagger) \ket{\Omega_{\mathrm{free}}}
     \pt
\end{align}
The ring of gauge-invariant polynomials of $N\times N$ matrices of homogeneous degree $n$ is upon restriction to diagonal matrices isomorphic to the ring of symmetric polynomials in $N$ variables of homogeneous degree $n$. On the latter, the Schur polynomials form a complete basis \cite{james1984Representation}. Therefore, the family $\{\ket{R}\}_R$ indexed by Young diagrams $R$ with $\ell(R) \leq N$ (at most $N$ rows) provides a complete basis for the physical Hilbert space of gauge-invariant states, too.

Since Young diagrams with $n$ boxes are in bijection with the partitions of $n$, the dimension of the singlet basis at excitation level $n$ is bounded by the partition number $p(n)$ (with equality when $N\geq n$), whose asymptotic growth is given by the Hardy--Ramanujan formula
\begin{align}
    \label{eq:dimension_center}
    \dim Z(\C[\sym{n}]) = p(n) \sim \frac{1}{4n\sqrt{3}} \exp\left(\pi \sqrt{\frac{2n}{3}}\right)  \pt
\end{align}

We now establish that this family is orthogonal. Specifically, we prove
\begin{align}
    \braket{R}{S} = \alpha_R \delta_{RS}  \comma
\end{align}
with the normalization constant given by
\begin{align}
    \label{eq:normalization_constant_one_matrix}
    \alpha_R = \frac{n! \mathrm{Dim}_R}{\dim_R} = \prod_{(i,j)\in R} (N - i + j)  \comma
\end{align}
where the product runs over all boxes $(i,j)$ of the Young diagram~$R$, see Appendix~\ref{app:young_diagrams}.

We first observe that if $R$ and $S$ have different numbers of boxes, the corresponding states $\ket{R}$ and $\ket{S}$ belong to distinct excitation sectors and are therefore automatically orthogonal. Henceforth, we assume $R$ and $S$ are diagrams with the same number $n$ of boxes.

The proof proceeds by expressing the inner product in terms of Young projectors. Substituting the representation~\eqref{eq:schur_polynomials} of the Schur polynomials into the overlap yields
\begin{multline}
    \braket{R}{S} 
    = \mel{\Omega_{\mathrm{free}}}{\chi_R(A) \chi_S(A^\dagger)}{\Omega_{\mathrm{free}}} \\
    = \frac{1}{\dim_R\dim_S} (P_R)_J^I (P_S)^K_L   \mel{\Omega_{\mathrm{free}}}{(A^{\otimes n})^J_I ((A^\dagger)^{\otimes n})_K^L}{\Omega_{\mathrm{free}}}  \pt
\end{multline}
The vacuum expectation value of creation and annihilation operators is evaluated using Wick's theorem (see Appendix~\ref{app:wick_formula})
\begin{align}
    \mel{\Omega_{\mathrm{free}}}{(A^{\otimes n})^J_I ((A^\dagger)^{\otimes n})_K^L}{\Omega_{\mathrm{free}}}
    = \sum_{\sigma \in \sym{n}} (\sigma^{-1})^L_I \sigma_K^J  \pt
\end{align}
Substituting this result, we obtain
\begin{align}
    \notag
    \braket{R}{S} 
    =&~ \frac{1}{\dim_R\dim_S} \sum_{\sigma \in \sym{n}} (P_R)_J^I (P_S)^K_L (\sigma^{-1})^L_I \sigma_K^J \\
    \notag
    =&~ \frac{1}{\dim_R\dim_S} \sum_{\sigma \in \sym{n}} (\sigma^{-1} P_R \sigma)_K^L (P_S)^K_L
    = \frac{1}{\dim_R\dim_S} \sum_{\sigma \in \sym{n}} (P_R)_K^L (P_S)^K_L \\
    =&~ \frac{n!}{\dim_R\dim_S} \tr(P_R P_S) 
    = \frac{n!}{\dim_R\dim_S} \tr(P_S) \delta_{RS} 
    = \frac{n!\mathrm{Dim}_R}{\dim_R} \delta_{RS}  \comma
\end{align}
where we have used: (i) the centrality of $P_R$ in $\C[\sym{n}]$~\eqref{eq:young_projectors_center}, so that $\sigma^{-1} P_R \sigma = P_R$; (ii) the orthogonality relation $P_R P_S = \delta_{RS} P_R$~\eqref{eq:young_projectors_orthonormality}; and (iii) the trace identity $\tr(P_R) = \dim_R \mathrm{Dim}_R$~\eqref{eq:young_projectors_trace}. This completes the proof of orthogonality. 

\paragraph{A graphical representation.}

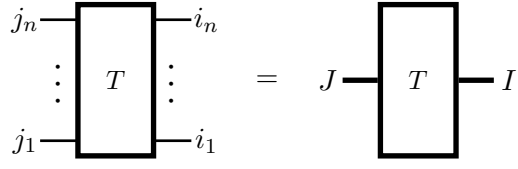
\begin{figure}[tp]
    \centering
    \begin{tikzpicture}[line width=2px]
    \begin{scope}[shift={(0,0)}]  
        \node[draw, rectangle, minimum width=1.0cm, minimum height=2.0cm, align=center] (Tleft) {\small$T$};
        \draw[-,line width=1px] (-1,.8) to (-.5,.8);
        \draw[-,line width=1px] (1,.8) to (.5,.8);
        \draw[-,line width=1px] (-1,-.8) to (-.5,-.8);
        \draw[-,line width=1px] (1,-.8) to (.5,-.8);
        \node[rotate=90] at (-.75,0) {\Large $\cdots$};
        \node[rotate=90] at (.75,0) {\Large $\cdots$};
        \node at (1.2,.8) {$i_n$};
        \node at (1.2,-.8) {$i_1$};
        \node at (-1.2,.8) {$j_n$};
        \node at (-1.2,-.8) {$j_1$};
    \end{scope}

    \node at (2,0) {$=$};

    \begin{scope}[shift={(4,0)}]  
        \node[draw, rectangle, minimum width=1.0cm, minimum height=2.0cm, align=center] (Tright) at (0,0) {\small$T$};
        \draw[-] (-1,0) to (Tright) to (1,0);
        \node at (1.2,0) {$I$};
        \node at (-1.2,0) {$J$};
    \end{scope}
    \end{tikzpicture}
    \caption{Graphical representation of a tensor $T_I^J \in \End{(\C^N)^{\otimes n}}$. The right legs carry the lower indices $I$, while the left legs carry the upper indices $J$.}
    \label{fig:tensor_diagram_1}
\end{figure}

\begin{figure}[tp]
    \centering
    \begin{tikzpicture}[line width=2px]
        \begin{scope}[shift={(-3,0)}]  
            \draw[-,line width=1px] (-1,.7) to[out=0,in=180] (1,0);
            \draw[-,line width=1px] (-1,0) to[out=0,in=180] (1,-.7);
            \draw[-,white,line width=4px] (-1,-.7) to[out=0,in=180] (1,.7);
            \draw[-,line width=1px] (-1,-.7) to[out=0,in=180] (1,.7);
        \end{scope}

        \node at (-1.5,0) {$=$};

        \begin{scope}[shift={(0,0)}]  
            \node[draw, rectangle, minimum width=1.0cm, minimum height=2.0cm, align=center] (sigma_left) at (0,0) {\small$(123)$};
            \draw[-,line width=1px] (-1,.7) to (-.5,.7);
            \draw[-,line width=1px] (1,.7) to (.5,.7);
            \draw[-,line width=1px] (-1,0) to (-.5,0);  
            \draw[-,line width=1px] (1,0) to (.5,0);  
            \draw[-,line width=1px] (-1,-.7) to (-.5,-.7);
            \draw[-,line width=1px] (1,-.7) to (.5,-.7);
        \end{scope}
  
        \node at (1.5,0) {$=$};
  
        \begin{scope}[shift={(3,0)}]  
            \node[draw, rectangle, minimum width=1.0cm, minimum height=2.0cm, align=center] (sigma_right) at (0,0) {\small$(123)$};
            \draw[-,line width=2px] (-1,0) to (sigma_right) to (1,0);
        \end{scope}
    \end{tikzpicture}
    \caption{Graphical representation of the permutation tensor $\sigma=(123)$, or written out $\sigma_{i_1i_2i_3}^{j_1j_2j_3} = \delta_{i_1}^{j_2} \delta_{i_2}^{j_3} \delta_{i_3}^{j_1}$.}
    \label{fig:sigma_diagram_2}
\end{figure}
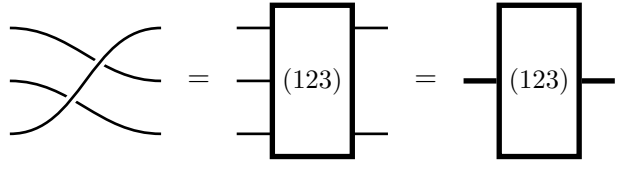

We now introduce a graphical notation for the tensors that appear in this work. A tensor $T_{i_1i_2\dots i_n}^{j_1j_2\dots j_n} = T_I^J \in \End{(\C^N)^{\otimes n}}$ is depicted as a box with $n$ upwards-oriented legs on the right (carrying the lower indices $I$) and $n$ upwards-oriented legs on the left (carrying the upper indices $J$). A multi-index is represented by a thick leg. Figure~\ref{fig:tensor_diagram_1} illustrates this in a general case and Figure~\ref{fig:sigma_diagram_2} illustrates this representation for the tensor corresponding to the cycle $\sigma =(123)$. There it also becomes clear that the diagrams are supposed to be read from right to left. 

\begin{figure}[tp]
    \centering
    
    \begin{tikzpicture}[line width=2px]
    \begin{scope}[shift={(0,0)}]  
        \draw[-,line width=1px] (-1,.7) to[out=0,in=180] (1,0);
        \draw[-,line width=1px] (-1,0) to[out=0,in=180] (1,-.7);
        \draw[-,white,line width=4px] (-1,-.7) to[out=0,in=180] (1,.7);
        \draw[-,line width=1px] (-1,-.7) to[out=0,in=180] (1,.7);
    \end{scope}

    \begin{scope}[shift={(-2.1,0)}]  
        \draw[-,line width=1px] (-1,-.7) to[out=0,in=180] (1,-.7);
        \draw[-,line width=1px] (-1,0) to[out=0,in=180] (1,.7);
        \draw[-,white,line width=4px] (-1,.7) to[out=0,in=180] (1,0);
        \draw[-,line width=1px] (-1,.7) to[out=0,in=180] (1,0);
    \end{scope}
    \begin{scope}[shift={(3.5,0)}]  
        \draw[-,line width=1px] (-1,.7) to[out=0,in=180] (1,-.7);
        \draw[-,line width=1px] (-1,0) to[out=0,in=180] (1,0);
        \draw[-,white,line width=4px] (-1,-.7) to[out=0,in=180] (1,.7);
        \draw[-,line width=1px] (-1,-.7) to[out=0,in=180] (1,.7);
    \end{scope}

    \begin{scope}[shift={(6.5,0)},line width=1px]  
        \draw (0,.5) ellipse (1cm and .3cm);
        \draw (0,-.5) ellipse (1cm and .3cm);
    \end{scope}

    \node at (1.5,0) {$=$};
    \node at (5,0) {$=$};
    \node at (8.25,0) {$=~N^2$};

    \node at (-3.5,0) {tr};
    
    \node at (2,0) {tr};
    \end{tikzpicture}
    \caption{Graphical representation of the contraction of $(123)$ and $(1)(23)$. The result is $(13)(2)$; taking the trace yields two loops, each corresponding to a trace in the defining representation of $\U{N}$ and contributing a factor of $N$.}
    \label{fig:contraction_diagram} 
\end{figure}
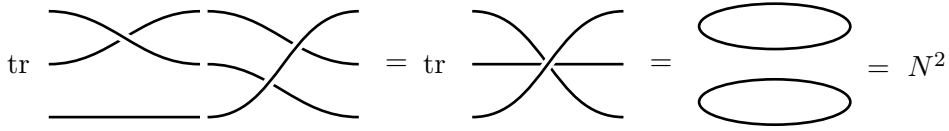

Each thin line carries the fundamental representation of $\U{N}$, and index contraction corresponds to joining the corresponding legs. Every tensor we encounter can be expressed as a linear combination of permutation tensors $\sigma^K_L$; for instance, the Young projector $P_R$ takes this form, as seen in~\eqref{eq:young_projector}. The contraction of two permutation tensors corresponds to multiplication in the symmetric group, which is represented graphically by concatenating and flattening the lines. Taking the trace produces closed loops, whose number equals the number of cycles in the resulting permutation. Each loop stands for a trace in the defining representation of $\U{N}$ and contributes a factor of $N$. These two rules are illustrated in Figure~\ref{fig:contraction_diagram}, which shows the contraction and subsequent trace of $(123)$ and $(1)(23)$ in $\sym{3}$.

We shall use this graphical notation throughout the computation of the Hamiltonian matrix elements in the gauge-invariant basis. As a further example, the essential steps in the evaluation of the inner product $\braket{R}{S}$ are:
\begin{equation}\nonumber
    \begin{tikzpicture}[line width=2px, scale=0.765]
    \begin{scope}
        \node[draw, rectangle, minimum width=1.0cm, minimum height=2.0cm, align=center] (PR) {$P_R$};
    \node[draw, rectangle, minimum width=1.0cm, minimum height=1.0cm, align=center, right of = PR, node distance=1.25cm] (sigma) {$\sigma$};
    \node[draw, rectangle, minimum width=1.0cm, minimum height=1.0cm, align=center, left of = PR, node distance=1.25cm] (invsigma) {$\sigma^{-1}$};
    \node[draw, rectangle, minimum width=1.0cm, minimum height=2.0cm, align=center, right of = sigma, node distance=1.25cm] (PS) {$P_S$};
        \draw(invsigma) to (PR) to (sigma) to (PS) to[out=0,in=0] (3.5,-2) to (-2,-2) to[out=180, in=180](invsigma);
    \end{scope}
    \node at (4.6,0) {$=$};
    \begin{scope}[xshift=6cm]
        \node[draw, rectangle, minimum width=1.0cm, minimum height=2.0cm, align=center] (PR) {$P_R$};
        \node[draw, rectangle, minimum width=1.0cm, minimum height=2.0cm, align=center, right of = PR, node distance=1.25cm] (PS) {$P_S$};
        \draw (PR) to (PS) to[out=0,in=0] (2,-2) to (-.5,-2) to[out=180, in=180](PR);
    \end{scope}
    \node at (9.45,0) {$=\delta_{RS}$};
    \begin{scope}[xshift=11.25cm]
        \node[draw, rectangle, minimum width=1.0cm, minimum height=2.0cm, align=center] (PR) {$P_R$};
        \node[right of = PR, node distance=2.5cm] {$=\delta_{RS}\text{Dim}_R \text{dim}_R$};
        \draw (PR) to[out=0,in=0,distance=1cm] (0,-2) to[out=180, in=180,distance=1cm](PR);
    \end{scope}
    \end{tikzpicture}
\end{equation}
The orthogonality of the basis then follows from the fact that the resulting diagram contains a loop if and only if $R=S$.

\subsection{Singlet states in multi-matrix models}
\label{sec:singlet_states_multi_matrices}

We follow the same strategy as in the one-matrix case, now keeping track of how the total excitation number is distributed among the $D$ matrices. Since the reasoning mainly follows the previous section, we do not present the details, and we mostly highlight the points that differ. 

Fix non-negative integers $\underline{\vb*{n}} = (n_1,\dots,n_D)$ and set $n = n_1 + \cdots + n_D$. After choosing the ordered tensor product
\begin{align}
    (A_1^\dagger)^{\otimes n_1} \otimes \cdots \otimes (A_D^\dagger)^{\otimes n_D}  \comma
\end{align}
any gauge-invariant state of this multidegree can be written in the form
\begin{align}
    \ket{Q}
    \coloneqq Q^K_L \left((A_1^\dagger)^{\otimes n_1} \otimes \cdots \otimes (A_D^\dagger)^{\otimes n_D}\right)^L_K \ket{\Omega_{\mathrm{free}}}  \comma
\end{align}
with $Q \in \End{(\C^N)^{\otimes n}}$. By the same Schur-Weyl argument as before, gauge invariance first forces $Q$ to lie in the group algebra $\C[\sym{n}]$. However, once the ordering of the excitations is fixed, only permutations that reshuffle slots carrying the same matrix leave the tensor product unchanged. These permutations form the Young subgroup
\begin{align}
    \sym{\underline{\vb*{n}}} \coloneqq \sym{n_1} \times \cdots \times \sym{n_D} \subset \sym{n} \pt
\end{align}
Therefore, the relevant coefficient algebra is the centralizer algebra~\cite{mattioli2016Permutation}
\begin{align}
    \C[\sym{n}]^{\sym{\underline{\vb*{n}}}} = \enstq{v \in \C[\sym{n}]}{\sigma v = v\sigma, ~\forall \sigma \in \sym{\underline{\vb*{n}}}}  \pt
\end{align}
As explained in Appendix~\ref{app:centralizer_group_algebra}, this centralizer is described by restricting irreducible representations $R \vdash n$ of $\sym{n}$ to $\sym{\underline{\vb*{n}}}$. The resulting basis is labelled by $R$, by irreducible $\sym{\underline{\vb*{n}}}$-types $\vb*{r} = (r_1,\dots,r_D)$ with $r_I \vdash n_I$, and by multiplicity labels $a,b$ associated with the multiplicity space $M^R_{\vb*{r}}$. Choosing the corresponding matrix units $P^{R,\vb*{r}}_{ab} \in \C[\sym{n}]^{\sym{\underline{\vb*{n}}}}$, one defines the restricted Schur polynomials~\cite{bhattacharyya2008Exact, bhattacharyya2008Exacta, koch2024Pedagogical} by contraction with the ordered tensor of matrices
\begin{align}
    \chi^{R,\vb*{r}}_{ab}(Y_1,\dots,Y_D)
    \coloneqq \frac{1}{\dim_R} \frac{n!}{\underline{\vb*{n}}!} (P^{R,\vb*{r}}_{ab})^K_L
    \left(Y_1^{\otimes n_1} \otimes \cdots \otimes Y_D^{\otimes n_D}\right)^L_K  \, ,
\end{align}
with the shorthand $\underline{\vb*{n}}! \coloneqq n_1!\cdots n_D!$. When these polynomials are evaluated at the creation operators, they produce states
\begin{align}\label{eq:restricted_schur_states}
    \ket{R,\vb*{r},a,b}
    \coloneqq \chi^{R,\vb*{r}}_{ab}(A_1^\dagger,\dots,A_D^\dagger)\ket{\Omega_{\mathrm{free}}} \comma
\end{align}
whose adjoints are $\bra{R,\vb*{r},a,b} = \bra{\Omega_{\mathrm{free}}} \chi^{R,\vb*{r}}_{ba}(A_1,\dots,A_D)$. These states form a complete orthogonal basis of singlet states for fixed $(n_1,\dots,n_D)$ (see Appendix \ref{app:ResSchurOrth})
\begin{align}
    \label{eq:orthogonality_restricted_schur}
    \braket{R,\vb*{r},a,b}{S,\vb*{s},c,d} = \alpha^{R}_{\vb*{r}}\delta_{RS} \delta_{\vb*{r}\vb*{s}} \delta_{ac} \delta_{bd}  \comma
\end{align}
with an explicit normalization given by
\begin{align}
    \label{eq:normalization_constant_multi_matrices}
    \alpha^{R}_{\vb*{r}} = \frac{n!}{\dim_R} \frac{\dim_{\vb*{r}}}{\underline{\vb*{n}}!} \alpha_R \qq{with} \dim_{\vb*{r}} \coloneqq \dim_{r_1} \cdots \dim_{r_D} \comma
\end{align}
and where $\alpha_R$ is given in~\eqref{eq:normalization_constant_one_matrix}. Thus, the role played by the central projectors $P_R$ and ordinary Schur polynomials in the one-matrix problem is taken over, in the multi-matrix case, by the centralizer basis $P^{R,\vb*{r}}_{ab}$ and the restricted Schur polynomials.

The number of singlet states at total excitation level $n$ grows significantly faster with $n$ than in the one-matrix case~\cite{bhattacharyya2008Exact, demellokoch2025Structure, demellokoch2026Symmetry}. When $N\geq n$, it can be computed via the centralizer dimension
\begin{align}
    \dim \C[\sym{n}]^{\sym{\underline{\vb*{n}}}} = \sum_{R, \vb*{r}} (\dim M^R_{\vb*{r}})^2 \comma
\end{align}
where the dimensions $\dim M^R_{\vb*{r}}$ of the multiplicity spaces are the generalized Littlewood--Richardson coefficients; they can be evaluated by iterating the Littlewood--Richardson rule, as explained in Appendix~\ref{app:dimension_counting}.

\begin{figure}[tp]
    \centering
    \begin{tikzpicture}[line width=2px]
    \begin{scope}[shift={(0,0)}]  
    \draw[-,line width=1px,blue] (.5,.85) to (1,.85) ;
    \node[blue,rotate=90] at (0.8,.625) {\small $\cdots$};
    \draw[-,line width=1px,blue] (.5,.35) to (1,.35) ;
    \draw[-,line width=1px,red] (.5,.25) to (1,.25);
    \node[red,rotate=90] at (0.8,0.025) {\small $\cdots$};
    \draw[-,line width=1px,red] (.5,-.25) to (1,-.25);
    \draw[-,line width=1px,ForestGreen] (.5,-.35) to (1,-.35);
    \node[ForestGreen,rotate=90] at (0.8,-0.575) {\small $\cdots$};
    \draw[-,line width=1px,ForestGreen] (.5,-.85) to (1,-.85);
    \draw[-,line width=1px,blue] (-.5,.85) to (-1,.85);
    \node[blue,rotate=90] at (-0.8,.625) {\small $\cdots$};
    \draw[-,line width=1px,blue] (-.5,.35) to (-1,.35);
    \draw[-,line width=1px,red] (-.5,.25) to (-1,.25);
    \node[red,rotate=90] at (-0.8,0.025) {\small $\cdots$};
    \draw[-,line width=1px,red] (-.5,-.25) to (-1,-.25);
    \draw[-,line width=1px,ForestGreen] (-.5,-.35) to (-1,-.35);
    \node[ForestGreen,rotate=90] at (-0.8,-0.575) {\small $\cdots$};
    \draw[-,line width=1px,ForestGreen] (-.5,-.85) to (-1,-.85);
    \node[draw, rectangle, minimum width=1.0cm, minimum height=2.0cm, align=center] (PS) {$T$};
    \end{scope}

    \node at (1.5,0) {$=$};
    
    \begin{scope}[shift={(3,0)}]  
    \draw[-,line width=1.5px,blue] (.5,.1) to (1,.1);
    \draw[-,line width=1.5px,red] (.5,0) to (1,0);
    \draw[-,line width=1.5px,ForestGreen] (.5,-.1) to (1,-.1);
    \draw[-,line width=1.5px,blue] (-.5,.1) to (-1,.1);
    \draw[-,line width=1.5px,red] (-.5,0) to (-1,0);
    \draw[-,line width=1.5px,ForestGreen] (-.5,-.1) to (-1,-.1);
    \node[draw, rectangle, minimum width=1.0cm, minimum height=2.0cm, align=center] (PS) {$T$};
    \end{scope}

    \node at (4.5,0) {$=$};
    
    \begin{scope}[shift={(6,0)}]  
    \draw[-,line width=2px] (.5,0) to (1,0);
    \draw[-,line width=2px] (-.5,0) to (-1,0);
    \node[draw, rectangle, minimum width=1.0cm, minimum height=2.0cm, align=center] (PS) {$T$};
    \end{scope}
    \end{tikzpicture}
    \caption{Graphical representation of a tensor $T_I^J \in \End{(\C^N)^{\otimes n}}$. The right legs carry the lower indices $I$, while the left legs carry the upper indices $J$. Different colors represent different matrix species.}
    \label{fig:tensor_diagram_multi}
\end{figure}
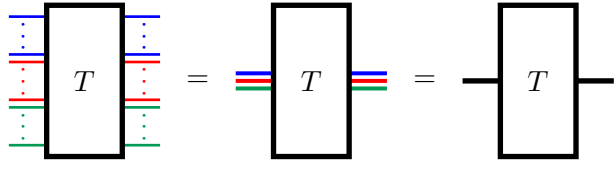

The graphical representation introduced in the one-matrix case can be adapted in a straightforward manner to the multi-matrix case. The only difference is that the legs of the tensors now carry an additional color label that corresponds to the matrix species. This color label is represented by a different color for the leg in the diagram. For instance, a leg carrying an index of the first matrix is represented by a blue line, while a leg carrying an index of a second matrix is represented by a red line, and so on. This is visualized in Figure~\ref{fig:tensor_diagram_multi} for a tensor with three matrix species. The rules for index contraction and trace evaluation remain unchanged, as they are determined by the underlying group structure and not by the specific matrix species. A single tensor may carry legs of several different species, but the species of each leg is fixed, and only legs of the same species may be contracted. The graphical representation thus continues to provide an intuitive and efficient way to visualize and compute with the tensors that arise in the multi-matrix case. 

\section{Computation of \texorpdfstring{$H$}{H} in the gauge-invariant basis}
\label{sec:computation_matrix_elements}

In Section~\ref{sec:singlet_states_multi_matrices} we reviewed how to construct a complete, orthogonal basis of gauge-invariant states for both one- and multi-matrix models. Each basis state belongs to a definite excitation sector labeled by $\underline{\vb*{n}} = (n_1,\dots,n_D)$. Gauge-invariant observables built from a finite number of creation and annihilation operators are sparse in this basis: a given interaction term couples only a few excitation sectors. In this section, we develop a systematic method to compute the Hamiltonian matrix elements for all non-vanishing excitation sectors up to a chosen cutoff $\Lambda$.

We present an algorithm that achieves this by exploiting the representation theory of the symmetric group. Its core idea is to express each matrix element as a sum over cosets of a suitable subgroup and to evaluate the resulting character sums by Fourier analysis on the center $Z(\C[\sym{n}])$ (one-matrix case) or on the centralizer algebra $\C[\sym{n}]^{\sym{\underline{\vb*{n}}}}$ (multi-matrix case). In the one-matrix problem, the relevant characters are the ordinary irreducible characters of $\sym{n}$, which can be computed efficiently via the Murnaghan--Nakayama rule. In the multi-matrix case, one needs the restricted characters of the symmetric group; their evaluation is substantially more involved and is deferred to future work.

The algorithm is exact: every Hamiltonian matrix element $H_{RS} = \mel{R}{H}{S}$ and the Gram matrix are obtained as a closed-form polynomial in $N$, the rank of the gauge group, with coefficients that depend only on the irreducible-representation labels of the basis states. Once the Hamiltonian and any observables of interest have been assembled in this gauge-invariant matrix form, standard numerical techniques -- exact diagonalization, time evolution, or thermal-state computations -- can be applied at any fixed truncation cutoff $\Lambda$; the only approximation is the finite truncation of the Hilbert space. 

We present the algorithm by working it through the physically relevant quartic interactions in the Hamiltonians~\eqref{eq:hamiltonian_form_1} (one matrix) and~\eqref{eq:hamiltonian_multi} (several matrices). The one-matrix case is developed in full detail and serves as a blueprint. We add complementary information where this example case differs from a general one-matrix gauge-invariant observable. The multi-matrix case is presented more concisely, as it follows the same conceptual steps; we highlight the points where the extension is nontrivial, most notably the replacement of ordinary symmetric-group characters by restricted characters.

In the one matrix case, once the parameters of the simulation are specified, we assemble the orthonormalized matrix $\hat{H}_{RS}=\tfrac{H_{RS}}{\sqrt{\alpha_R\alpha_S}}$ up to some cutoff and pass it to the eigensolver to obtain gauge invariant eigenstates. These operations are inexpensive, since all the matrices involved are sparse. In the code, we use the sparse Hermitian eigensolver from SciPy, which wraps the ARPACK routines SSEUPD and DSEUPD; these implement the Implicitly Restarted Lanczos Method to compute eigenvalues and eigenvectors.

The algorithm applies to any set of gauge-invariant observables. Once the corresponding matrices have been assembled up to a given cutoff $\Lambda$, physical quantities such as energy levels, entropy, and two-point correlation functions reduce to standard sparse linear algebra.

For validation, we have implemented the one-matrix algorithm in a publicly available library; the implementation is described in the subsequent section.

\subsection{The one-matrix case}
\label{sec:one_matrix_case}

We now walk through the algorithm for the single-matrix Hamiltonian
\begin{align}
    H = \frac{1}{2} \tr( P^2 + m^2 X^2) + \frac{g^2}{2} \tr(X^4)
     = \sum_{\ell,k=1}^N m(N^k_\ell + \tfrac{1}{2}) + \frac{g^2}{2} \tr(X^4) \pt
\end{align}
Our goal is to evaluate the matrix elements $\mel{R}{H}{S}$ for all pairs of Young diagrams $(R,S)$ with respectively $n$ and $n'$ boxes and no more than $N$ rows: $\ell(R),\ell(S)\leq N$. Remember that the states $\ket{R}$ are not normalized. Decompose $H = H_{\mathrm{free}} + H_{\mathrm{int}}$ into free and interacting parts. Since each basis state $\ket{R}$ has a definite excitation number $n$, the free Hamiltonian is diagonal:
\begin{align}
    \mel{R}{H_\mathrm{free}}{S} = \alpha_R m (n + N^2/2)\delta_{RS}  \comma
\end{align}
where $\alpha_R$ is the normalization constant~\eqref{eq:normalization_constant_one_matrix}. 

We now describe the procedure for the quartic term $\tr(X^4)$, which is however general and applies to any gauge-invariant observable; we point out the relevant modifications as we go.

\paragraph{Antinormal ordering.}

The first step is to antinormally order $\tr(X^4)$, shifting all creation operators to the right of the annihilation operators. To this end, we introduce a compact notation for traces of products of operators~\cite{koch2024Pedagogical}. For a permutation $\sigma \in \sym{n}$ and matrices $Y_1,\dots,Y_n$, define
\begin{align}
    \ttr{\sigma}{Y_1 \otimes\cdots \otimes Y_n}
    \coloneqq \sigma^L_K (Y_1 \otimes \cdots \otimes Y_n)_L^K 
    = (Y_1)_{\ell_1}^{\ell_{\sigma_1}} \cdots (Y_n)_{\ell_n}^{\ell_{\sigma_n}}  \comma
\end{align}
where $\sigma^L_K \coloneqq \sigma_{k_1\cdots k_n}^{\ell_1\cdots\ell_n} = \delta_{k_1}^{\ell_{\sigma_1}}\cdots\delta_{k_n}^{\ell_{\sigma_n}}$ is the permutation tensor acting on multi-indices. After antinormal ordering (see Appendix~\ref{sec:decomposition_power_trace} for the explicit decomposition), $\tr(X^4)$ decomposes into a linear combination of 13 terms of the form $\ttr{\sigma}{B}$, where
\begin{align}
    \label{eq:B_operators}
    B\in \left\{\mathbb{I}, A^{\otimes 2}, A \otimes A^\dagger, (A^\dagger)^{\otimes 2}, A^{\otimes 4}, A^{\otimes 3} \otimes A^\dagger,  A^{\otimes 2} \otimes (A^\dagger)^{\otimes 2}, A \otimes (A^\dagger)^{\otimes 3}, (A^\dagger)^{\otimes 4}\right\} \comma
\end{align}
and $\sigma \in \sym{p}$ acts on the $p$ tensor factors of $B$ ($p=0,2,$ or $4$). For a general gauge-invariant observable $\mathcal{O}$, the antinormal decomposition involves operator products whose length does not exceed the maximum order of the traces appearing in $\mathcal{O}$.

\paragraph{Reduction to double cosets.}

Only 12 of the 13 terms are nontrivial; the constant term ($B=\mathbb{I}$) gives a diagonal contribution thanks to the orthogonality of the basis. For each of the remaining terms we need to compute
\begin{align}
    \label{eq:mel_R_S_sigma}
    \mel{R}{\ttr{\sigma}{B}}{S}  \pt
\end{align}
A necessary condition for this matrix element to be non-zero is excitation-number conservation:
\begin{align}
    \label{eq:non_vanishing_condition}
    \mel{R}{\ttr{\sigma}{B}}{S} \neq 0 \implies n + n_A = n' + n_{A^\dagger}  \comma
\end{align}
where $n_A$ and $n_{A^\dagger}$ are the numbers of annihilation and creation operators in $B$. This selection rule provides a useful first filter. In the case of $\tr(X^4)$, only the excitation sectors $(n,n')$ such that $\abs{n-n'} \in \{0,2,4\}$ contribute.

Following the orthogonality proof, we express the Schur polynomials via Young projectors and expand the matrix element. Introduce multi-index notation for the additional operator slots:
\begin{align}
     I_A = (i_{n+1}, \dots, i_{n+n_A}) ,\qquad I_{A^\dagger} = (i_{n'+1}, \dots, i_{n'+n_{A^\dagger}}) \comma
\end{align}
and analogously for $J$, $K$, and $L$. Evaluating the vacuum expectation value with Wick's theorem (Appendix~\ref{app:wick_formula}) yields
\begin{align}
    \dim_R  \dim_S \mel{R}{\ttr{\sigma}{B}}{S} 
    = \sum_{\xi \in \sym{n+n_A}} (P_R)_{J}^{I} (P_S)^{K}_{L} \sigma_{J_A L_{A^\dagger} }^{I_A K_{A^\dagger} }  (\xi^{-1})^{LL_{A^\dagger}}_{II_{A}}  \xi_{KK_{A^\dagger}}^{JJ_{A}} \pt
\end{align}
The summand, which we denote by $\mathcal{W}(\xi; R, S, \sigma)$, has the graphical representation
\begin{equation}
    \label{eq:diagram_woven_contraction}
    \hbox{\begin{tikzpicture}[line width=2px,baseline={([yshift=-.5ex]current bounding box.center)}]
    \node[draw, rectangle, minimum width=1.0cm, minimum height=2.0cm, align=center] (PR) {$P_R$};
    \node [below of = PR, node distance= 4cm, draw, rectangle, minimum width=1.0cm, minimum height=2.0cm, align=center] (PS) {$P_S$};
    \draw [-] (.5,0) to[out=0,in=180] (3.5,-.5);
    \draw [-] (-.5,0) to[out=180,in=180] (-.5,-5.5) to[out=0,in=180] (4.5,-5.5) to[out=0,in=0,distance=1cm] (4.5,-4.5);
    \draw [-] (-.5,-4) to[out=180,in=180] (-0.5,-2) to[out=0,in=180] (4.5,-2) to[out=0,in=0]  (4.5,-.5);
    \draw [-] (4.5,-3.5) to[out=0,in=180] (6,-2.5);
    \draw [-] (4.5,.5) to[out=0,in=180] (6,-1.5);
    \draw [-] (.5,-4) to[out=0,in=180] (3.5,-4.5);
    \draw [-] (3.5,.5) to[out=180,in=180] (3.5,1.5) to[out=0,in=180] (7,1.5) to[out=0,in=0]  (7,-2.5);
    \draw [-,white, line width=4] (3.5,-3.5) to[out=180,in=180] (3.5,-6) to[out=0,in=180] (7,-6) to[out=0,in=0]  (7,-1.5);
    \draw [-] (3.5,-3.5) to[out=180,in=180] (3.5,-6) to[out=0,in=180] (7,-6) to[out=0,in=0]  (7,-1.5);
    \node [right of = PR, node distance= 4cm, draw, rectangle, minimum width=1.0cm, minimum height=2.0cm, align=center] (XI) {$\xi$};
    \node [right of = PS, node distance= 4cm, draw, rectangle, minimum width=1.0cm, minimum height=2.0cm, align=center] (XI-1) {$\xi^{-1}$};
    \node [draw, rectangle, minimum width=1.0cm, minimum height=2.0cm, align=center] at (6.5,-2) (s) {$\sigma$};
    \end{tikzpicture}} \pt
\end{equation}
We observe that $\mathcal{W}(\xi; R, S, \sigma)$ is constant on double cosets. Indeed, because $P_R$ and $P_S$ belong to the center of $\C[\sym{n}]$ (equation~\eqref{eq:young_projectors_center}), which for $P_R$ is graphically represented as
\begin{align}
    \begin{tikzpicture}[line width=2px,baseline={([yshift=-.5ex]current bounding box.center)}]
        \node[draw, rectangle, minimum width=1.0cm, minimum height=2.0cm, align=center] (PR) {$P_R$};
        \node[left of = PR, node distance= 2cm, draw, rectangle, minimum width=1.0cm, minimum height=1.0cm, align=center] (nu) {$\nu$};
        \node[right of = PR, node distance= 2cm, draw, rectangle, minimum width=1.0cm, minimum height=1.0cm, align=center] (nu-1) {$\nu^{-1}$};
        \draw[-] (nu) -- (PR);
        \draw[-] (PR) -- (nu-1);
        \draw[-] (-3,0) -- (nu);
        \draw[-] (3,0) -- (nu-1);
        \node[right of = nu-1, node distance= 1.5cm] (eq) {$=$};
        \node[right of = eq, node distance= 1.5cm, draw, rectangle, minimum width=1.0cm, minimum height=2.0cm, align=center] (PR2) {$P_R$};
        \draw [-] (4,0) -- (PR2);
        \draw [-] (6,0) -- (PR2);
    \end{tikzpicture}\, \comma
\end{align}
left multiplication of $\xi$ by elements of $\sym{n}$ and right multiplication by elements of $\sym{n'}$ leave $\mathcal{W}$ unchanged. Hence, $\mathcal{W}(\xi; R, S, \sigma)$ is constant along the double cosets
\begin{align}
    \sym{n} \xi \sym{n'} = \enstq{\nu^{-1} \xi \mu}{\nu \in \sym{n}, \mu \in \sym{n'}} \in \sym{n} \backslash  \sym{n+n_A} / \sym{n'} \comma
\end{align}
where $\sym{n}$ and $\sym{n'}$ are understood as subgroups of $\sym{n+n_A}$ via the natural embeddings. This reduces the sum over $(n+n_A)!$ permutations to a sum over double coset representatives $\zeta$, each weighted by its double coset size $\#_\zeta$:
\begin{align}
    \label{eq:double_coset_decomposition}
    \mel{R}{\ttr{\sigma}{B}}{S} = \frac{1}{\dim_R \dim_S }
    \sum_{\zeta} \#_\zeta  \mathcal{W}(\zeta; R, S, \sigma)  \pt 
\end{align}
Representatives and double coset sizes are computed efficiently with computer algebra systems such as GAP~\cite{GAP}.

\paragraph{Size of the double coset space.}

In the decomposition of $\tr(X^4)$, the largest number of double cosets occurs when $n=n'$ and $n_A=2$, and is bounded:
\begin{align}
    \label{eq:bound_double_cosets}
    \abs{\sym{n} \backslash \sym{n+2} / \sym{n}} \leq 7 \comma
\end{align}
with equality for all $n \geq 4$.

More generally, an interaction term may involve double cosets of the form $\sym{n} \backslash \sym{p} / \sym{n'}$ with $n,n'\leq p=n+n_A$. These are a special case of double cosets of Young subgroups $\sym{\underline{\vb*{n}}} \backslash \sym{p} / \sym{\underline{\vb*{m}}}$, where $\underline{\vb*{n}} = (n_1, \dots, n_{D_1})$ and $\underline{\vb*{m}} = (m_1,\dots, m_{D_2})$ are tuples of nonnegative integers satisfying
\begin{align}
    \sum_{k=1}^{D_1} n_k = \sum_{k=1}^{D_2} m_k = p \comma
\end{align}
and the Young subgroup is the direct product
\begin{align}
    \sym{\underline{\vb*{n}}} = \sym{n_1} \times \cdots \times  \sym{n_{D_1}} \subset \sym{p} \pt
\end{align}
The number of double cosets in $\sym{\underline{\vb*{n}}} \backslash \sym{p} / \sym{\underline{\vb*{m}}}$ is known to equal the number of nonnegative integer matrices with row sums $\underline{\vb*{n}}$ and column sums $\underline{\vb*{m}}$. Estimates for this quantity can be found in the literature; see for instance~\cite{barvinok2009Asymptotic, barvinok2012Asymptotic}.

\paragraph{Expansion of double coset contributions.}

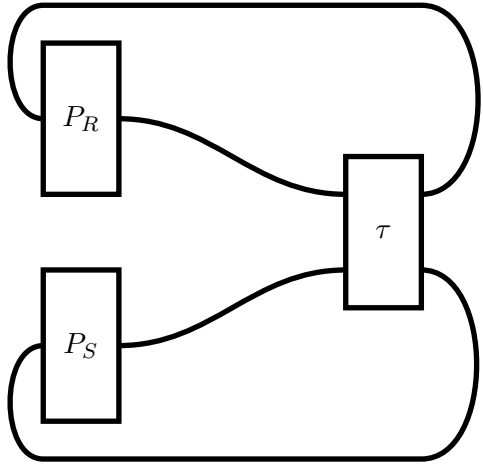
\begin{figure}
    \centering
    \begin{tikzpicture}[line width=2px]
        \node[draw, rectangle, minimum width=1.0cm, minimum height=2.0cm, align=center] (PR) {$P_R$};
        \node [below of = PR, node distance= 3cm, draw, rectangle, minimum width=1.0cm, minimum height=2.0cm, align=center] (PS) {$P_S$};
        \node[draw, rectangle, minimum width=1.0cm, minimum height=2.0cm, align=center] (tau) at (4,-1.5) {$\tau$};
        \draw [-] (.5,0) to[out=0,in=180] (3.5,-1);
        \draw [-] (.5,-3) to[out=0,in=180] (3.5,-2);
        \draw [-] (-.5,0) to[out=180,in=180] (-.5,1.5) to[out=0,in=180] (4.5,1.5) to[out=0,in=0,distance=1cm] (4.5,-1);
        \draw [-] (-.5,-3) to[out=180,in=180] (-0.5,-4.5) to[out=0,in=180] (4.5,-4.5) to[out=0,in=0]  (4.5,-2);
    \end{tikzpicture}
    \caption{Graphical representation of the contraction $\tr((P_R \otimes P_S) \cdot \tau)$.}
    \label{fig:contraction_diagram_2_1}
\end{figure}

\begin{figure}
    \centering
   \begin{tikzpicture}[line width=2px]
            \node at (-2,-2.4) {$N^k$};
            \node[draw,rectangle, minimum width=1.0cm, minimum height=2.0cm, align=center] at (-.5,-2.5) {$\tau$};
            \draw [-] (-1.5,-3) to[out=0,in=180] (-1,-3);
            \draw [-] (-1.5,-2) to[out=0,in=180] (-1,-2);
            \draw [-] (0,-3) to[out=0,in=180] (.5,-3);
            \draw [-] (0,-2) to[out=0,in=180] (.5,-2);
            \node at (1.5,-2.5) {=};
            \draw [-] (2,-.5) to[out=0,in=180] (3.5,-.5);
            \draw [-] (9.5,-.5) to[out=180,in=0] (7.5,-4.5) to (4.5,-4.5);
            \draw [-] (4.5,-3.5) to[out=0,in=180] (6,-2.5);
            \draw [-] (4.5,.5) to[out=0,in=180] (6,-1.5);
            \draw [-] (2,-4.5) to[out=0,in=180] (3.5,-4.5);
            \draw [-] (3.5,.5) to[out=180,in=180] (3.5,1.5) to[out=0,in=180] (7,1.5) to[out=0,in=0]  (7,-2.5);
            \draw [-,white, line width=4] (9.5,-4.5) to[out=180,in=0] (7.5,-.5) to (4.5,-.5);
            \draw [-] (9.5,-4.5) to[out=180,in=0] (7.5,-.5) to (4.5,-.5);
            \draw [-,white, line width=4] (3.5,-3.5) to[out=180,in=180] (3.5,-5.5) to[out=0,in=180] (7,-5.5) to[out=0,in=0]  (7,-1.5);
            \draw [-] (3.5,-3.5) to[out=180,in=180] (3.5,-5.5) to[out=0,in=180] (7,-5.5) to[out=0,in=0]  (7,-1.5);
            \node [draw, rectangle, minimum width=1.0cm, minimum height=2.0cm, align=center] at (4,0) (XI) {$\zeta$};
            \node [draw, rectangle, minimum width=1.0cm, minimum height=2.0cm, align=center] at (4,-4) (XI-1) {$\zeta^{-1}$};
            \node [draw, rectangle, minimum width=1.0cm, minimum height=2.0cm, align=center] at (6.5,-2) (s) {$\sigma$};
        \end{tikzpicture}    
    \caption{Graphical representation the permutation $\tau$ and loop count $k$. It shows the contraction that leads to $N^k\tau$. Each closed loop in the contraction contributes a factor of $N$.}
    \label{fig:contraction_diagram_2_2}
\end{figure}
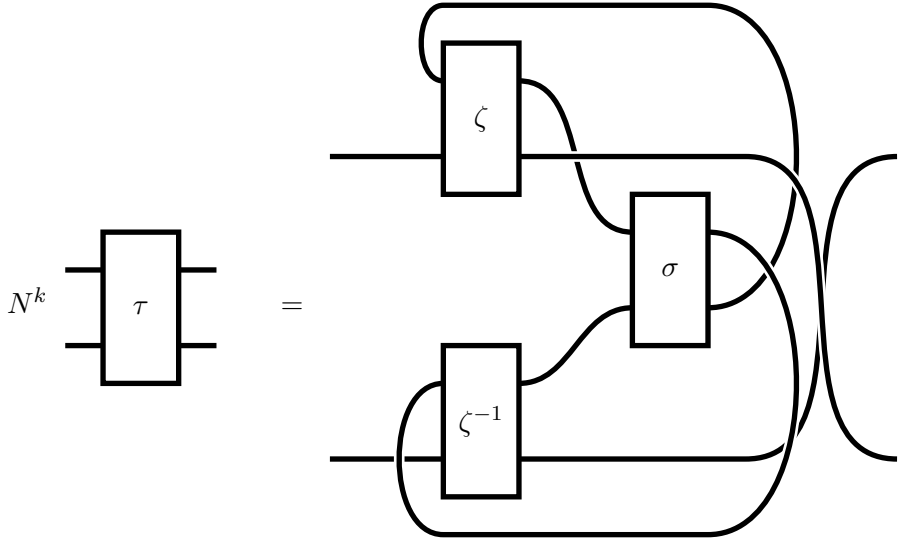

To compute $\mathcal{W}(\zeta; R, S, \sigma)$ for each double coset representative, we first contract the permutation tensors $\zeta$, $\zeta^{-1}$, and $\sigma$. Graphically, the corresponding diagram~\eqref{eq:diagram_woven_contraction} is flattened by composing permutations along each closed path. The result is a permutation $\tau \equiv \tau_{\zeta,\sigma,n,n'} \in \sym{n+n'}$ together with a loop count $k \equiv k_{\zeta,\sigma,n,n'} \in \N$:
\begin{align}
    \mathcal{W}(\zeta; R, S, \sigma) = N^{k} \tr((P_R \otimes P_S) \cdot \tau) \comma
\end{align}
where $P_R \otimes P_S \in \C[\sym{n}] \otimes \C[\sym{n'}] \subset \C[\sym{n+n'}]$ acts block-diagonally. The diagrammatic representation of the contraction in the trace is shown in Figure~\ref{fig:contraction_diagram_2_1} and the contraction that defines $N^k \tau$ in Figure~\ref{fig:contraction_diagram_2_2}.

\begin{figure}
    \centering
    \begin{tikzpicture}[line width=2px]
    \begin{scope}[shift={(0,0)}]
        \node[draw, rectangle, minimum width=1.0cm, minimum height=1.0cm, align=center] at (-1.5,0) (rho) {$\rho^{-1}$};
        \node[draw, rectangle, minimum width=1.0cm, minimum height=2.0cm, align=center] (alpha) {$\rho_\alpha$};
        \node[draw, rectangle, minimum width=1.0cm, minimum height=1.0cm, align=center] at (1.5,0) (rho-1) {$\rho$};
        \node [draw, rectangle, minimum width=1.0cm, minimum height=2.0cm, align=center] at (0,-3) (PS) {$P_S$};
        \node[draw, rectangle, minimum width=1.0cm, minimum height=2.0cm, align=center] (tau) at (4,-1.5) {$\tau$};
        \draw [-] (rho) to (alpha) to (rho-1) to[out=0,in=180] (3.5,-1);
        \draw [-] (.5,-3) to[out=0,in=180] (3.5,-2);
        \draw [-] (rho) to[out=180,in=180] (-2,1.5) to[out=0,in=180] (4.5,1.5) to[out=0,in=0,distance=1cm] (4.5,-1);
        \draw [-] (-.5,-3) to[out=180,in=180] (-0.5,-4.5) to[out=0,in=180] (4.5,-4.5) to[out=0,in=0]  (4.5,-2);
    \end{scope}
    \begin{scope}[shift={(1,-6.5)}]
        \node[draw, rectangle, minimum width=1.0cm, minimum height=1.0cm, align=center] at (-1.5,0) (omega-1) {$\omega^{-1}$};
        \node[draw, rectangle, minimum width=1.0cm, minimum height=2.0cm, align=center] (alpha) {$\rho_\alpha$};
        \node[draw, rectangle, minimum width=1.0cm, minimum height=1.0cm, align=center] at (1.5,0) (omega) {$\omega$};
        \node[draw, rectangle, minimum width=1.0cm, minimum height=2.0cm, align=center] at (0,-3) (PS) {$P_S$};
        \node[draw, rectangle, minimum width=1.0cm, minimum height=1.0cm, align=center] at (3,-.9) (rho-omega) {$\nu$};
        \node[draw, rectangle, minimum width=1.0cm, minimum height=1.0cm, align=center] at (6,-.9) (rho-omega-1) {$\nu^{-1}$};
        \node[draw, rectangle, minimum width=1.0cm, minimum height=2.0cm, align=center] (tau) at (4.5,-1.5) {$\tau$};
        \node[draw, rectangle, minimum width=1.0cm, minimum height=1.0cm, align=center] at (3,-2.1) (mu) {$\mu$};
        \node[draw, rectangle, minimum width=1.0cm, minimum height=1.0cm, align=center] at (6,-2.1) (mu-1) {$\mu^{-1}$};
        \draw [-] (omega-1) to (alpha) to (omega) to[out=0,in=180] (rho-omega) to[out=0,in=180] (4,-1);
        \draw [-] (.5,-3) to[out=0,in=180] (mu) to[out=0,in=180] (4,-2);
        \draw [-] (omega-1) to[out=180,in=180] (-2,1.5) to[out=0,in=180] (6.5,1.5) to[out=0,in=0,distance=1cm] (rho-omega-1) to[out=180,in=0] (5,-1);
        \draw [-] (-.5,-3) to[out=180,in=180] (-0.5,-4.5) to[out=0,in=180] (6.5,-4.5) to[out=0,in=0] (mu-1) to[out=180,in=0] (5,-2);
    \end{scope}
    \begin{scope}[shift={(1,-13)}]
        \node[draw, rectangle, minimum width=1.0cm, minimum height=1.0cm, align=center] at (-1.5,0) (rho) {$\omega^{-1}$};
        \node[draw, rectangle, minimum width=1.0cm, minimum height=2.0cm, align=center] (alpha) {$\rho_\alpha$};
        \node[draw, rectangle, minimum width=1.0cm, minimum height=1.0cm, align=center] at (1.5,0) (rho-1) {$\omega$};
        \node [draw, rectangle, minimum width=1.0cm, minimum height=2.0cm, align=center] at (0,-3) (PS) {$P_S$};
        \node[draw, rectangle, minimum width=1.0cm, minimum height=2.0cm, align=center] (tau) at (4,-1.5) {$\tau$};
        \draw [-] (rho) to (alpha) to (rho-1) to[out=0,in=180] (3.5,-1);
        \draw [-] (.5,-3) to[out=0,in=180] (3.5,-2);
        \draw [-] (rho) to[out=180,in=180] (-2,1.5) to[out=0,in=180] (4.5,1.5) to[out=0,in=0,distance=1cm] (4.5,-1);
        \draw [-] (-.5,-3) to[out=180,in=180] (-0.5,-4.5) to[out=0,in=180] (4.5,-4.5) to[out=0,in=0]  (4.5,-2);
    \end{scope}
    \begin{scope}[shift={(7,-11.5)}]
        \node [draw, rectangle, minimum width=1.0cm, minimum height=2.0cm, align=center] (PS) at (2,-3)  {$P_S$};
        \node[draw, rectangle, minimum width=1.0cm, minimum height=2.0cm, align=center] (rho) at (4,-3) {$\tilde{\rho}_{\omega,\alpha,\tau}$};
        \draw [-] (PS) to (rho) to[out=0,in=0] (4.5,-4.5) to (1.5,-4.5) to[out=180,in=180] (PS);
    \end{scope}
    \node at (-2,-8) {$=$};
    \node at (-2,-14.5) {$=$};
    \node at (7,-14.5) {$=$};

    \end{tikzpicture}
\caption{Graphical representation of the essential steps in the factorization of $\tr((P_R \otimes P_S) \cdot \tau)$ into a sum over cosets of $G^1_{\tau,n}$. Every element $\rho \in \sym{n}$ can be written as $\rho = \omega \nu$ with $\omega$ a coset representative and $\nu \in G^1_{\tau,n}$. Using the centrality of $P_S$ and the definition of $G^1_{\tau,n}$, one can move $\nu$ through the contraction with $\tau$, which leaves the trace invariant. The sum over $\rho$ then becomes a sum over $\nu$ that gives a factor of $|G^1_{\tau,n}|$ and only the sum over $\omega$ remains explicitly. The contraction of $\rho_\alpha$ with $\tau$ produces $\tilde{\rho}_{\omega,\alpha,\tau} = N^{c_{\omega,\alpha,\tau}} \cdot \rho_{\omega,\alpha,\tau}$, where $\rho_{\omega,\alpha,\tau} \in \sym{n'}$, and $c_{\omega,\alpha,\tau}$ counts the loops in the contraction, which gives rise to the factor $N^{c_{\omega,\alpha,\tau}}$.}
    \label{fig:factorization}   
\end{figure}
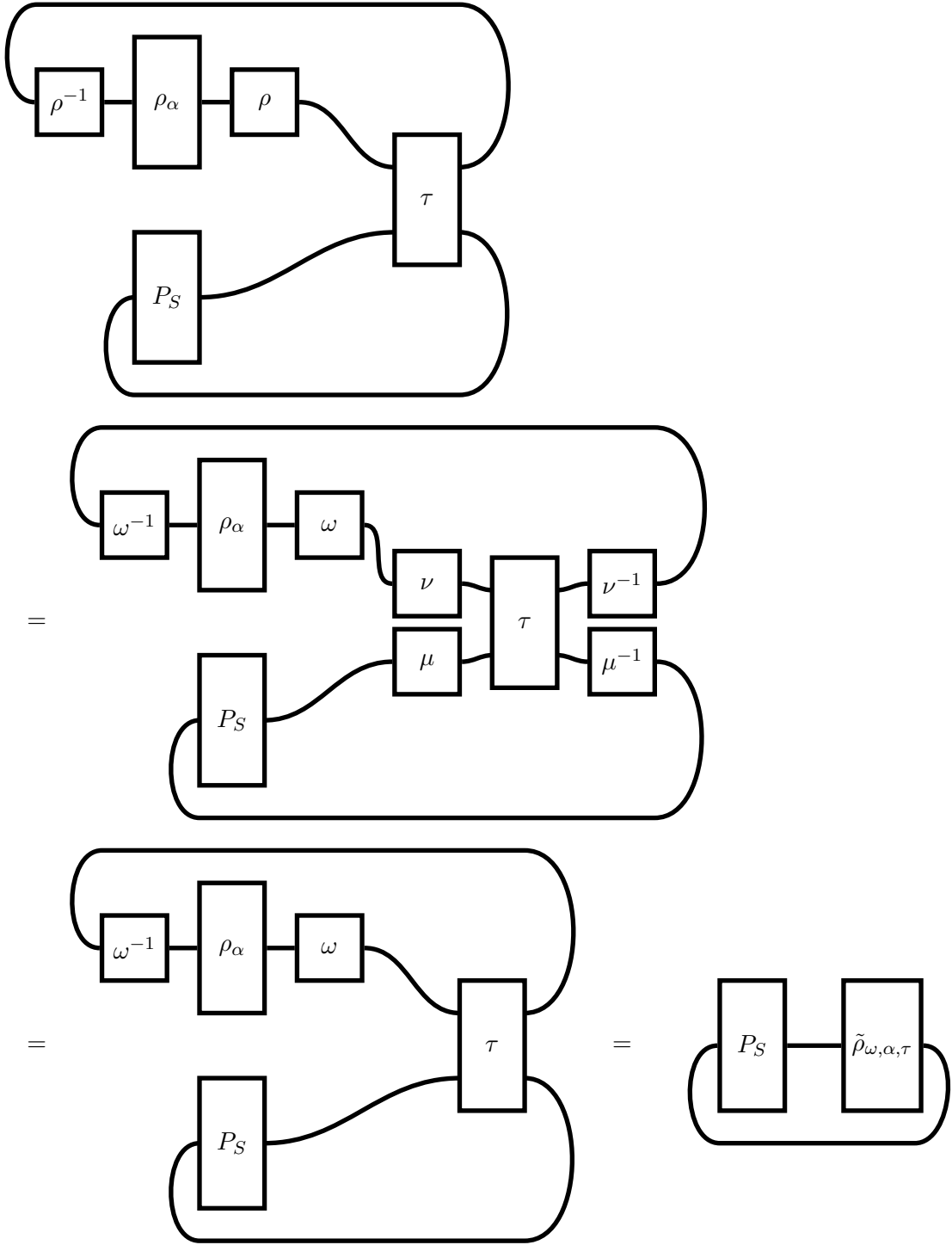

To evaluate $\tr((P_R \otimes P_S) \cdot \tau)$, we expand $P_R$ in the basis of class sums of $\C[\sym{n}]$. Introduce the class sum
\begin{align}
    [\alpha] \coloneqq \sum_{\rho\in \mathcal{C}_\alpha} \rho \in Z(\C[\sym{n}]) \comma
\end{align}
so that the character expansion of $P_R$ reads
\begin{align}
    \label{eq:young_projectors_expansion}
    P_R = \frac{\dim_R}{n!} \sum_{\rho \in \sym{n}} \chi_R (\rho) \rho
    = \frac{\dim_R}{n!}\sum_{\alpha \vdash n} \chi_R(\alpha) [\alpha] \comma 
\end{align} 
where $\chi_R(\alpha)$ denotes the character of $R$ on the conjugacy class $\mathcal{C}_{\alpha}$. Expanding each class sum as the orbit of a fixed representative $\rho_\alpha \in \mathcal{C}_\alpha$ under conjugation gives
\begin{align}
    P_R = \frac{\dim_R}{n!}\sum_{\alpha \vdash n} \chi_R(\alpha) \frac{|\mathcal{C}_\alpha|}{n!} \sum_{\rho\in\sym{n}} \rho^{-1}\rho_\alpha \rho \pt
\end{align}
Substituting this expansion into the trace yields a sum over all $\rho \in \sym{n}$,
\begin{align}
    \tr((P_R \otimes P_S) \cdot \tau) = \frac{\dim_R}{(n!)^2}\sum_{\alpha \vdash n} \chi_R(\alpha) |\mathcal{C}_\alpha| \sum_{\rho\in\sym{n}} \mathcal{W}'(\rho; \alpha, S, \tau)\comma 
\end{align}
where 
\begin{align}
    \label{eq:woven_contraction_prime}
    \mathcal{W}'(\rho; \alpha, S, \tau) \coloneqq \tr(((\rho^{-1}\rho_\alpha \rho) \otimes P_S) \cdot \tau) \pt
\end{align}
A direct evaluation of the sum over $\sym{n}$ is prohibitive for large $n$. However, $\mathcal{W}'$ is invariant under right multiplication of $\rho$ by elements of a suitable subgroup. To identify this subgroup, consider the stabilizer of $\tau$ in $\sym{n} \times \sym{n'}$,
\begin{align}
    C_{\sym{n} \times \sym{n'}}(\tau)
    \coloneqq \enstq{(\nu, \mu) \in \sym{n} \times \sym{n'}}{(\nu \otimes \mu) \cdot \tau = \tau \cdot (\nu \otimes \mu)} \comma
\end{align}
and let $G^1_{\tau,n} \subset \sym{n}$ be its projection onto the first factor:
\begin{align}
    G^1_{\tau,n} \coloneqq \enstq{\nu \in \sym{n}}{\exists \mu \in \sym{n'},~ (\nu, \mu) \in C_{\sym{n} \times \sym{n'}}(\tau)} \pt
\end{align}
The subgroup $G^1_{\tau,n}$ is the projection onto the first component of a Cartesian product of wreath products (see Appendix~\ref{app:stabilizer_subgroup}). For any $\nu \in G^1_{\tau,n}$, pick $\mu \in \sym{n'}$ with $(\nu \otimes \mu)\,\tau = \tau\,(\nu \otimes \mu)$. Using the centrality of $P_S$ in $\C[\sym{n'}]$,
\begin{align}
    \notag
    \mathcal{W}'(\rho\nu; \alpha, S, \tau)
    &= \tr(((\nu^{-1}\rho^{-1}\rho_\alpha \rho \nu) \otimes P_S) \cdot \tau) \\
    & = \tr(((\rho^{-1}\rho_\alpha \rho) \otimes P_S) \cdot (\nu \otimes \mu)\,\tau\,(\nu^{-1} \otimes \mu^{-1}))
    = \mathcal{W}'(\rho; \alpha, S, \tau) \pt
\end{align}
Thus $\mathcal{W}'$ is constant on left cosets of $G^1_{\tau,n}$. Decomposing $\sym{n}$ into left cosets $\sym{n} = \bigcup_\omega \omega G^1_{\tau,n}$, the sum over $\sym{n}$ factorizes:
\begin{align}
    \label{eq:coset_reduction}
    \tr((P_R \otimes P_S) \cdot \tau)
    = \frac{\dim_R |G^1_{\tau,n} |}{(n!)^2}\sum_{\alpha \vdash n} \sum_\omega \chi_R(\alpha) |\mathcal{C}_\alpha| \tr(P_S \cdot \tilde{\rho}_{\omega,\alpha,\tau}) \comma
\end{align}
where $\tilde{\rho} \equiv \tilde{\rho}_{\omega,\alpha,\tau} = N^{c_{\omega,\alpha,\tau}} \cdot \rho_{\omega,\alpha,\tau}$, and $\rho_{\omega,\alpha,\tau} \in \sym{n'}$ together with the loop count $c_{\omega,\alpha,\tau} \in \N$ are obtained by contracting the class representative $\rho_\alpha$ (conjugated by $\omega$) with $\tau$. The essential steps of this calculation are shown diagrammatically in Figure~\ref{fig:factorization}. Again, the representatives $\omega$ and the stabilizer size $|G^1_{\tau,n}|$ can be efficiently precomputed with GAP.

\paragraph{Evaluation of $\tr(P_S\rho)$.}

It remains to evaluate $\tr(P_S \rho)$ for a given $\rho\in \sym{n'}$. To this aim, we expand $P_S$ in the class sum basis of $\C[\sym{n'}]$, analogously to the expansion~\eqref{eq:young_projectors_expansion} of $P_R$:
\begin{align}
    P_S = \frac{\dim_S}{n'!}\sum_{\beta \vdash n'} \chi_S(\beta) [\beta] \pt
\end{align}
Since each $[\beta]$ is central in $\C[\sym{n'}]$, we have $\tr([\beta]\rho) = \tr([\beta]\,\kappa\rho\kappa^{-1})$ for any $\kappa\in\sym{n'}$. In what follows we slightly abuse notation: when a permutation appears as a subscript of a class sum or a class size (for instance $[\rho]$ or $|\mathcal{C}_\rho|$), it stands for its cycle type. With this convention, averaging over $\kappa$ replaces $\rho$ by its conjugacy class sum:
\begin{align}
    \tr(P_S \rho) 
    = \frac{\dim_S}{n'!}\sum_{\beta \vdash n'} \chi_S(\beta) \, \frac{1}{n'!}\sum_{\kappa\in\sym{n'}} \tr([\beta]  \kappa \rho \kappa^{-1})
    = \frac{\dim_S}{n'!} \sum_{\beta \vdash n'} \frac{\chi_S(\beta)}{|\mathcal{C}_\rho|} \tr([\beta] [\rho]) \pt
\end{align}
The product of two class sums expands as (see Appendix~\ref{app:center_group_algebra})
\begin{equation}
    [\beta] [\rho] = \sum_{\gamma \vdash n'} C_{\beta\rho}^{\gamma} [\gamma] \qq{where} C_{\beta\rho}^{\gamma} = \frac{|\mathcal{C}_\beta||\mathcal{C}_\rho|}{n'!} \sum_{\theta \vdash n'} \frac{\chi_\theta(\beta)\chi_\theta(\rho)\chi_\theta(\gamma)}{\dim_\theta} \comma
\end{equation} 
with $\theta$ running over all irreducible representations of $\sym{n'}$. The trace of a class sum is $\tr([\gamma]) = |\mathcal{C}_\gamma| N^{\ell(\gamma)}$, where $\ell(\gamma)$ counts the cycles of any permutation in $\mathcal{C}_\gamma$. Substituting these expansions and applying the character orthogonality relation
\begin{align}
    \frac{1}{n'!}\sum_\beta |\mathcal{C}_\beta| \chi_S(\beta)\chi_\theta(\beta) = \delta_{S\theta} 
\end{align}
eliminates the sums over $\beta$ and $\theta$. Altogether we obtain the final formula
\begin{align}
    \label{eq:woven_contraction_one_matrix_formula_1}
    \mathcal{W}(\zeta; R, S, \sigma)
    = \frac{N^k \dim_R |G^1_{\tau,n}|}{(n!)^2 n'!} \left(\sum_{\alpha \vdash n} \sum_{\omega}  N^{c}  |\mathcal{C}_\alpha| \chi_R(\alpha)  \chi_S(\rho)  \right)  \left(\sum_{\gamma \vdash n'} N^{\ell(\gamma)}  |\mathcal{C}_\gamma| \chi_S(\gamma) \right) \comma
\end{align}
with $k = k_{\zeta,\sigma, n, n'}$ and $\tau = \tau_{\zeta,\sigma, n, n'}$, where the sum over $\omega$ runs over a set of representatives of the left cosets $\sym{n} / G^1_{\tau,n}$, and with $c = c_{\alpha,\omega, \tau}$ and $\rho = \rho_{\alpha,\omega,\tau}$. By exchanging the roles of $R$ and $S$ in the derivation, one obtains a symmetric formula
\begin{align}
    \label{eq:woven_contraction_one_matrix_formula_2}
    \mathcal{W}(\zeta; R, S, \sigma)
    = \frac{N^k \dim_S |G^2_{\tau,n}|}{(n'!)^2 n!} \left(\sum_{\beta \vdash n'} \sum_{\omega}  N^{c}  |\mathcal{C}_\beta| \chi_S(\beta)  \chi_R(\rho)  \right)  \left(\sum_{\gamma \vdash n} N^{\ell(\gamma)}  |\mathcal{C}_\gamma| \chi_R(\gamma) \right) \comma
\end{align}
where $k$ and $\tau$ are unchanged, $G^2_{\tau,n}$ denotes the projection of $C_{\sym{n} \times \sym{n'}}(\tau)$ onto its second component $\sym{n'}$, the sum over $\omega$ runs over a set of representatives of the left cosets $\sym{n'} / G^2_{\tau,n}$, and $c$ and $\rho$ are obtained by the same contraction procedure.

We are free to use whichever of the two formulas yields the smaller number of cosets. For the interaction term $\tr(X^4)$, this number is at most $n(n-1)$ when $n=n'$, at most $\min(n,n')$ when $\abs{n-n'}=2$, and at most $1$ when $\abs{n-n'}=4$.

\paragraph{Factorization of the matrix element.}

Combining the explicit formula~\eqref{eq:woven_contraction_one_matrix_formula_1} (or its symmetric counterpart~\eqref{eq:woven_contraction_one_matrix_formula_2}) with the double-coset expansion~\eqref{eq:double_coset_decomposition} and the antinormal ordering of $\tr(X^4)$ listed in Table~\ref{tab:contributions-interactions} from Appendix~\ref{sec:decomposition_power_trace} yields each matrix element $\mel{R}{\tr(X^4)}{S}$ as an exact polynomial in $N$. Note that the constant term $B=\mathbb{I}$ also contributes polynomially, because the normalization $\alpha_R$ is itself a polynomial in $N$.

Within a fixed excitation sector $(n,n')$, the only quantities that carry the dependence on the Young diagrams $R$ and $S$ are the traces $\tr((P_R \otimes P_S) \cdot \tau)$ and the overall prefactor $(\dim_R \dim_S)^{-1}$. In practice, we precompute and factor out the latter and expand the remaining expression as a linear combination of such traces, with weights that are polynomials in $N$ and independent of $R$ and $S$ (but dependent on $n$ and $n'$):
\begin{align}
    \label{eq:decomposition_trace_tau}
    \mel{R}{\tr(X^4)}{S} = \frac{1}{\dim_R \dim_S }\sum_{\tau \in \sym{n+n'}} w(N; \tau, n, n') \tr((P_R \otimes P_S) \cdot \tau) \pt
\end{align}
Obtaining this decomposition symbolically is fast: for $\tr(X^4)$ and all excitation sectors up to $\Lambda = 20$, the computation takes about two seconds on a single CPU core of a consumer desktop, using Mathematica. For $\tr(X^4)$, at most five distinct $\tau$ appear in any given sector $(n,n')$. The method generalizes immediately to an arbitrary gauge-invariant observable: only the weights in~\eqref{eq:decomposition_trace_tau} change.

\paragraph{Complexity analysis.}

We now estimate the computational cost of evaluating all the traces $\tr((P_R \otimes P_S) \cdot \tau)$ in a fixed excitation sector $(n,n')$ and for fixed $N$ and a fixed $\tau$ appearing in the decomposition~\eqref{eq:decomposition_trace_tau} of $\tr(X^4)$.

Let $p_N(n)$ denote the number of Young diagrams with $n$ boxes and at most $N$ rows, i.e. the dimension of admissible singlet states at excitation level $n$ (see Appendix~\ref{app:dimension_counting}); note that, whenever $N \geq n$, we have $p_N(n) = p(n)$ where we recall that $p(n)$ is the number of partitions of $n$. Since only the excitation sectors such that $\abs{n-n'} \in \{0,2,4\}$ contribute, we have $p_N(n) \sim p_N(n')$ when $n \to \infty$.

Most of the data entering~\eqref{eq:woven_contraction_one_matrix_formula_1} and \eqref{eq:woven_contraction_one_matrix_formula_2} depend only on the excitation numbers $(n,n')$ and on the interaction, not on the Young diagrams $R$ and $S$. Concretely, the following quantities are precomputed once per sector:
\begin{itemize}
    \item the subgroup sizes $|G^1_{\tau,n}|$, $|G^2_{\tau,n}|$ and the coset spaces $\sym{n} / G^1_{\tau,n}$, $\sym{n'} / G^2_{\tau,n}$ together with their representatives $\omega$;
    \item the characters $\chi_R(\alpha)$ of the symmetric group for all $R\vdash n$ and all conjugacy classes $\alpha \vdash n$, and analogously for $\sym{n'}$.
\end{itemize}
Closed formulas also exist for $\dim_R$, $\dim_S$, and the class sizes $|\mathcal{C}_\alpha|$, so their evaluation is negligible. The cost of the two precomputation steps above is likewise small. The subgroup and coset data are obtained via GAP~\cite{GAP} once per $\tau$ (recall there is at most five distinct such $\tau$ to consider per excitation sector). The cost of enumerating the coset representatives is at most quartic in $n$ (see Appendix~\ref{app:stabilizer_subgroup}). The character table of $\sym{n}$ is built incrementally via the Murnaghan--Nakayama rule~\cite{james1984Representation}: assuming the tables of all $\sym{k}$ with $k<n$ have already been precomputed, each entry costs $\bigO{n}$, for a total of $\bigO{n p(n)^2}$ per level. 

Consequently, within a given excitation sector the only quantities that must be recomputed for each pair $(R,S)$ are the character sums appearing in~\eqref{eq:woven_contraction_one_matrix_formula_1}--\eqref{eq:woven_contraction_one_matrix_formula_2}.

In~\eqref{eq:woven_contraction_one_matrix_formula_1}, the factor $\sum_{\gamma \vdash n'} N^{\ell(\gamma)} |\mathcal{C}_\gamma| \chi_S(\gamma)$ depends on $S$ alone (and on $N$, $n'$), not on $R$. It can therefore be precomputed once for every $S \vdash n'$ at a total cost of $\bigO{p(n')p_N(n')}=\bigO{p(n)p_N(n)}$ operations. The corresponding factor in~\eqref{eq:woven_contraction_one_matrix_formula_2} can be precomputed in the same way at the cost of $\bigO{p(n)p_N(n)}$ operations.

The number of cosets for a given $\tau$ appearing in the decomposition of $\tr(X^4)$ can be chosen such that it is at most quadratic in $n$ (see Appendix~\ref{app:stabilizer_subgroup}). Evaluating the remaining expression for a given pair $(R,S)$ requires $\bigO{n^2 p(n)}$ operations. Summing over all $p_N(n) p_N(n')$ pairs and adding the precomputation cost yields
\begin{align}
    \bigO{p(n)(n p(n) + n^2 p_N(n)^2)}
\end{align}
operations for the full excitation sector. In the regime $N \geq n$, where every partition of $n$ labels an admissible $\U{N}$ representation and $p_N(n) = p(n)$, the complexity estimate reduces to $\bigO{n^2 p(n)^3}$.  Using the Hardy--Ramanujan asymptotic formula~\eqref{eq:dimension_center}, the total complexity in this regime thus scales as $\bigO{n^{-1} \exp(\pi\sqrt{6n})}$.

\paragraph{Implementation details and numerical results.}

\begin{figure}[tp]
    \centering
    \includegraphics[width=0.81\textwidth]{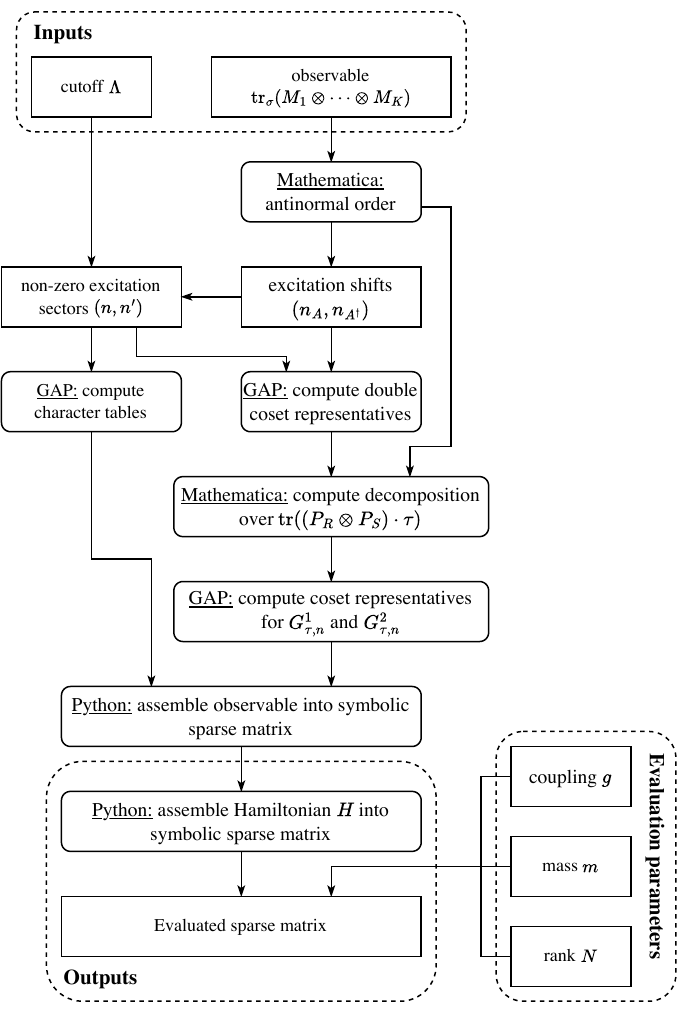}
    \caption{Overview of the computation pipeline. GAP precomputes group-theoretic data (character tables, double cosets, coset representatives). Mathematica antinormally orders the interaction and outputs the decomposition~\eqref{eq:decomposition_trace_tau}. GAP provides the representatives for the coset-reduction~\eqref{eq:coset_reduction}. Then, a Python script reads all precomputed data, evaluates the contraction coefficients via the coset-reduction plus character-theory algorithm using equation~\eqref{eq:woven_contraction_one_matrix_formula_1} or~\eqref{eq:woven_contraction_one_matrix_formula_2}, and assembles the Hamiltonian into a sparse matrix with symbolic coefficients. Finally, the symbolic matrix can be evaluated  for the model parameters: the gauge symmetry rank $N$, the mass $m$, and the coupling $g$.}
    \label{fig:algorithm_pipeline}
\end{figure}

We have implemented the algorithm described above and validated it against the exact fermion mapping, see Appendix~\ref{app:fermion_mapping}. Before presenting the numerical results, we briefly summarize the implementation, an overview of which is shown in Figure~\ref{fig:algorithm_pipeline}.

The computation is distributed across three languages:
\begin{itemize}
    \item \textbf{GAP}~\cite{GAP} handles all group-theoretic precomputation: character tables of $\sym{n}$ (via Murnaghan--Nakayama rule), double coset enumeration, and coset representative generation. It provides native symmetric-group primitives.
    \item \textbf{Mathematica}~\cite{mathematica14.3} antinormally orders gauge-invariant observable and outputs the decomposition of the form~\eqref{eq:decomposition_trace_tau}, leveraging Wolfram Language's symbolic computational capabilities.
    \item \textbf{Python} implements the core contraction algorithm, Hamiltonian assembly, and caching, leveraging Python's numerical ecosystem: NumPy for vectorized character sums, Numba for JIT-compiling large inner loops to near-C speed, and SciPy for sparse linear algebra.
\end{itemize}

GAP and Mathematica are run once per cutoff $\Lambda$ and observable; their outputs depend only on group theory and the chosen set of observables -- not on the physical couplings, the mass $m$, or the rank $N$. The Python stage reads these precomputed files and can be re-run instantly when only the couplings change.

The simulations were performed on a desktop computer equipped with an Intel\textsuperscript{\textregistered} Core\textsuperscript{\texttrademark} Ultra 7 265KF processor (20 cores, up to 5.5~GHz frequency) and 32~GB of DDR5 RAM, running Ubuntu 24.04.3 LTS.

Building the symbolic (i.e. $N$-dependent) one-matrix Hamiltonian~\eqref{eq:hamiltonian_form_1} from scratch -- including all GAP and Mathematica precomputations -- takes approximately 4~minutes and 21~seconds at $\Lambda=20$. Once the symbolic Hamiltonian is built, specifying $N$, filtering the basis, and computing the first $k$ eigenvalues and eigenfunctions via sparse solvers is fast: for $N=\Lambda=k=20$, this takes roughly one second.

The present implementation is limited to $\Lambda\leq 21$. This bound is not algorithmic but technical. First, the basis dimension grows with the partition function $p(\Lambda)$, and the memory required to store dense intermediate arrays becomes prohibitive. Second, certain intermediate quantities -- factorials and conjugacy class sizes -- overflow 64-bit integers at large $\Lambda$. Both issues could be mitigated with a more careful implementation: memory usage by fully exploiting the sparsity, and integer overflow by switching to Python's arbitrary-precision integers. These improvements are left for future work.

To validate the algorithm, we study the convergence of the low-lying eigenvalues of the one-matrix Hamiltonian~\eqref{eq:hamiltonian_form_1} as a function of the cutoff $\Lambda$. The interaction term is $\tr(X^4)$, the mass is fixed to $m=2$, and the interaction strength is set to $g^2/2 = m^2 / N$ (mean-field regime). This choice ensures the eigenvalue scales quadratically when $N\to \infty$.

For each $N\in\{12,15,20,30\}$, we compute the first $k_{\max}=5$ scaled eigenvalues $E_k/N^2$ of the Hamiltonian matrix assembled in the gauge-invariant basis, for cutoffs ranging from $\Lambda=10$ to $\Lambda=20$. The results are compared against the exact eigenvalues (dashed horizontal lines) obtained from the mapping to $N$ non-interacting fermions reviewed in Appendix~\ref{app:fermion_mapping}, solved with a one-body cutoff of $120$, which is large enough to guarantee convergence to machine precision for the energy range considered here.

\begin{figure}[tp]
\centering
\begin{tabular}{cc}
    \includegraphics[width=0.472\textwidth]{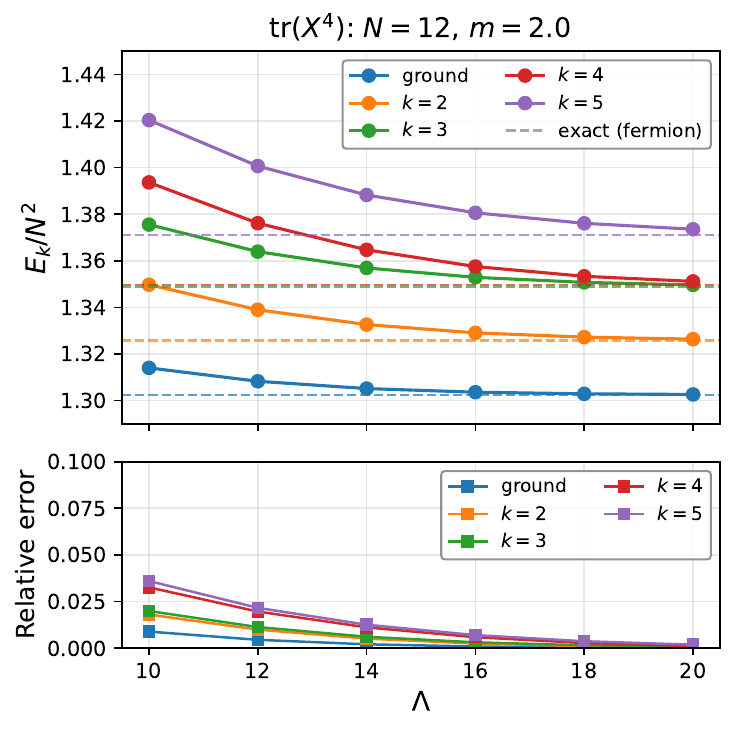} &
    \includegraphics[width=0.472\textwidth]{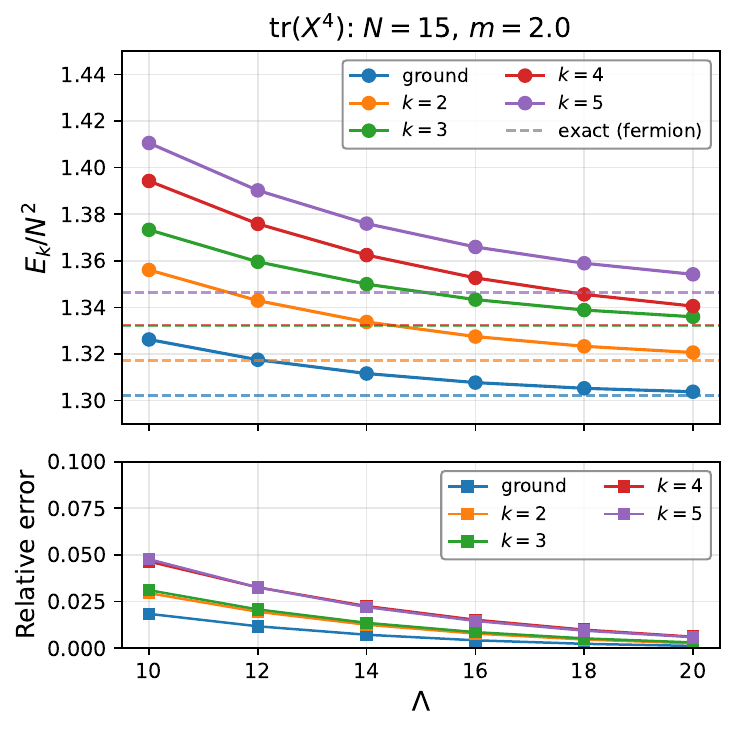} \\
    \includegraphics[width=0.472\textwidth]{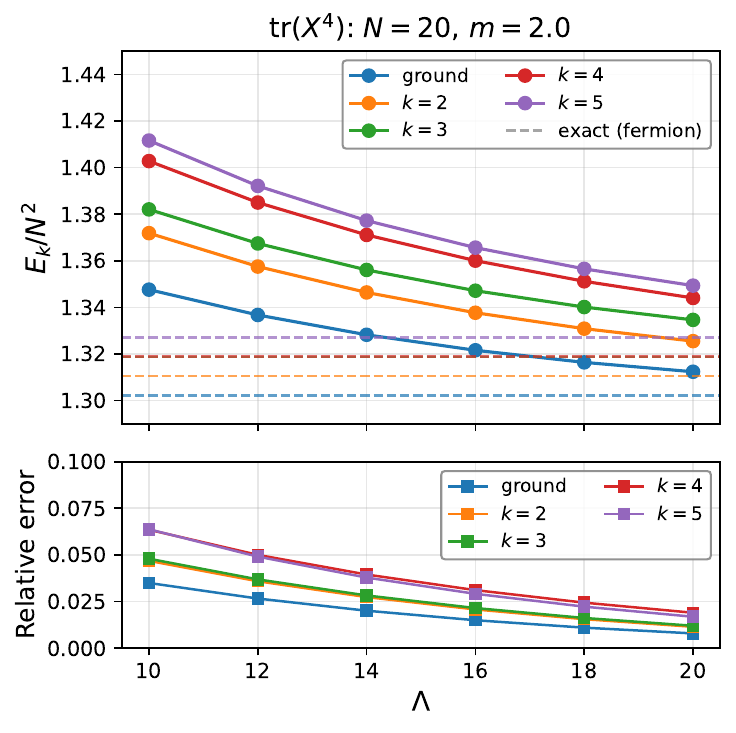} &
    \includegraphics[width=0.472\textwidth]{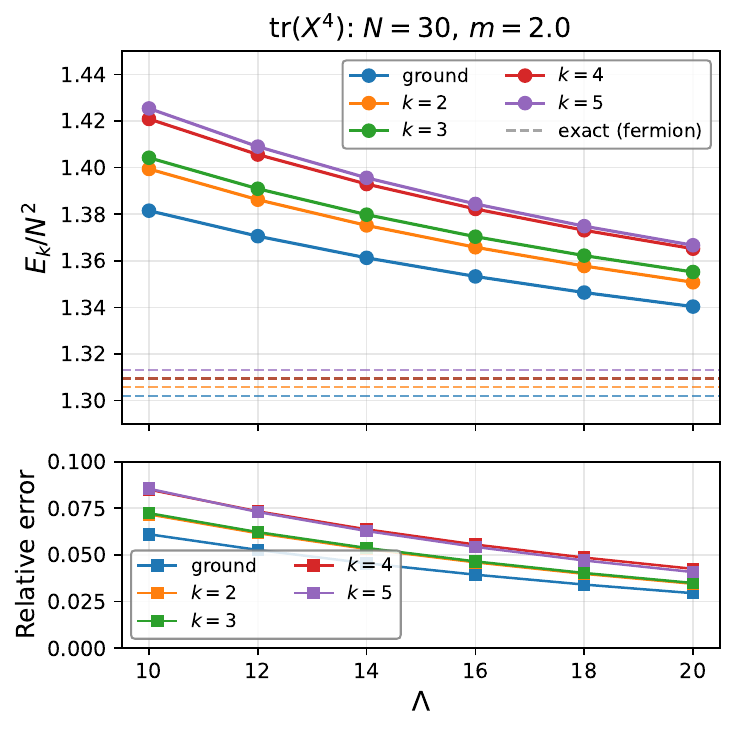} \\
\end{tabular}
\caption{Convergence of the first five scaled eigenvalues of the one-matrix Hamiltonian with interaction $\tr(X^4)$, $m=2$, and coupling $g^2/2=m^2/N$, as a function of the cutoff $\Lambda$. Solid lines with markers show the eigenvalues computed in the gauge-invariant basis for $\Lambda=10,\dots,20$; dashed horizontal lines mark the exact eigenvalues obtained from the fermion mapping with a one-body cutoff of $120$. Each panel corresponds to a different value of $N$.}
\label{fig:eigenvalue_convergence_grid}    
\end{figure}

Figure~\ref{fig:eigenvalue_convergence_grid} collects the convergence plots for all values of $N$. As expected, the relative error grows with $N$ and decreases as $\Lambda$ increases, for all eigenvalues. At $\Lambda=20$, the first five eigenvalues are all within $0.8\%$ of the exact result for $N\leq 15$, and the ground state remains within this accuracy up to $N=20$. For $N=30$, the cutoff is not sufficiently large as the relative errors range between $2.9\%$ and $4.2\%$. Overall, the agreement with the exact fermion mapping confirms that the algorithm correctly reproduces the spectrum and converges toward it as $\Lambda$ grows, thereby validating the method.

\subsection{The multiple-matrix case}

The computation of the Hamiltonian matrix elements in this basis proceeds along similar lines as in the one-matrix case. The main difference is that, due to the presence of multiple matrices, the relevant symmetry group is not $\sym{n}$ but rather the Young subgroup $\sym{\underline{\vb*{n}}} = \sym{n_1}\times \cdots \times \sym{n_D} \subset \sym{n}$ where $n_1 + \cdots + n_D = n$. Recall that our primary example in the multi-matrix case is the commutator-squared Hamiltonian
\begin{align}
    H = \frac{1}{2} \sum_{I=1}^D \tr( P_I^2 + m^2 X_I^2) - \frac{g^2}{2} \sum_{I,J=1}^D \tr([X_I,X_J]^2) \comma
\end{align}
for which we now outline how the single-matrix algorithm extends. As before, the method applies to any gauge-invariant observable.

We seek to evaluate matrix elements $\mel{R,\vb*{r},a,b}{H}{S,\vb*{s},c,d}$ for the restricted Schur polynomial basis. We reuse the notation of Section~\ref{sec:singlet_states_multi_matrices}:
\begin{align}
    R \vdash n,~ \vb*{r} = (r_1,\dots,r_D),~ r_i \vdash n_i, ~\sum_i n_i = n, ~ 1 \leq a,b \leq m^R_{\vb*{r}} \comma \\
    S \vdash n',~ \vb*{s} = (s_1,\dots,s_D),~ s_i \vdash n_i', ~\sum_i n_i' = n', ~1 \leq c,d \leq m^S_{\vb*{s}} \pt
\end{align}
As in the single-matrix case we decompose $H = H_{\text{free}} + H_{\text{int}}$ into a free and interacting part. Again, the free Hamiltonian is diagonal in the basis $\ket{R,\vb*{r},a,b}$ and 
\begin{equation}
    \label{eq:free_matrix_elements}
    \mel{R,\vb*{r},a,b}{H_\text{free}}{S,\vb*{s},c,d} = \alpha_{\vb*{r}}^R m (n + D N^2/2) \delta_{RS}\delta_{\vb*{rs}}\delta_{bc}\delta_{ad} \comma
\end{equation}
where the normalization constant $\alpha_{\vb*{r}}^R$ is defined in~\eqref{eq:normalization_constant_multi_matrices}.

For the interaction Hamiltonian $H_{\text{int}}$, it suffices to focus on a single commutator-squared term $\tr([X_1,X_2]^2)$; the matrix elements for other index pairs $(I, J)$ are obtained by the same computation with the species labels permuted. This term decomposes into a sum of contributions of the form $\ttr{\sigma}{B}$, where $B$ is a tensor product of up to four creation and annihilation operators, with at most two factors from matrix $I=1$ and two from matrix $J=2$ (see Table~\ref{tab:contributions-interactions-multi} in Appendix~\ref{sec:decomposition_power_trace}, where all the terms of this decomposition are collected). We introduce a graphical notation for such a contribution:
\begin{equation}
    \begin{tikzpicture}[line width=2px,baseline={([yshift=-.5ex]current bounding box.center)}]
        \draw [-,blue,line width=1.5px] (5,-2.45) to (6,-2.45);
        \draw [-,red,line width=1.5px] (5,-2.55) to (6,-2.55);
        \draw [-,blue,line width=1.5px] (5,-1.45) to (6,-1.45);
        \draw [-,red,line width=1.5px] (5,-1.55) to (6,-1.55);
        \draw [-,blue, line width=1.5]  (8,-1.45) to  (7,-1.45);
        \draw [-,red, line width=1.5] (8,-1.55) to  (7,-1.55);
        \draw [-,blue,line width = 1.5px] (8,-2.45) to  (7,-2.45);
        \draw [-,red,line width = 1.5px] (8,-2.55) to  (7,-2.55);
        \node [draw, rectangle, minimum width=1.0cm, minimum height=2.0cm, align=center] at (6.5,-2) (s) {$\sigma$};
    \end{tikzpicture}\,,
\end{equation}
where the color of each line indicates which matrix the corresponding operator belongs to -- blue for matrix $I=1$ and red for matrix $J=2$ in this example. 

The matrix element $\mel{R,\vb*{r},a,b}{\ttr{\sigma}{B}}{S,\vb*{s},c,d}$ vanishes unless excitation number is conserved separately for each species:
    \begin{align}
        \label{eq:excitation_condition_multi_matrices}
        n_I + n_{A_I} = n_I' + n_{A_I^\dagger},\quad \forall I =1, \dots, D \comma
    \end{align}
    where $n_{A_I}$ and $n_{A_I^\dagger}$ denote respectively the numbers of annihilation and creation operators from matrices of species $I$ appearing in $B$. For the interaction term $\tr([X_1,X_2]^2)$ this forces $n_I = n_I'$ for all $I\geq 3$; otherwise the matrix element vanishes. From now on, we assume the conditions~\eqref{eq:excitation_condition_multi_matrices} are satisfied.

The computation of $\mel{R,\vb*{r},a,b}{\ttr{\sigma}{B}}{S,\vb*{s},c,d}$ then proceeds along the same lines as in the single-matrix case, with the following differences: the Young projectors $P_R$ and $P_S$ are replaced by the centralizer basis elements $P^{R,\vb*{r}}_{ab}$ and $P^{S,\vb*{s}}_{cd}$; the Wick contractions now range over
    \begin{align}
        \xi \in \sym{\underline{\vb*{n}} + \underline{\vb*{n}}_A} \coloneq \sym{n_1+ n_{A_1}} \times \sym{n_2+ n_{A_2}} \times \sym{n_3} \times \cdots \times \sym{n_D} \pt
    \end{align} 
Thus, the matrix element takes the form
\begin{equation}
    \label{eq:matrix_element_multi}
    \mel{R,\vb*{r},a,b}{\ttr{\sigma}{B}}{S,\vb*{s},c,d}
    = \frac{1}{\dim_R \dim_S} \frac{n!}{\underline{\vb*{n}}!} \frac{n'!}{\underline{\vb*{n}}'!} \sum_{\xi} \mathcal{W}\comma
\end{equation}
with the summand $\mathcal{W} \equiv \mathcal{W}(\xi;R,S,\vb*{r},\vb*{s},a,c,b,d,\sigma)$. The summand can then be graphically represented as 
\begin{equation}
    \begin{tikzpicture}[line width=2px,baseline={([yshift=-.5ex]current bounding box.center)}]
        \draw [-,blue] (.5,.1) to[out=0,in=180] (3.5,.7);
        \draw [-,red] (.5,0) to[out=0,in=180] (3.5,-.05);
        \draw [-,ForestGreen] (.5,-.1) to[out=0,in=180] (3.5,-.75);
        \draw [-,blue,line width=1.5px] (4.5,-3.15) to[out=0,in=180] (6,-2.45);
        \draw [-,red,line width=1.5px] (4.5,-3.9) to[out=0,in=180] (6,-2.55);
        \draw [-,white,line width=4] (-.5,.1) to[out=180,in=180,distance=2cm] (-.5,-5.6) to[out=0,in=180] (4.5,-5.6) to[out=0,in=0,distance=.9cm] (4.5,-3.3);
        \draw [-,blue] (-.5,.1) to[out=180,in=180,distance=2cm] (-.5,-5.6) to[out=0,in=180] (4.5,-5.6) to[out=0,in=0,distance=.9cm] (4.5,-3.3);
        \draw [-,red] (-.5,0) to[out=180,in=180,distance=1.85cm] (-.5,-5.5) to[out=0,in=180] (4.5,-5.5) to[out=0,in=0,distance=.7cm] (4.5,-4.05);
        \draw [-,ForestGreen] (-.5,-.1) to[out=180,in=180,distance=1.7cm] (-.5,-5.4) to[out=0,in=180] (4.5,-5.4) to[out=0,in=0,distance=.5cm] (4.5,-4.7);
        \draw [-,blue] (-.5,-3.9) to[out=180,in=180] (-0.5,-2.1) to[out=0,in=180] (4.58,-2.1) to[out=0,in=0,distance=.8cm]  (4.5,.7);
        \draw [-,red] (-.5,-4) to[out=180,in=180] (-0.5,-2) to[out=0,in=180] (4.5,-2) to[out=0,in=0,distance=.7cm]  (4.5,-.05);
        \draw [-,ForestGreen] (-.5,-4.1) to[out=180,in=180] (-0.5,-1.9) to[out=0,in=180] (4.5,-1.9) to[out=0,in=0,distance=.5cm]  (4.5,-.75);
        \draw [-,white,line width=3px] (4.5,.8) to[out=0,in=170] (6,-1.45);
        \draw [-,white,line width=3px] (4.5,.1) to[out=0,in=180] (6,-1.55);
        \draw [-,blue,line width=1.5px] (4.5,.8) to[out=0,in=170] (6,-1.45);
        \draw [-,red,line width=1.5px] (4.5,.1) to[out=0,in=180] (6,-1.55);
        \draw [-,blue] (.5,-3.9) to[out=0,in=180] (3.5,-3.3);
        \draw [-,red] (.5,-4) to[out=0,in=180] (3.5,-4.05);
        \draw [-,ForestGreen] (.5,-4.1) to[out=0,in=180] (3.5,-4.75);
        \draw [-,blue,line width = 1.5px] (3.5,.8) to[out=180,in=180] (3.5,1.4) to[out=0,in=180] (7,1.4) to[out=0,in=0]  (7,-2.45);
        \draw [-,white,line width = 3px] (3.5,.1) to[out=180,in=180] (3.5,1.5) to[out=0,in=180] (7.05,1.5) to[out=0,in=0]  (7,-2.55);
        \draw [-,red,line width = 1.5px] (3.5,.1) to[out=180,in=180] (3.5,1.5) to[out=0,in=180] (7.05,1.5) to[out=0,in=0]  (7,-2.55);
        \draw [-,white, line width=3] (3.5,-3.15) to[out=180,in=180] (3.5,-6) to[out=0,in=180] (7.05,-6) to[out=0,in=0]  (7,-1.45);
        \draw [-,blue, line width=1.5] (3.5,-3.15) to[out=180,in=180] (3.5,-6) to[out=0,in=180] (7.05,-6) to[out=0,in=0]  (7,-1.45);
        \draw [-,white, line width=3] (3.5,-3.9) to[out=180,in=180] (3.5,-5.9)to[out=0,in=180] (7,-5.9) to[out=0,in=0]  (7,-1.55);
        \draw [-,red, line width=1.5] (3.5,-3.9) to[out=180,in=180] (3.5,-5.9) to[out=0,in=180] (7,-5.9) to[out=0,in=0]  (7,-1.55);
        \node[draw, rectangle, minimum width=1.0cm, minimum height=2.0cm, align=center] (PR) {$P^{R,r}_{ab}$};
        \node [below of = PR, node distance= 4cm, draw, rectangle, minimum width=1.0cm, minimum height=2.0cm, align=center] (PS) {$P^{S,s}_{cd}$};
        \node [draw, rectangle, minimum width=1.0cm, minimum height=0.5cm, align=center, blue] at (4,.75) (xi1) {$\xi_1$};
        \node [draw, rectangle, minimum width=1.0cm, minimum height=0.5cm, align=center, red] at (4,0) (xi2) {$\xi_2$};
        \node [draw, rectangle, minimum width=1.0cm, minimum height=0.5cm, align=center, ForestGreen  ] at (4,-.75) (xi3) {$\xi_3$};
        \node [draw, rectangle, minimum width=1.0cm, minimum height=0.5cm, align=center, blue] at (4,-3.25) (xi12) {$\xi_1^{-1}$};
        \node [draw, rectangle, minimum width=1.0cm, minimum height=0.5cm, align=center, red] at (4,-4) (xi22) {$\xi_2^{-1}$};
        \node [draw, rectangle, minimum width=1.0cm, minimum height=0.5cm, align=center, ForestGreen] at (4,-4.75) (xi32) {$\xi_3^{-1}$};
        \node [draw, rectangle, minimum width=1.0cm, minimum height=2.0cm, align=center] at (6.5,-2) (s) {$\sigma$};
    \end{tikzpicture} \comma
\end{equation}
where we collect in $\xi_3$ all the terms that are not mixed up with species 1 and 2. Since the centralizer basis elements commute with every element of the Young subgroup, $\nu P^{R,\vb*{r}}_{ab}=P^{R,\vb*{r}}_{ab}\nu$ for all $\nu\in\sym{\underline{\vb*{n}}}$, $\mathcal{W}$ is constant on the double cosets
\begin{align}
    \sym{\underline{\vb*{n}}} \backslash \sym{\underline{\vb*{n}} + \underline{\vb*{n}}_A} / \sym{\underline{\vb*{n}}'} \comma
\end{align}
where the subgroups embed via suitable inclusions. For the interaction $\tr([X_1,X_2]^2)$, only the first two species carry nontrivial operator insertions; the remaining species are spectators, and the double coset factorizes as
\begin{align}
    \sym{\underline{\vb*{n}}} \backslash \sym{\underline{\vb*{n}} + \underline{\vb*{n}}_A} / \sym{\underline{\vb*{n}}'}
    \simeq \left(\sym{n_1} \backslash \sym{n_1 + n_{A_1}} / \sym{n_1'} \right) \times  \left(\sym{n_2} \backslash \sym{n_2 + n_{A_2}} / \sym{n_2'} \right) \times \{1\} \pt
\end{align} 
Thus, the sum over $\xi$ can be reduced to a sum over double coset representatives $\zeta$ weighted by their sizes. For an arbitrary multi-matrix interaction the double coset space always factorizes as a product of single-species double coset factors (one per matrix species that carries operator insertions) together with trivial factors for the spectator species. Consequently, the number of double cosets in the general case is simply the product of the corresponding one-matrix counts, each of which is bounded by the estimates of Section~\ref{sec:one_matrix_case}. For the commutator-squared interaction, this product yields at most four double cosets. 

\begin{figure}
    \centering
    \begin{tikzpicture}[line width=2px]
        \begin{scope}[shift={(0,0)}]
        \draw [-,blue] (.5,.1) to[out=0,in=180] (3.5,-.9);
        \draw [-,red] (.5,0) to[out=0,in=180] (3.5,-1);
        \draw [-,ForestGreen] (.5,-.1) to[out=0,in=180] (3.5,-1.1);
        \draw [-,blue] (.5,-2.9) to[out=0,in=180] (3.5,-1.9);
        \draw [-,red] (.5,-3) to[out=0,in=180] (3.5,-2);
        \draw [-,ForestGreen] (.5,-3.1) to[out=0,in=180] (3.5,-2.1);
        \draw [-,blue] (-.5,.1) to[out=180,in=180,distance=.85cm] (-.5,1.4) to[out=0,in=180] (4.5,1.4) to[out=0,in=0,distance=.85cm] (4.5,-.9);
        \draw [-,red] (-.5,0) to[out=180,in=180,distance=1cm] (-.5,1.5) to[out=0,in=180] (4.5,1.5) to[out=0,in=0,distance=1cm] (4.5,-1);
        \draw [-,ForestGreen] (-.5,-.1) to[out=180,in=180,distance=1.15cm] (-.5,1.6) to[out=0,in=180] (4.5,1.6) to[out=0,in=0,distance=1.15cm] (4.5,-1.1);
        \draw [-,blue] (-.5,-2.9) to[out=180,in=180,distance=1.15cm] (-0.5,-4.6) to[out=0,in=180] (4.5,-4.6) to[out=0,in=0,distance=1.15cm]  (4.5,-1.9);
        \draw [-,red] (-.5,-3) to[out=180,in=180,distance=1cm] (-0.5,-4.5) to[out=0,in=180] (4.5,-4.5) to[out=0,in=0,distance=1cm]  (4.5,-2);
        \draw [-,ForestGreen] (-.5,-3.1) to[out=180,in=180,distance=.85cm] (-0.5,-4.4) to[out=0,in=180] (4.5,-4.4) to[out=0,in=0,distance=.85cm]  (4.5,-2.1);
        \node[draw, rectangle, minimum width=1.0cm, minimum height=2.0cm, align=center] (PR) {$P^{R,r}_{ab}$};
        \node [below of = PR, node distance= 3cm, draw, rectangle, minimum width=1.0cm, minimum height=2.0cm, align=center] (PS) {$P^{S,s}_{cd}$};
        \node[draw, rectangle, minimum width=1.0cm, minimum height=2.0cm, align=center] (tau) at (4,-1.5) {$\tau$};
        \end{scope}

        \begin{scope}[shift={(7.6,0)}]
        \draw [-,blue] (.5,.1) to[out=0,in=180] (3.5,-.9);
        \draw [-,red] (.5,0) to[out=0,in=180] (3.5,-1);
        \draw [-,blue] (.5,-2.9) to[out=0,in=180] (3.5,-1.4);
        \draw [-,red] (.5,-3) to[out=0,in=182] (3.5,-1.5);
        \draw [-,blue] (-.5,.1) to[out=180,in=180,distance=.85cm] (-.5,1.4) to[out=0,in=180] (4.5,1.4) to[out=0,in=0,distance=.85cm] (4.5,-.9);
        \draw [-,red] (-.5,0) to[out=180,in=180,distance=1cm] (-.5,1.5) to[out=0,in=180] (4.5,1.5) to[out=0,in=0,distance=1cm] (4.5,-1);
        \draw [-,blue] (-.5,-2.9) to[out=180,in=180,distance=1.15cm] (-0.5,-4.6) to[out=0,in=180] (4.5,-4.6) to[out=0,in=0,distance=1.15cm]  (4.5,-1.4);
        \draw [-,red] (-.5,-3) to[out=180,in=180,distance=1cm] (-0.5,-4.5) to[out=0,in=180] (4.5,-4.5) to[out=0,in=0,distance=1cm]  (4.5,-1.5);
        \draw [-,white,line width=3.5pt] (-.5,-3.1) to[out=180,in=180,distance=.85cm] (-0.5,-4.4) to[out=0,in=180] (4.5,-4.4) to[out=0,in=0,distance=.75cm]  (4.5,-2.5) to[out=180,in=0] (3.5,-2.1) to[out=180,in=0] (.5,-.1);
        \draw [-,ForestGreen] (-.5,-3.1) to[out=180,in=180,distance=.85cm] (-0.5,-4.4) to[out=0,in=180] (4.5,-4.4) to[out=0,in=0,distance=.75cm]  (4.5,-2.5) to[out=180,in=0] (3.5,-2.1) to[out=180,in=0] (.5,-.1);
        \draw [-,white,line width=3.5pt] (-.5,-.1) to[out=180,in=180,distance=1.15cm] (-.5,1.6) to[out=0,in=180] (4.55,1.6) to[out=0,in=0,distance=1.2cm] (4.5,-2.1) to[out=180,in=0] (3.5,-2.5) to[out=180,in=0] (.5,-3.1);
        \draw [-,ForestGreen] (-.5,-.1) to[out=180,in=180,distance=1.15cm] (-.5,1.6) to[out=0,in=180] (4.55,1.6) to[out=0,in=0,distance=1.2cm] (4.5,-2.1) to[out=180,in=0] (3.5,-2.5) to[out=180,in=0] (.5,-3.1);
        \node[draw, rectangle, minimum width=1.0cm, minimum height=2.0cm, align=center] (PR) {$P^{R,r}_{ab}$};
        \node [below of = PR, node distance= 3cm, draw, rectangle, minimum width=1.0cm, minimum height=2.0cm, align=center] (PS) {$P^{S,s}_{cd}$};
        \node[draw, rectangle, minimum width=1.0cm, minimum height=1.4cm, align=center] (tau) at (4,-1.2) {$\tau'$};
        \end{scope}
        \node at (6,-1.5) {$=$};    
    \end{tikzpicture}
    \caption{Diagrammatic representation of the contraction structure of the trace for the multi-matrix interaction. The left-hand side shows the contraction of the centralizer basis elements $P^{R,\vb*{r}}_{ab}$ and $P^{S,\vb*{s}}_{cd}$ with the permutation $\tau$ that arises from the Wick contractions. The right-hand side shows a more fine-grained decomposition of $\tau$ into a part $\tau'$ that acts only on species 1 and 2, while it acts as a swap for species 3, which is the structure of every interaction term that only involves two species and particularly the case for the commutator-squared interaction $\tr([X_1,X_2]^2)$.}
    \label{fig:tau_multi_matrices}
\end{figure}
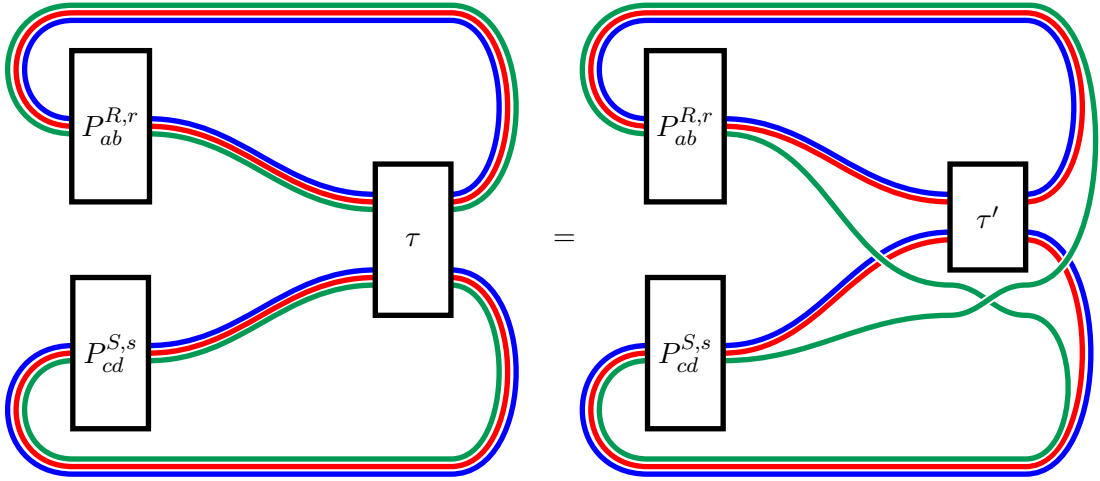

The evaluation of $\mathcal{W}$ for each orbit proceeds analogously to the single-matrix case. By contracting the elements $\zeta$, $\zeta^{-1}$, and $\sigma$, we have 
\begin{align}
    \label{eq:tau_multi_matrices}
    \mathcal{W}(\zeta;R,S,\vb*{r},\vb*{s},a,c,b,d,\sigma) = N^k \tr\left((P^{R,\vb*{r}}_{ab} \otimes P^{S,\vb*{s}}_{cd}) \cdot \tau \right) \comma
\end{align}
for some $k\in\N$ and $\tau \in \sym{n+n'}$. The contraction for evaluating the trace is shown in Figure~\ref{fig:tau_multi_matrices}. To evaluate such traces, we expand the centralizer basis elements using the character expansion:
\begin{align}
    P^{R,\vb*{r}}_{ab} = \frac{\dim_R}{n!} \sum_{\rho \in \sym{n}}\chi^{R,\vb*{r}}_{ab}(\rho) \rho 
    &= \frac{\dim_R}{n!} \sum_{\Omega}\chi^{R,\vb*{r}}_{ab}(\Omega) [\Omega] \\
    &=  \frac{\dim_R}{n!} \sum_{\Omega} \chi^{R,\vb*{r}}_{ab}(\Omega) \frac{\abs{\mathcal{C}_{\Omega}}}{\underline{\vb*{n}}!}  \sum_{\rho\in \sym{\underline{\vb*{n}}}} \rho^{-1}\rho_\Omega \rho \pt
\end{align}
Here the sum runs over labels $\Omega$ of restricted conjugacy classes $\mathcal{C}_\Omega$ (the orbits under the action by conjugation of the corresponding Young subgroup, see Appendix~\ref{app:centralizer_group_algebra}), and $[\Omega]$ (resp. $\rho_\Omega$) denotes the corresponding class sum (resp. a representative). Inserting this expansion in~\eqref{eq:tau_multi_matrices}, we obtain:
\begin{align}
    \tr\left(\left(P^{R,\vb*{r}}_{ab} \otimes P^{S,\vb*{s}}_{cd}\right)\cdot\tau\right)
    = \frac{\dim_R}{n! \underline{\vb*{n}}!}
    \sum_{\Omega} \chi^{R,\vb*{r}}_{ab}(\Omega)
    \left|\mathcal{C}_{\Omega}\right|
    \sum_{\rho\in \sym{\underline{\vb*{n}}}}
    \tr\left(\left(\rho^{-1}\rho_\Omega\rho \otimes P^{S,\vb*{s}}_{cd}\right)\cdot\tau\right) \pt
\end{align}
The sum over $\sym{\underline{\vb*{n}}}$ can be reduced, as in the single-matrix case, by noting that the summand is invariant under right multiplication by elements of a suitable stabilizer subgroup. Define the subgroup
\begin{align}
    \label{eq:multi_stabilizer}
    C_{\sym{\underline{\vb*{n}}} \times \sym{\underline{\vb*{n}}'}}(\tau) \coloneqq \enstq{(\nu,\mu) \in \sym{\underline{\vb*{n}}}\times\sym{\underline{\vb*{n}}'}}{(\nu \otimes \mu)\cdot\tau
        = \tau \cdot (\nu \otimes \mu)}\comma
\end{align}
and its projection onto the first factor
\begin{align}
    G^1_{\tau,\underline{\vb*{n}},\underline{\vb*{n}}'} \coloneqq \enstq{\nu \in \sym{\underline{\vb*{n}}}}{\exists \mu \in \sym{\underline{\vb*{n}}'},~ (\nu, \mu) \in C_{\sym{\underline{\vb*{n}}} \times \sym{\underline{\vb*{n}}'}}(\tau)} \pt
\end{align}
The summand is then constant on the left cosets of $G^1_{\tau,\underline{\vb*{n}},\underline{\vb*{n}}'}$ in $\sym{\underline{\vb*{n}}}$.

For the interaction term $\tr([X_1,X_2]^2)$ a simplification occurs because $\tau$ does not mix the matrix species~$3$ with the other two: the stabilizer factorizes as
\begin{align}
     G^1_{\tau,\underline{\vb*{n}},\underline{\vb*{n}}'} \simeq G^1_{\tau',n_1,n_2,n_1',n_2'} \times \sym{n_3} \times \cdots \times \sym{n_D} \comma
\end{align}
where $G^1_{\tau',n_1,n_2,n_1',n_2'}$ is the projection onto the first factor of $C_{\sym{n_1} \times \sym{n_2} \times \sym{n_1'} \times \sym{n_2'}}(\tau')$, defined analogously to~\eqref{eq:multi_stabilizer}. In that case, it can be seen that the number of left cosets in $\sym{\underline{\vb*{n}}}$ is at most $\bigO{n_1 n_2}$ for this interaction.

Decomposing $\sym{\underline{\vb*{n}}}$ into left cosets indexed by representatives $\omega$ collapses the sum to the evaluation of terms of the form $N^c \tr(P^{S,\vb*{s}}_{cd} \rho)$, represented graphically as 
\begin{equation}
    \begin{tikzpicture}[line width=2px,baseline={([yshift=-.5ex]current bounding box.center)}]
        \draw[-,line width=1.5px,blue] (.5,.1) to (1,.1);
        \draw[-,line width=1.5px,red] (.5,0) to (1,0);
        \draw[-,line width=1.5px,ForestGreen] (.5,-.1) to (1,-.1);
        \draw[-,line width=1.5px,blue] (-.5,.1) to[out=180,in=180] (-.55,-1.6) to (3.05,-1.6) to[out=0,in=0] (3,.1);
        \draw[-,line width=1.5px,red] (-.5,0) to[out=180,in=180] (-.5,-1.5) to (3,-1.5) to[out=0,in=0] (3,0);
        \draw[-,line width=1.5px,ForestGreen] (-.5,-.1) to[out=180,in=180] (-.45,-1.4) to (2.95,-1.4) to[out=0,in=0] (3,-.1);
        \node[draw, rectangle, minimum width=1.0cm, minimum height=2.0cm, align=center] (PS) {$P^{S,\vb*{s}}_{cd}$};
        \node[draw, rectangle, minimum width=2.0cm, minimum height=2.0cm, align=center, right of = PS, node distance=2cm] (tau) {$\rho$};
    \end{tikzpicture} \,.
\end{equation}
Here $\rho \equiv \rho_{\Omega,\omega,\tau} \in \sym{n'}$ and $c\equiv c_{\Omega,\omega,\tau}\in \N$ are obtained from the contraction
\begin{equation}
    \begin{tikzpicture}[line width=2px,baseline={([yshift=-.5ex]current bounding box.center)}]
        \draw[blue, line width=1.5pt] (-2,.75) to[out=0,in=180] (-.5,.75);
        \draw[red, line width=1.5pt] (-2,-.75) to[out=0,in=180] (-.5,-.75);

        \draw[blue, line width=1.5pt] (.5,.75) to[out=0,in=180] (1.5,-.2);
        \draw[red, line width=1.5pt] (.5,-.75) to[out=0,in=180] (1.5,-1);
        
        \draw[blue, line width=1.5pt] (2.5,-.2) to[out=0,in=180] (3.5,-.2);
        \draw[red, line width=1.5pt] (2.5,-1) to[out=0,in=180] (3.5,-1);
        
        \draw[blue, line width=1.5pt] (4.5,-.2) to[out=0,in=180] (5.5,-.2);
        \draw[red, line width=1.5pt] (4.5,-1) to[out=0,in=180] (5.5,-1);
        
        \draw[blue, line width=1.5pt] (6.5,-.2) to[out=0,in=180] (7.5,.75);
        \draw[red, line width=1.5pt] (6.5,-1) to[out=0,in=180] (7.5,-.75);

        \draw[blue, line width=1.5pt] (8.5,.75) to[out=0,in=180] (10,.75);
        \draw[red, line width=1.5pt] (8.5,-.75) to[out=0,in=180] (10,-.75);

        \draw[blue, line width=1pt] (.5,.85) to[out=0,in=180] (3.5,1.3);
        \draw[white, line width=2pt] (.5,-.65) to[out=0,in=-90] (1,-.2) to[out=90,in=180] (3.5,1.2);
        \draw[red, line width=1pt] (.5,-.65) to[out=0,in=-90] (1,-.2) to[out=90,in=180] (3.5,1.2);

        \draw[blue, line width=1pt] (7.5,.85) to[out=180,in=0] (4.5,1.3);
        \draw[white, line width=2pt] (7.5,-.65) to[out=180,in=-90] (7,-.2) to[out=90,in=0] (4.5,1.2);
        \draw[red, line width=1pt] (7.5,-.65) to[out=180,in=-90] (7,-.2) to[out=90,in=0] (4.5,1.2);

        \draw[white, line width=2pt] (-.5,.85) to[out=180,in=180] (0,2.4) to[out=0,in=180] (4.5,2.4) to[out=0,in=0] (4.5,1.8);
        \draw[blue, line width=1pt] (-.5,.85) to[out=180,in=180] (0,2.4) to[out=0,in=180] (4.5,2.4) to[out=0,in=0] (4.5,1.8);
        \draw[white, line width=2pt] (-.5,-.65) to[out=180,in=180] (0,2.5) to[out=0,in=180] (4.5,2.5) to[out=0,in=0] (4.5,1.7);
        \draw[red, line width=1pt] (-.5,-.65) to[out=180,in=180] (0,2.5) to[out=0,in=180] (4.5,2.5) to[out=0,in=0] (4.5,1.7);

        \draw[white, line width=2pt] (8.5,.85) to[out=0,in=0] (8,2.6) to[out=180,in=0] (3.5,2.6) to[out=180,in=180] (3.5,1.8);
        \draw[blue, line width=1pt] (8.5,.85) to[out=0,in=0] (8,2.6) to[out=180,in=0] (3.5,2.6) to[out=180,in=180] (3.5,1.8);
        \draw[white, line width=2pt] (8.5,-.65) to[out=0,in=0] (8,2.7) to[out=180,in=0] (3.5,2.7) to[out=180,in=180] (3.5,1.7);
        \draw[red, line width=1pt] (8.5,-.65) to[out=0,in=0] (8,2.7) to[out=180,in=0] (3.5,2.7) to[out=180,in=180] (3.5,1.7);

        \draw[ForestGreen, line width=1.5pt] (-2,-2) to[out=0,in=180] (3.5,-2);
        \draw[ForestGreen, line width=1.5pt] (4.5,-2) to[out=0,in=180] (10,-2);
        
        \node [draw, blue, rectangle, minimum width=1.0cm, minimum height=1.0cm, align=center] at (0,.8) (invzeta1) {$\zeta_1^{-1}$}; 
        \node [draw, red, rectangle, minimum width=1.0cm, minimum height=1.0cm, align=center] at (0,-.7) (invzeta2) {$\zeta_2^{-1}$}; 
        \node [draw, blue, rectangle, minimum width=1.0cm, minimum height=.7cm, align=center] at (2,-.2) (invomega1) {$\omega_1^{-1}$};
        \node [draw, red, rectangle, minimum width=1.0cm, minimum height=.7cm, align=center] at (2,-1) (invomega2) {$\omega_2^{-1}$};
        \node [draw, rectangle, minimum width=1.0cm, minimum height=3.0cm, align=center] at (4,-1) (alpha) {$\rho_\Omega$}; 
        \node [draw, blue, rectangle, minimum width=1.0cm, minimum height=.7cm, align=center] at (6,-.2) (omega1) {$\omega_1$}; 
        \node [draw, red, rectangle, minimum width=1.0cm, minimum height=.7cm, align=center] at (6,-1) (omega2) {$\omega_2$}; 
        \node [draw, blue, rectangle, minimum width=1.0cm, minimum height=1.0cm, align=center] at (8,.8) (invzeta1) {$\zeta_1$}; 
        \node [draw, red, rectangle, minimum width=1.0cm, minimum height=1.0cm, align=center] at (8,-.7) (invzeta2) {$\zeta_2$}; 
        \node [draw, rectangle, minimum width=1.0cm, minimum height=1.0cm, align=center] at (4,1.5) (sigma) {$\sigma$}; 
    \end{tikzpicture}\comma
\end{equation}
where $c$ counts the closed loops in that contraction. This trace is evaluated by Fourier analysis on the centralizer algebra, exactly as in the one-matrix case but with restricted characters and class sums replacing ordinary ones (see Appendix~\ref{app:centralizer_group_algebra}). One finds
\begin{align}
    \label{eq:tr_restricted_projector}
    \tr(P^{S,\vb*{s}}_{cd} \rho)
    = \frac{\dim_S}{n'!\dim_{\vb*{s}}} \sum_{e} \chi^{S,\vb*{s}}_{ed}(\Omega^{-1}_\rho)
        \left( \sum_{\Omega'} N^{\ell(\Omega')} \abs{\mathcal{C}_{\Omega'}} \chi^{S,\vb*{s}}_{ce}(\Omega') \right) \comma
\end{align}
where $e$ runs over a basis of the multiplicity space $M^S_{\vb*{s}}$, $\Omega_\rho$ labels the restricted conjugacy class of $\rho$ under $\sym{\underline{\vb*{n}}'}$, and the class $\Omega_\rho^{-1}$ is defined by $\mathcal{C}_{\Omega_\rho^{-1}} = \mathcal{C}_{\Omega_\rho}^{-1}$.

Altogether we obtain the multi-matrix analog of~\eqref{eq:woven_contraction_one_matrix_formula_1}:
\begin{multline}
    \label{eq:woven_contraction_multi_matrix_formula_1}
    \mathcal{W}(\zeta; R,S,\vb*{r},\vb*{s},a,c,b,d,\sigma) \\
    = \frac{N^k \dim_R \dim_S \left|G^1_{\tau,\underline{\vb*{n}},\underline{\vb*{n}}'}\right|}{n! n'! \underline{\vb*{n}}! \dim_{\vb*{s}}}
     \sum_{e} 
    \left(\sum_{\Omega, \omega}
        N^{c} \left|\mathcal{C}_{\Omega}\right|
        \chi^{R,\vb*{r}}_{ab}(\Omega)
        \chi^{S,\vb*{s}}_{ed}(\Omega_\rho^{-1})
    \right) \\
    \times
    \left(\sum_{\Omega'}
        N^{\ell(\Omega')} \left|\mathcal{C}_{\Omega'}\right|
        \chi^{S,\vb*{s}}_{ce}(\Omega')
    \right) \pt
\end{multline}
Here $k = k_{\zeta,\sigma,\underline{\vb*{n}},\underline{\vb*{n}}'}$ and $\tau = \tau_{\zeta,\sigma, \underline{\vb*{n}},\underline{\vb*{n}}'}$ are obtained from the contraction of $\zeta$, $\zeta^{-1}$, and $\sigma$ as in the single-matrix case; the sum over $\omega$ runs over a set of representatives of the left cosets $\sym{\underline{\vb*{n}}} / G^1_{\tau,\underline{\vb*{n}},\underline{\vb*{n}}'}$; and $c = c_{\Omega,\omega,\tau}$ together with $\rho = \rho_{\Omega,\omega,\tau} \in \sym{n'}$ are obtained by contracting the restricted class representative $\Omega$ (conjugated by $\omega$) with $\tau$ and absorbing the resulting loop factor. The last factor depends only on $(S,\vb*{s},c,d)$ and is precomputable. By expanding $P^{S,\vb*{s}}_{cd}$ first instead of $P^{R,\vb*{r}}_{ab}$ one obtains a symmetric formula with the roles of $(R,\vb*{r})$ and $(S,\vb*{s})$ exchanged, involving the projection $G^2_{\tau,\underline{\vb*{n}},\underline{\vb*{n}}'}$ of $C_{\sym{\underline{\vb*{n}}} \times \sym{\underline{\vb*{n}}'}}$ onto its second factor. As in the one-matrix case, one chooses whichever expansion yields the smaller number of cosets.

\paragraph{Complexity analysis.}

We outline the complexity for the interaction $\tr(\comm{X_1}{X_2}^2)$ in a fixed excitation sector $(\underline{\vb*{n}},\underline{\vb*{n}}')$, following the same logic as in the one-matrix case. Let $p_N(\underline{\vb*{n}})$ denote the number of admissible basis states $\ket{R,\vb*{r},a,b}$ at finite $N$, and let $p(\underline{\vb*{n}})$ be the number of restricted conjugacy classes of $\sym{n}$ under $\sym{\underline{\vb*{n}}}$, which also equals the unrestricted number of basis states; hence $p_N(\underline{\vb*{n}}) \leq p(\underline{\vb*{n}})$ with equality when $N \geq n$. For this interaction only species $1$ and $2$ carry operator insertions, so $n_I=n_I'$ for $I\geq 3$ and $|n_I-n_I'|\in\{0,2\}$ for $I=1,2$.

Several ingredients entering~\eqref{eq:woven_contraction_multi_matrix_formula_1} are cheaply precomputed once per sector: the subgroup and coset data are obtained via GAP (the number of double cosets is $\bigO{1}$, and the number of left cosets per $\tau$ is at most $\bigO{n_1n_2}$), while closed formulas give $\dim_R$, $\dim_{\vb*{r}}$, and the restricted class sizes $|\mathcal{C}_\Omega|$.

The computational bottleneck is the precomputation of the restricted characters. Unlike ordinary $\sym{n}$ characters, no algorithm comparable to the Murnaghan--Nakayama rule is known; we denote the cost of computing one restricted characters for the sector $\underline{\vb*{n}}$ by $C_\chi(\underline{\vb*{n}})$. At present, the most advanced method~\cite{padellaro2025Eigenvalue,ramgoolam2026Finitedimensional} reformulates the problem as a simultaneous diagonalization of commuting central actions within the centralizer algebra and has been carried out for two matrices up to cutoff $\Lambda=14$. Leveraging the action of the dilatation operator, large-$N$ formulas for these characters were developed in~\cite{decomarmond2011Surprisingly,demellokoch2012Double} within the distant corners limit, where the Young diagram of the ambient representation $R$ has a fixed number of rows whose pairwise size differences are $\Omega(n)$.

Once the restricted characters are available, the last sum in~\eqref{eq:woven_contraction_multi_matrix_formula_1} can be evaluated for every basis state at a cost proportional to $p_N(\underline{\vb*{n}}')\,p(\underline{\vb*{n}}')$.

Evaluating the remaining expression for a given pair of states involves a sum over $\bigO{p(\underline{\vb*{n}})}$ restricted classes and $\bigO{n_1n_2}$ coset representatives, costing $\bigO{n_1n_2\,p(\underline{\vb*{n}})}$ operations per pair. Summing over all $p_N(\underline{\vb*{n}})\,p_N(\underline{\vb*{n}}')$ pairs and adding the precomputation yields
\begin{align}
    \bigO{C_\chi(\underline{\vb*{n}})p(\underline{\vb*{n}})^{2} + C_\chi(\underline{\vb*{n}}')p(\underline{\vb*{n}}')^{2} + n_1n_2 p_N(\underline{\vb*{n}}) p_N(\underline{\vb*{n}}')  p(\underline{\vb*{n}})}
\end{align}
operations for the full excitation sector. In the regime $N\geq n$ the truncation $p_N$ reduces to the unrestricted count $p$, giving
\begin{align}
    \bigO{C_\chi(\underline{\vb*{n}})p(\underline{\vb*{n}})^{2} + C_\chi(\underline{\vb*{n}}')p(\underline{\vb*{n}}')^{2} + n_1n_2 p(\underline{\vb*{n}}')  p(\underline{\vb*{n}})^{2}} \pt
\end{align}

To make the complexity estimate more quantitative we need the asymptotic growth of $p(\underline{\vb*{n}})$, the number of restricted Schur polynomial states in a given excitation sector. By P\'{o}lya enumeration theory, the generating function is~\cite{bhattacharyya2008Exact, pasukonis2011Counting}
\begin{align}
    \sum_{\underline{\vb*{n}}} p(\underline{\vb*{n}}) x^{\underline{\vb*{n}}} = \prod_{i=1}^{\infty} \frac{1}{1 - (x_1^i + \cdots + x_D^i)} \comma
\end{align}
where $x^{\underline{\vb*{n}}} \coloneqq x_1^{n_1} \cdots x_D^{n_D}$. Closed asymptotic formulas for $p(\underline{\vb*{n}})$ are known only for small $D=2$ and in specific scaling regimes.

For $D=2$, this counting coincides with the enumeration of quarter-BPS operators in free $\mathcal{N}=4$ super Yang--Mills theory~\cite{dolan2008Counting,dhoker2003Systematics}. Set $n=n_1+n_2$ and let $n_1/n\to\lambda\in[0,1]$ as $n\to\infty$. The asymptotic growth of $p(n_1,n_2)$ is then~\cite{ramgoolam2020Quiver}
\begin{align}
    p(n\lambda, n(1-\lambda)) \sim \frac{1}{\sqrt{2\pi n}} 
    \left(\frac{1}{\lambda}\right)^{n\lambda + \frac{1}{2}} 
    \left(\frac{1}{1-\lambda}\right)^{(1-\lambda)n + \frac{1}{2}} 
    \mathcal{Z}_2(\lambda, 1-\lambda) \comma
\end{align}
where $\mathcal{Z}_2$ is a known function (an infinite product) of the ratio $\lambda$. This leading-order behavior can be recovered by noticing that the contributions are naturally organized by the size of the minimal cycles in the contributing permutations~\cite{lei2026Critical}.

For $D>2$, closed asymptotic formulas for $p(\underline{\vb*{n}})$ remain an open challenge; recent progress extends the combinatorial small-cycle interpretation to the all-orders asymptotic expansion for general $D$~\cite{lei2026Critical}.

\section{Summary, conclusions, and outlook}
\label{sec:conclusion}

In this work we have presented a new algorithm for computing the spectrum of quantum mechanical models of bosonic Hermitian matrices in the $\U{N}$-singlet sector. The gauge-invariant states we are using are labeled by irreducible representations of the symmetric group together with multiplicity indices and can be explicitly constructed from Schur polynomials \eqref{eq:Schurr_singlets} in the single matrix case and restricted Schur polynomials \eqref{eq:restricted_schur_states} in the multi-matrix case. These states are orthogonal and complete, provide a natural truncation scheme for the Hilbert space, and diagonalize the free Hamiltonian. The algorithm is based on a group-theoretic evaluation of the matrix elements of the interaction Hamiltonian in the respective basis. The method is exact at finite $N$ and can be systematically improved by increasing the cutoff on the excitation number. The results for matrix elements are polynomial expressions in $N$ and, thus, can be evaluated for arbitrary $N$ once the group-theoretic data are precomputed. Larger $N$ usually requires a larger cutoff to reach convergence, but the group-theoretic data can be reused. We have provided a detailed complexity analysis, and for reasonable cutoffs, the algorithm is efficient and the reduction to singlet states rather than the full Fock space allows for a significant reduction of computational resources. We have implemented the algorithm in a publicly available code and demonstrated its performance on the one-matrix model. For multi-matrix models, we lack a systematic method to compute the restricted characters, which is the main bottleneck for the efficient evaluation of the matrix elements. The existing methods to compute the restricted characters are limited to small cutoffs, and we hope that our work will motivate further research in this direction. 

Beyond spectral computations, the framework developed here enables the systematic study of transition amplitudes between excited singlet states with the full interaction Hamiltonian taken into account. Since the Hamiltonian matrix elements $\mel{R}{H}{S}$ (or their restricted Schur counterparts $\mel{R,\vb*{r},a,b}{H}{S,\vb*{s},c,d}$) are assembled, one has direct access to time evolution of arbitrary initial states, real-time scattering processes, and the response of the system to perturbations. This makes it possible to compute, in a controlled and systematically improvable approximation, scattering amplitudes between states corresponding to specific multi-excitation configurations in the singlet sector.

A particularly natural setting for these ideas is the connection to $\U{N}$ gauge theories and the gauge/gravity duality. In the context of $\mathcal{N}=4$ super Yang--Mills theory, Schur polynomials labeled by Young diagrams $R$ have been identified with half-BPS operators, which, via the AdS/CFT correspondence, are dual to specific graviton states and giant graviton configurations in the dual geometry~\cite{corley2002Exact, corley2002Finite}. The restricted Schur polynomials carry composite labels $(R, \vb*{r})$ that admit a natural physical interpretation: the outer label $R$ encodes the total excitation content, while the inner label $\vb*{r}$ captures the distribution among the constituent matrix species, so that $\ket{R, \vb*{r}, a, b}$ can be thought of as describing a system of giant gravitons and their open-string excitations~\cite{bhattacharyya2008Exact, bhattacharyya2008Exacta, demellokoch2011Giant}. The study of the dynamical properties of these states -- their scattering, mixing, and energy renormalization under interactions -- has been actively pursued in the free-field and large-$N$ planar limits. Our recipe provides a complementary approach at finite $N$ and at finite coupling: the same Schur polynomial basis that labels graviton states is used to assemble the fully interacting Hamiltonian matrix, and the resulting framework directly addresses questions about the non-planar and strongly coupled dynamics of these configurations. In this way, the present work extends existing ideas and results beyond the free-field regime.

A natural and technically accessible next extension of the framework is the inclusion of fermionic degrees of freedom. For fermionic matrices transforming in the adjoint representation of $\U{N}$, the gauge-invariant states are constructed from Grassmann-odd creation operators, and the relevant polynomial algebra is antisymmetric rather than symmetric. As a consequence, the Pauli exclusion principle severely restricts the admissible Young diagrams, and the singlet state space at each excitation level is significantly smaller than in the bosonic case. This makes the fermionic sector computationally less demanding and provides a favorable testing ground for extensions of the algorithm. Restricted Schur polynomials for fermionic fields have been introduced in~\cite{demellokoch2013Restricted}, where it was shown that they diagonalize the free two-point function to all orders in $1/N$ and allow the computation of the one-loop dilatation operator in certain sectors. Incorporating the fermionic sector into the present algorithm is therefore a natural and achievable next step. A further natural generalization is to supersymmetric matrix models, where bosonic and fermionic degrees of freedom are paired by supersymmetry. The singlet sector of such models is spanned by restricted Schur polynomials built jointly from both types of fields~\cite{demellokoch2013Restricted}, and the computational machinery developed here carries over directly once the appropriate antinormal ordering and Wick contractions for the fermionic modes are implemented.

Perhaps the most important direction for future work is the extension of the method from $\U{N}$ to $\SU{N}$ gauge symmetry. The BFSS and BMN matrix models, which are the physically most relevant examples in the context of M-theory and holography~\cite{Banks:1996vh, Berenstein:2002jq}, are formulated with $\SU{N}$ gauge group. The central difficulty lies in the construction of an orthonormal basis for the $\SU{N}$-singlet sector. For $\U{N}$, Schur polynomials form an orthogonal and complete basis of gauge-invariant states, as established in this work. For $\SU{N}$, however, the generators are required to be traceless, and this additional constraint introduces nontrivial linear relations among the states constructed from $\SU{N}$-invariant matrices. As a result, the set of Schur polynomial states is no longer linearly independent: it becomes overcomplete, and additional steps must be taken to obtain a proper basis. The extension of the Schur-basis approach to $\SU{N}$ was initiated in~\cite{koch2004Giant}, and further developed in~\cite{brown2008HalfBPS, alcock-zeilinger2019Compact} using a recursive approach; however, such constructions are not computationally efficient for large cutoffs due to the rapid growth of the state space and the complexity of enforcing the tracelessness conditions at each level. An alternative approach -- and potentially a more practical one -- is to retain the overcomplete set of $\SU{N}$ singlet states and to impose the tracelessness constraint implicitly, either through a penalty term added to the Hamiltonian or through a Lagrange multiplier formulation of the eigenvalue problem. This strategy avoids the explicit construction of an orthonormal basis and can be combined with sparse linear algebra to handle the constrained system. Exploring these avenues will be the subject of future work.

\clearpage
\appendix

\section{Overview to the appendices}
\label{app:notation}

This appendix collects the mathematical background and technical derivations that support the main text. Appendix~\ref{app:representation_finite_groups} reviews the representation theory of finite groups and the symmetric group, introduces the center $Z(\C[\sym{n}])$ and the centralizer algebra $\C[\sym{n}]^{\sym{\underline{\vb*{n}}}}$, and estimates the dimensions of the singlet sectors. Appendix~\ref{app:schur-weyl-duality} summarizes the Schur--Weyl duality, which underlies the correspondence between $\U{N}$-invariant tensors and symmetric-group representations. Appendix~\ref{app:bosonic_fock_space} presents the bosonic Fock space formalism, gauge-invariant observables, generators of gauge transformations, Wick's formula, and the antinormal ordering of the interaction terms. Appendix~\ref{app:fermion_mapping} reviews the mapping from the one-matrix singlet sector to a system of $N$ (non-)interacting fermions.

Throughout this work, we employ the Einstein summation convention: repeated implicit upper and lower indices are summed over their full range. For coefficients, lower indices label column positions (inputs), while upper indices label row positions (outputs).

Let $A$ be an $N\times N$ matrix and $X$ a column vector:
\begin{align}
    A = \mqty(A_1^1 & \cdots & A^1_N \\ \vdots & & \vdots \\ A_1^N & \cdots & A_N^N) \qq{and} X = \mqty(X^1 \\ \vdots \\ X^N)  \pt
\end{align} 
Then the matrix-vector product $Y = AX$ has coefficients
\begin{align}
    Y^\ell = A_k^\ell X^k  \comma
\end{align}
where the summation over $k = 1, \dots, N$ is implicit. For tensors with multi-indices $I$, $J$, $K$, the composition of two tensors $T_1$ and $T_2$ satisfies
\begin{align}
    (T_1)^K_J (T_2)^J_I = (T_1T_2)_I^K  \comma
\end{align}
where $T_1T_2$ denotes the composition obtained by first applying $T_2$, then $T_1$.

\section{Elements of representation theory for finite groups}
\label{app:representation_finite_groups}

This section introduces the concepts from the representation theory of finite groups and their group algebras that are used throughout the paper. We begin by recalling how irreducible representations of a finite-dimensional algebra are encoded by primitive idempotents, and then specialize this correspondence to finite groups through the group algebra $\C[G]$. We next introduce Young diagrams and standard Young tableaux as a concrete tool for constructing the irreducible representations of the symmetric group $\sym{n}$. Then, we describe the two algebras that play a central role in the main text, namely the center $Z(\C[\sym{n}])$ relevant to the one-matrix model and the centralizer algebra $\C[\sym{n}]^{\sym{\underline{\vb*{n}}}}$ relevant to the multi-matrix model. We briefly discuss the dimension of the singlet sectors. We conclude by describing the wreath product structure of the subgroups $G^\tau_{1,n}$ and $G^\tau_{2,n}$.

Reference textbooks on the representation theory of finite groups include Fulton and Harris~\cite{fulton2004Representation}. For the symmetric group specifically, we also refer to~\cite{james1984Representation, sagan2010Symmetric}. Additional treatments can be found in~\cite{jacobson1985Basic, tung1985Group, goodman2009Symmetry, kosmann-schwarzbach2022Groups}, as well as in the lecture notes~\cite{alcock-zeilingerSpecial,alcock-zeilingerSymmetric, grinberg2025Introduction}. The use of restricted characters and restricted Schur polynomials to describe singlets was initiated in~\cite{brown2008Diagonal, bhattacharyya2008Exacta, bhattacharyya2008Exact}. See also~\cite{koch2024Pedagogical} for a pedagogical introduction. The centralizer algebra (also called permutation centralizer algebra in the literature) is described in detail in~\cite{mattioli2016Permutation}.

\subsection{Representation theory for algebras: the irreducible representations are given by the primitive idempotents}

The material in this appendix subsection is based on~\cite[Appendix III]{tung1985Group} and~\cite[Section I.4]{assem2006Elements}.

Consider a finite-dimensional (unital associative) algebra $\mathcal{A}$ over the field $\C$, with unit element $1_\mathcal{A}$. A left \emph{$\mathcal{A}$-module} is a vector space equipped with a left multiplication operation by elements of $\mathcal{A}$. A \emph{representation of $\mathcal{A}$} is an algebra homomorphism from $\mathcal{A}$ to $\End{V}$, for some complex vector space $V$, that maps the identity of $\mathcal{A}$ to the identity of $\End{V}$. These two notions, representations of $\mathcal{A}$ and $\mathcal{A}$-modules, are equivalent. 

A \emph{submodule} of an $\mathcal{A}$-module $\mathcal{M}$ is a subspace of $\mathcal{M}$ which is invariant by left multiplication by the elements of $\mathcal{A}$. The \emph{irreducible modules} are those such that their only submodules are either $\{0\}$ or themselves. As expected, there is a one-to-one correspondence between equivalence classes of irreducible submodules and equivalence classes of irreducible representation of $\mathcal{A}$.   

The \emph{left regular $\mathcal{A}$-module} is the vector space $\mathcal{A}$ itself, where the left multiplication is the algebra multiplication. A fundamental result states that the left regular module decomposes as a direct sum of all irreducible submodules, each appearing with multiplicity equal to its dimension as a complex vector space. Consequently, studying the irreducible representations of $\mathcal{A}$ reduces to studying the irreducible submodules of the left regular $\mathcal{A}$-module (also known as minimal left ideals of $\mathcal{A}$).

These minimal submodules are characterized by primitive idempotents. An \emph{idempotent} element $e \in \mathcal{A}$ satisfies $e^2=e$. Two idempotents $e_1, e_2$ are \emph{orthogonal} if $e_1e_2=e_2e_1=0$. An idempotent $e$ is \emph{primitive} if it cannot be written as $e=e_1+e_2$ with $e_1$, $e_2$ nonzero orthogonal idempotents. The key result is that $e$ is a primitive idempotent if and only if $\mathcal{A}e$ is an irreducible submodule. Moreover, every irreducible submodule of the left regular $\mathcal{A}$-module have the form $\mathcal{A}e$ for $e$ some primitive idempotent.

Finding a decomposition in indecomposables of $\mathcal{A}$ seen as an $\mathcal{A}$-module amounts to finding a complete set of primitive orthogonal idempotents of $\mathcal{A}$, that is pairwise orthogonal idempotents $e_1,\dots,e_k \in \mathcal{A}$ such that
\begin{align}
    1_{\mathcal{A}} = e_1 + \cdots + e_k  \pt
\end{align}
In that case, we have the direct sum decomposition
\begin{align}
    \mathcal{A} = \mathcal{A}e_1 \oplus \cdots \oplus \mathcal{A}e_k  \pt
\end{align}
As a consequence of Schur's lemma, we have the following characterization of the primitive idempotents~\cite[Theorem III.3]{tung1985Group}: an idempotent $e$ is primitive if and only if
\begin{align}
    \forall r\in \mathcal{A} ,~ \exists \lambda_r \in \C , ~ e r e = \lambda_r e  \pt
\end{align}
Besides (see~\cite[Theorem III.4]{tung1985Group}): two primitive idempotents $e_1$ and $e_2$ generate equivalent representations if and only if
\begin{align}
    \exists r \in \mathcal{A} , ~ e_1 r e_2 \neq 0  \pt
\end{align}

\subsection{Group algebra of a finite group}
\label{app:group_algebra}

Consider a finite group $G$ of order $n$. Its \emph{group algebra} over the field $\C$ is the vector space
\begin{align}
    \C[G] = \enstq{\sum_{g \in G} \lambda_g g}{\lambda_g \in \C}  \comma
\end{align}
where the multiplication operation is inherited from the group structure. Notice that each element can be viewed as a $\C$-valued function on $G$.

Representations of $G$ and representations of its group algebra $\C[G]$ are in natural correspondence. If $\varphi_G: G \to \GL{V}$ is a representation of $G$ on a vector space $V$, then extending by linearity defines a representation of the group algebra:
\begin{align}
    \varphi_{\C[G]}\left(\sum_{g\in G} \lambda_g g\right) = \sum_{g\in G} \lambda_g \varphi_G(g)  \pt
\end{align}
Conversely, each representation of $\C[G]$ restricts to a representation of the underlying group $G$. Moreover, $\varphi_G$ is irreducible if and only if $\varphi_{\C[G]}$ is irreducible.

The left regular representation $R_G:G \to \Aut{G}$ of $G$ is defined by left multiplication:
\begin{align}
    R_G(g)h = g\cdot h  \pt
\end{align}
A fundamental result states that the left regular representation contains all irreducible representations of $G$, each appearing with multiplicity equal to its dimension. The corresponding object at the level of the group algebra is $\C[G]$ itself, viewed as a left $\C[G]$-module via the left regular representation:
\begin{align}
    R_{\C[G]}\left(\sum_{g\in G} \lambda_g g\right) = \sum_{g\in G} \lambda_g g  \comma
\end{align}
where the module multiplication is the algebra multiplication of $\C[G]$.

Therefore, studying the irreducible representations of $G$ reduces to studying the irreducible submodules of the left regular $\C[G]$-module. As established in the previous section, this is equivalent to determining the primitive idempotents of $\C[G]$.

The correspondence between primitive idempotents and irreducible representations proceeds as follows. Let $\varphi_G$ be an irreducible subrepresentation of the left regular representation of $G$, realized on a vector space $V \subseteq \C[G]$. Then $V$ is a minimal left ideal of $\C[G]$, generated by some primitive idempotent $e \in \C[G]$: $V = \C[G] e$. Conversely, given a primitive idempotent $e$, the corresponding irreducible representation is supported on $\C[G] e$ with the group action inherited from left multiplication in the group algebra. Explicitly, for $g \in G$ and $v = \left(\sum_h \mu_h h\right)e \in V$, the representation acts as
\begin{align}
    \varphi_V(g) v = g\left(\sum_h \mu_h h\right)e \in V  \pt
\end{align}

\subsection{Young diagrams and the representation theory of the symmetric group}
\label{app:young_diagrams}

In this section, we explain how to construct a basis of primitive idempotents when $G = \sym{n}$ is the symmetric group.

A \emph{partition} $\lambda$ of a positive integer $n$ (denoted $\lambda \vdash n$) is a sequence $\lambda = (\lambda_1, \lambda_2, \dots, \lambda_k)$ of positive integers satisfying $\lambda_1 \geq \lambda_2 \geq \cdots \geq \lambda_k$ and $\sum_{i=1}^k \lambda_i = n$. The \emph{Young diagram} $R_\lambda$ associated with partition $\lambda$ is a graphical representation consisting of $k$ left-aligned rows, where the $i$-th row contains $\lambda_i$ boxes. For instance, for $n=6$, we have:
\begin{align*}
    R_{(3,2,1)} = \ydiagram{3,2,1} \qq{and} R_{(4,2)} = \ydiagram{4,2}  \pt
\end{align*}
The set of Young diagrams with $n$ boxes is denoted by $\mathcal{D}_n$. We also often write $R \vdash n$ to indicate that $R\in \mathcal{D}_n$.

Young diagrams play a fundamental role in the representation theory of the symmetric group: they label the irreducible representations of $\sym{n}$. More precisely, there exists a one-to-one explicit correspondence between partitions of $n$ and irreducible representations of $\sym{n}$:
\begin{align}
    \{\text{irreps of } \sym{n}\} \overset{\sim}{\longleftrightarrow} \{\lambda \vdash n\}  \overset{\sim}{\longleftrightarrow}  \mathcal{D}_n  \pt
\end{align}
In the remainder of this section, we describe how to effectively realize the irreps of $\sym{n}$.

A \emph{Young tableau} of shape $R\in \mathcal{D}_n$ is a filling of the boxes of $R$ with the integers $1, 2, \dots, n$ (each used exactly once). A \emph{standard Young tableau} of shape $R$ is a Young tableau of shape $R$ with its entries increasing along each row and down each column. For instance, for the partition $(3,2,1) \vdash 6$,
\begin{align*}
    \Theta = \ytableausetup{centertableaux}
    \begin{ytableau}
    1 & 2 & 4 \\
    3 & 5  \\
    6
    \end{ytableau}
    \qq{is a valid standard Young tableau,}
\end{align*}
while the Young tableau
\begin{align*}
    \Theta = \ytableausetup{centertableaux}
    \begin{ytableau}
    1 & 5 & 6 \\
    2 & 4  \\
    3
    \end{ytableau}
    \qq{is not,}
\end{align*}
as the entry 4 in the second row is smaller than 5 in the first row of the same column. We denote the set of all standard Young tableaux of shape $R$ by $\mathrm{SYT}(R)$.

From a (not necessarily standard) Young tableau $\Theta$, one can construct an explicit primitive idempotent $Y_\Theta$ of $\C[\sym{n}]$. Let $R(\Theta)$ be the subgroup of $\sym{n}$ consisting of permutations that preserve the set of numbers in each row of $\Theta$, and let $C(\Theta)$ be the subgroup that preserves the numbers in each column. We introduce two elements in the group algebra $\C[\sym{n}]$: the row symmetrizer $S_\Theta \coloneqq \sum_{\sigma \in R(\Theta)} \sigma$ and the column anti-symmetrizer $A_\Theta \coloneqq \sum_{\sigma \in C(\Theta)} \mathrm{sgn}(\sigma) \sigma$. Then, the \emph{Young projector operator}
\begin{align}
    Y_\Theta \coloneqq a_\Theta S_\Theta A_\Theta \comma
\end{align}
with $a_\Theta$ an explicit constant, is a primitive idempotent. The minimal left ideal
\begin{align}
    V_\Theta \coloneq \C[\sym{n}]Y_\Theta  \comma
\end{align}
is an irreducible representation of $\sym{n}$ corresponding to the shape $R$. The Young projector operator $Y_\Theta$ acts, by right multiplication, as a projector from $\C[\sym{n}]$ onto $V_\Theta$. The irreps $V_\Theta$ and $V_\Phi$ are equivalent if and only if the tableaux $\Theta$ and $\Phi$ have the same shape $R$. We can denote by $V_R$ the equivalence class of irreps generated by tableaux of shape $R$. Furthermore, there are no others: every irreducible representation of $\C[\sym{n}]$ arises as $V_R$ for some $R \in \mathcal{D}_n$.

The dimension of $V_R$ is equal to the number of standard Young tableaux of shape $R$~\cite[Theorem 2.6.5]{sagan2010Symmetric}, which can be computed efficiently using the \emph{hook-length formula}:
\begin{align}
    \label{eq:hook_length_formula}
    \dim_R = \abs{\mathrm{SYT}(R)} = \frac{n!}{\prod_{(i,j)\in R} h_R(i,j)}  \comma
\end{align}
where $h_R(i,j)$ denotes the hook length of the cell $(i,j)$ in diagram $R$, defined as the number of cells directly to the right of $(i,j)$, plus the number of cells directly below $(i,j)$, plus one (for the cell itself). The formula for the constant $a_\Theta$ follows:
\begin{align}
    a_\Theta = a_R = \dim_R / n! = \prod_{(i,j)\in R} h_R(i,j)^{-1}  \pt
\end{align}

The idempotents $Y_\Theta$, where $\Theta$ ranges over all Young tableaux of size $n$, span the entire group algebra $\C[\sym{n}]$ but are linearly dependent. Restricting to standard Young tableaux yields a complete and linearly independent set of primitive idempotents~\cite[Theorem 5.8]{tung1985Group}. However, these idempotents fail to be mutually orthogonal for $n \geq 5$.

\subsection{The center of \texorpdfstring{$\C[\sym{n}]$}{C[Sn]}}
\label{app:center_group_algebra}

For the one-matrix problem, the relevant commutative algebra is the \emph{center} of the group algebra of the symmetric group:
\begin{align}
    Z(\C[\sym{n}]) = \enstq{v \in \C[\sym{n}]}{\forall \sigma \in \sym{n},~ \sigma v = v \sigma}  \pt
\end{align}
In this subsection, we describe two bases for this subalgebra.

Recall that two permutations $\sigma,\tau \in \sym{n}$ are \emph{conjugate} if there exists $g \in \sym{n}$ such that $\tau = g\sigma g^{-1}$. The conjugacy classes, as the irreps of $\sym{n}$, are indexed by partitions $\alpha \vdash n$, and we denote by $\mathcal{C}_\alpha$ the class of cycle type $\alpha$.

If $R \vdash n$ labels an irreducible representation $\Gamma_R\colon \sym{n} \to \End{V_R}$, its \emph{character} is the class function
\begin{align}
    \chi_R(\sigma) \coloneqq \tr_{V_R}\!\left(\Gamma_R(\sigma)\right)  \pt
\end{align}
When evaluated on the identity, the character gives the dimension of the corresponding irrep:
\begin{align}
    \chi_R(\id) = \dim(V_R) = \dim_R  \pt
\end{align}
Since $\chi_R$ is constant on conjugacy classes, we also write $\chi_R(\alpha)$ for its value on $\mathcal{C}_\alpha$. 

A first basis for the center $Z(\C[\sym{n}])$ is given by the (conjugacy-)\emph{class sums}
\begin{align}
    [\alpha] \coloneqq \sum_{\sigma \in \mathcal{C}_\alpha} \sigma  , \quad \alpha \vdash n  \pt
\end{align}
While straightforward to define, this basis is not orthogonal. The multiplication law
\begin{align}
    [\alpha][\beta] = \sum_{\gamma \vdash n} C^\gamma_{\alpha\beta} [\gamma]
\end{align}
defines the \emph{structure constants} $C^\gamma_{\alpha\beta}$ of the class algebra.

A second basis of the center, indexed by Young diagrams, enjoys better properties. First, recall that the Fourier transform identifies the center with the commutative algebra of scalar endomorphisms on the irreducible blocks~\cite[Proposition 4.3.8]{goodman2009Symmetry}
\begin{align}
    Z(\C[\sym{n}]) \simeq \bigoplus_{R\vdash n} \operatorname{End}_{\sym{n}}(V_R) \simeq \bigoplus_{R\vdash n} \C   \comma
\end{align}
where $\operatorname{End}_{\sym{n}}(V_R)$, the algebra of $\sym{n}$-intertwining operators on $V_R$, is isomorphic to $\C$ by Schur's lemma. We denote by $P_R$ the inverse image of the natural basis vector supported on the irrep $R$, namely the element acting as $1$ on the $R$-block and as $0$ on all other blocks. Explicitly, for all $R\vdash n$, we have
\begin{align}
    P_R
    \coloneqq \frac{\dim_R}{n!} \sum_{\sigma \in \sym{n}} \chi_R(\sigma)\sigma
    = \frac{\dim_R}{n!} \sum_{\alpha \vdash n} \chi_R(\alpha) [\alpha]  \pt
\end{align}
Then, the family $\{P_R\}_{R \vdash n}$ is a basis of $Z(\C[\sym{n}])$ and enjoys orthogonality: 
\begin{align}
    P_R P_S = \delta_{RS} P_R  \pt
\end{align}
The idempotent $P_R$ is also, up to a prefactor, the average of the Young projector operators $Y_\Theta$ over the Young tableaux $\Theta$ with shape $R$, see~\cite[Proposition 9.3.14]{goodman2009Symmetry}.

Starting from the basis $\{P_R : R\vdash n \}$, one can easily obtain a closed formula for the structure constants $C_{\alpha\beta}^\gamma$. First, inverting the character transform~\cite[Section 4.3]{goodman2009Symmetry} gives
\begin{align}
    [\alpha] = \sum_{R \vdash n} \frac{|\mathcal{C}_\alpha|\chi_R(\alpha)}{\dim_R} P_R  \pt
\end{align}
Then, using the orthogonality $P_R P_S = \delta_{RS} P_R $, we obtain:
\begin{align}
    \label{eq:center_structure_constants}
    C^\gamma_{\alpha\beta}
    =
    \frac{|\mathcal{C}_\alpha||\mathcal{C}_\beta|}{n!}
    \sum_{R \vdash n}
    \frac{\chi_R(\alpha)\chi_R(\beta)\chi_R(\gamma)}{\dim_R}  \pt
\end{align}

\subsection{The centralizer algebra \texorpdfstring{$\C[\sym{n}]^{\sym{\underline{\vb*{n}}}}$}{C[Sn]Sn}}
\label{app:centralizer_group_algebra}

For the several-matrix problem, the relevant algebra is no longer the center, but the centralizer algebra associated with the Young subgroup
\begin{align}
    \sym{\underline{\vb*{n}}} \coloneqq \sym{n_1} \times \cdots \times \sym{n_D} \subset \sym{n}  ,
    \quad n_1 + \cdots + n_D = n  \comma
\end{align}
where we have denoted $\underline{\vb*{n}} = (n_1,\dots,n_D)$. Two permutations $\sigma,\tau \in \sym{n}$ are said to be \emph{$\sym{\underline{\vb*{n}}}$-conjugate} if there exists $h \in \sym{\underline{\vb*{n}}}$ such that $\tau = h\sigma h^{-1}$. The corresponding $\sym{\underline{\vb*{n}}}$-orbits are the \emph{restricted conjugacy classes}. We use the symbol $\Omega$ to label a generic restricted conjugacy class, and we denote by $\mathcal{C}_{\Omega}$ the associated class. Then restricted class sums
\begin{align}
    [\Omega]\coloneqq \sum_{\sigma \in \mathcal{C}_\Omega} \sigma
\end{align}
form a basis of the \emph{centralizer algebra}
\begin{align}
    \C[\sym{n}]^{\sym{\underline{\vb*{n}}}} \coloneqq \enstq{x \in \C[\sym{n}]}{hxh^{-1} = x,\ \forall h \in \sym{\underline{\vb*{n}}}}  \pt
\end{align}
Like the single-matrix basis, this basis is not orthogonal. The multi-matrix structure constants are defined by the product
\begin{align}
    \label{eq:restricted_structure_constants}
    [\Omega] [\Omega'] = \sum_{\Omega''} f_{\Omega,\Omega'}^{\Omega''} [\Omega'']  \pt
\end{align}
We now construct an orthogonal basis that yields, analogously to the single-matrix case, an explicit formula for these structure constants.

First, we describe the irreducible representations of $\sym{\underline{\vb*{n}}} = \sym{n_1} \times \cdots \times \sym{n_D}$. They are given by tensor products of irreducible representations of the factors. Equivalently, they are indexed by $D$-tuples of Young diagrams
\begin{align}
    V_{\vb*{r}} = V_{r_1} \otimes \cdots \otimes V_{r_D} \qq{with} \vb*{r} = (r_1,\dots,r_D), \qquad r_i \vdash n_i  \comma
\end{align}
and where $(\sigma_1,\dots,\sigma_D) \in \sym{\underline{\vb*{n}}}$ acts as $\Gamma_{r_1}(\sigma_1) \otimes \cdots \otimes \Gamma_{r_D}(\sigma_D)$.

Let $R \vdash n$ be an irreducible representation of $\sym{n}$. Upon restriction to $\sym{\underline{\vb*{n}}}$, one has the following decomposition into isotypic components~\cite[Section 4.1.6]{goodman2009Symmetry}
\begin{align}
    \label{eq:decomposition_isotypic_centralizer}
    \operatorname{Res}_{\sym{\underline{\vb*{n}}}}^{\sym{n}} V_R \simeq \bigoplus_{\vb*{r}} V_{\vb*{r}} \otimes M^R_{\vb*{r}}  \comma
\end{align}
where $\vb*{r}$ runs over irreducible representations of $\sym{\underline{\vb*{n}}}$ and $M^R_{\vb*{r}}$ is the corresponding multiplicity space. This space has dimension the multiplicity of $V_{\vb*{r}}$ in $\operatorname{Res}_{\sym{\underline{\vb*{n}}}}^{\sym{n}} V_R$ and is isomorphic to
\begin{align}
    \operatorname{End}_{\sym{\underline{\vb*{n}}}}(V_{\vb*{r}}, \operatorname{Res}_{\sym{\underline{\vb*{n}}}}^{\sym{n}} V_R) \comma
\end{align}
the algebra of intertwining maps from $V_{\vb*{r}}$ to $\operatorname{Res}_{\sym{\underline{\vb*{n}}}}^{\sym{n}} V_R$, with respect to the action of $\sym{\underline{\vb*{n}}}$. The Fourier transform~\cite[Theorem 4.3.1]{goodman2009Symmetry} provides an algebra isomorphism
\begin{align}
    \C[\sym{n}] \simeq \bigoplus_{R\vdash n} \End{V_R} \comma
\end{align}
from which the subalgebra $\C[\sym{n}]^{\sym{\underline{\vb*{n}}}}$ is obtained by taking in each $R$-block the commutant of the restricted $\sym{\underline{\vb*{n}}}$-action. By Schur's lemma and the decomposition~\eqref{eq:decomposition_isotypic_centralizer}, this commutant is precisely
\begin{align}
    \operatorname{End}_{\sym{\underline{\vb*{n}}}}(\operatorname{Res}_{\sym{\underline{\vb*{n}}}}^{\sym{n}} V_R ) \simeq \bigoplus_{\vb*{r}} \End{M^R_{\vb*{r}}}  \comma
\end{align}
so collecting all irreducible $R$ gives the full centralizer algebra. Choosing matrix units $E^{R,\vb*{r}}_{ab} \in \End{M^R_{\vb*{r}}}$, one obtains the algebra decomposition
\begin{align}
    \label{eq:centralizer_decomposition}
    \C[\sym{n}]^{\sym{\underline{\vb*{n}}}} \simeq \bigoplus_{R,\vb*{r}} \End{M^R_{\vb*{r}}}  \pt
\end{align}
Then, the inverse Fourier transform~\cite[Theorem 4.3.4]{goodman2009Symmetry} provides a basis $P^{R,\vb*{r}}_{ab}$ of $\C[\sym{n}]^{\sym{\underline{\vb*{n}}}}$:
\begin{align}
    \label{eq:p_restricted}
    P^{R,\vb*{r}}_{ab}
    = \frac{\dim_R}{n!} \sum_{\sigma \in \sym{n}} \chi^{R,\vb*{r}}_{ab}(\sigma)\sigma \comma
\end{align}
where the coefficients are the \emph{restricted characters}, defined as 
\begin{align}\label{eq:restricted_character_definition}
    \chi^{R,\vb*{r}}_{ab}(\sigma)
    \coloneqq \tr_{V_R}\!\left(E^{R,\vb*{r}}_{ab}\Gamma_R(\sigma^{-1})\right)  ,
    \qquad \sigma \in \sym{n}  \pt
\end{align}
Note that these restricted characters are constant over the restricted conjugacy classes $\mathcal{C}_\Omega$. Denoting by $\chi^{R,\vb*{r}}_{ab}(\Omega)$ the value of $\chi^{R,\vb*{r}}_{ab}$ over one such class, we have
\begin{align}
    P^{R,\vb*{r}}_{ab} = \frac{\dim_R}{n!} \sum_{\Omega} \chi^{R,\vb*{r}}_{ab}(\Omega)[\Omega]  \pt
\end{align}
The $P^{R,\vb*{r}}_{ab}$'s are the several-matrix analogs of the central projectors $P_R$, and they satisfy the same multiplication law as the $E^{R,\vb*{r}}_{ab}$ basis:
\begin{align}
    \label{eq:multiplication_centralizer_basis}
    P^{R,\vb*{r}}_{ab} P^{S,\vb*{s}}_{cd}
    = \delta_{RS}\delta_{\vb*{r}\vb*{s}}\delta_{bc}P^{R,\vb*{r}}_{ad}  \pt
\end{align}
It implies that the restricted characters satisfy the analogs of the usual first and second orthogonality relations. More precisely, writing $\dim_{\vb*{r}} \coloneqq \dim V_{\vb*{r}} = \prod_{I=1}^D \dim_{r_I}$, one has
\begin{align}
    \label{eq:restricted_character_orthogonality_1}
    \frac{1}{n!}\sum_{\sigma\in\sym{n}}
    \chi^{R,\vb*{r}}_{ab}(\sigma)\chi^{S,\vb*{s}}_{cd}(\sigma^{-1})
    = \frac{1}{n!}\sum_{\Omega}
    \abs{\mathcal{C}_\Omega }\chi^{R,\vb*{r}}_{ab}(\Omega)\chi^{S,\vb*{s}}_{cd}(\Omega^{-1}) 
    = \frac{\dim_{\vb*{r}}}{\dim_R}
    \delta_{RS}\delta_{\vb*{r}\vb*{s}}\delta_{ad}\delta_{bc} \comma
\end{align}
where $\Omega^{-1}$ is the label such that $\mathcal{C}_{\Omega^{-1}} = \mathcal{C}_\Omega^{-1}$. For two labels $\Omega_1$ and $\Omega_2$, one has
\begin{align}
    \label{eq:restricted_character_orthogonality_2}
    \sum_{R,\vb*{r}}\sum_{a,b}
    \frac{\dim_R}{\dim_{\vb*{r}}}
    \chi^{R,\vb*{r}}_{ab}(\Omega_1)
    \chi^{R,\vb*{r}}_{ba}(\Omega_2^{-1})
    = \frac{n!}{\abs{\mathcal{C}_{\Omega_1}}}\delta_{\Omega_1\Omega_2} \pt
\end{align}
The expansion of the restricted class sums with respect to this basis then follows from the second orthogonality relation:
\begin{align}
    [\Omega] = \sum_{R,\vb*{r}} \sum_{a,b} \frac{\abs{\mathcal{C}_\Omega}}{\dim_{\vb*{r}}} \chi^{R,\vb*{r}}_{ba}(\Omega^{-1}) P^{R,\vb*{r}}_{ab}  \pt
\end{align}
If we define the matrix $F_{\Omega,R,\vb*{r}}$ as
\begin{align}
    \left(F_{\Omega,R,\vb*{r}}\right)_{a,b} \coloneqq \chi^{R,\vb*{r}}_{ba}(\Omega^{-1})  \comma
\end{align}
then the structure constants in~\eqref{eq:restricted_structure_constants} are given by
\begin{align}
    \label{eq:restricted_class_sum_structure_constants}
    f_{\Omega,\Omega'}^{\Omega''}
    = \frac{\abs{\mathcal{C}_\Omega} \abs{\mathcal{C}_{\Omega'}}}{n!}
    \sum_{R,\vb*{r}} \frac{\dim_R}{\dim_{\vb*{r}}^2} \tr(F_{\Omega,R,\vb*{r}} F_{\Omega',R,\vb*{r}} F_{(\Omega'')^{-1},R,\vb*{r}})  \pt
\end{align}
When $\sym{\underline{\vb*{n}}} = \sym{n}$, we recover the structure constants of the center~\eqref{eq:center_structure_constants}. This is the restricted analog of the ordinary Fourier analysis of the center: ordinary characters are replaced by restricted characters, and the scalar projectors $P_R$ are replaced by the matrix units $P^{R,\vb*{r}}_{ab}$ acting on the multiplicity spaces. 

\subsection{Dimension counting}
\label{app:dimension_counting}

For the one-matrix model, the dimension of the singlet sector at excitation level $n$ equals
\begin{align}
    \dim Z(\C[\sym{n}]) = p(n)  \comma
\end{align}
where $p(n)$ is the number of partitions of $n$. Imposing a cutoff at excitation level $\Lambda$, the total dimension of the truncated singlet basis becomes
\begin{align}
    \mathcal{P}_1(\Lambda) = \sum_{n=0}^{\Lambda} p(n)  \pt
\end{align}
This formula applies in the regime $N \geq \Lambda$, where every partition of every $n \leq \Lambda$ labels an admissible $\U{N}$ representation. For $N < \Lambda$, one must truncate the basis by retaining only the Young diagrams $R \vdash n$ with $\ell(R) \leq N$.

Although $p(n)$ grows exponentially,
\begin{align}
    p(n) \sim \frac{1}{4n\sqrt{3}} \exp\left(\pi \sqrt{\frac{2n}{3}}\right)  \comma
\end{align}
this growth remains moderate from a practical point of view: $\mathcal{P}_1(20) = 2714$, $\mathcal{P}_1(25) = 9296$, and $\mathcal{P}_1(30) = 28629$.

For several matrices, one fixes a decomposition $n=n_1+\cdots+n_D$ and uses \eqref{eq:centralizer_decomposition} to write
\begin{align}
    \dim \C[\sym{n}]^{\sym{\underline{\vb*{n}}}} = \sum_{R,\vb*{r}} (m^R_{\vb*{r}})^2  \comma
\end{align}
where $m^R_{\vb*{r}} = \dim M^R_{\vb*{r}}$ is the dimension of the multiplicity space. Summing over all compositions of $n$ then gives the dimension of the singlet sector at excitation level $n$:
\begin{align}
    \mathcal{P}_D(\Lambda) = \sum_{n_1,\dots,n_D} \sum_{R,\vb*{r}} (m^R_{\vb*{r}})^2  \comma
\end{align}
where the first sum spans $(n_1,\dots, n_D) \in \N^D$ such that $n_1+\dots+n_D \leq \Lambda$. Again, this expression is valid in the regime $N \geq \Lambda$. For $N < \Lambda$, the finite-$N$ singlet basis is obtained by keeping only the labels where all the diagrams have less than $N$ rows.

The multiplicities $m^R_{\vb*{r}}$ are the generalized Littlewood-Richardson coefficients associated with the restriction of the irreducible $\sym{n}$-representation $R$ to the Young subgroup $\sym{\underline{\vb*{n}}}=\sym{n_1}\times\cdots\times\sym{n_D}$. They can be computed using Frobenius reciprocity:
\begin{align}
    m^R_{\vb*{r}} = \langle \operatorname{Res}_{\sym{\underline{\vb*{n}}}}^{\sym{n}}\chi_R, \chi_{\vb*{r}} \rangle_{\sym{\underline{\vb*{n}}}} = \left\langle \chi_R, \operatorname{Ind}_{\sym{\underline{\vb*{n}}}}^{\sym{n}} \chi_{\vb*{r}} \right\rangle_{\sym{n}}  \pt
\end{align}
For two matrices, this reduces to the usual Littlewood-Richardson rule, which describes the product of two Schur polynomials,
\begin{align}
    s_{\vb*{r}_1}s_{\vb*{r}_2} = \sum_{R\vdash n} c^{R}_{\vb*{r}_1,\vb*{r}_2} s_R  \comma
\end{align}
so that the multiplicities are precisely the Littlewood-Richardson coefficients $m^R_{(\vb*{r}_1,\vb*{r}_2)} = c^{R}_{\vb*{r}_1,\vb*{r}_2}$. For more than two matrices, one may then iterate this rule by combining the factors successively. Although this recursive strategy is not optimal from an algorithmic point of view, it is fully sufficient for the range of parameters relevant here. To compute them in practice, we use SageMath, the Sage Mathematics Software System (Version 10.7)~\cite{thesagemathdevelopers2025SageMath}, which can evaluate the Littlewood-Richardson coefficients \emph{out-of-the-box}.

\begin{table}[h!]
\centering
\begin{tabular}{c|ccccccccc}
    \toprule
    \diagbox{$D$}{$N$} & \textbf{$2$} & \textbf{$3$} & \textbf{$4$} & \textbf{$5$} & \textbf{$6$} & \textbf{$7$} & \textbf{$8$} & \textbf{$9$} & \textbf{$10$} \\
    \midrule
    $1$ & $1997$ & $326$ & $149$ & $99$ & $78$ & $68$ & $61$ & $58$ & $55$ \\
    $2$ & $58$ & $27$ & $20$ & $18$ & $17$ & $17$ & $17$ & $17$ & $17$ \\
    $3$ & $21$ & $14$ & $12$ & $11$ & $11$ & $11$ & $11$ & $11$ & $11$ \\
    $4$ & $14$ & $10$ & $9$ & $9$ & $9$ & $9$ & $9$ & $9$ & $9$ \\
    $5$ & $11$ & $8$ & $8$ & $8$ & $8$ & $8$ & $8$ & $8$ & $8$ \\
    $6$ & $9$ & $7$ & $7$ & $7$ & $7$ & $7$ & $7$ & $7$ & $7$ \\
    $7$ & $8$ & $7$ & $6$ & $6$ & $6$ & $6$ & $6$ & $6$ & $6$ \\
    $8$ & $7$ & $6$ & $6$ & $6$ & $6$ & $6$ & $6$ & $6$ & $6$ \\
    $9$ & $7$ & $6$ & $6$ & $6$ & $6$ & $6$ & $6$ & $6$ & $6$ \\
    \bottomrule
\end{tabular}
\caption{Maximal excitation cutoff $\Lambda$ compatible with a budget of $10^6$ singlet states}
\label{tab:budget_cutoffs}
\end{table}

From a practical perspective, what matters is not only the size of the basis but also the sparsity of the Hamiltonian matrices: each interaction term connects only a small fraction of the singlet states. This sparsity makes it possible to handle bases that would be prohibitive in a dense representation. With this in mind, we adopt a working budget of $10^6$ singlet states, and show in Table~\ref{tab:budget_cutoffs} the maximal excitation cutoffs $\Lambda$ compatible with this budget for gauge-group ranks $N=2,\dots,10$ and matrix numbers $D=1,\dots,9$.

\subsection{Structure of the stabilizer subgroup}
\label{app:stabilizer_subgroup}

In Section~\ref{sec:computation_matrix_elements}, we introduced the subgroups $G^1_{\tau,n} \subset \sym{n}$ and $G^2_{\tau,n} \subset \sym{n}$, defined as the projection onto the first and second factor of the centralizer of the permutation $\tau \in \sym{n+n'}$ inside $\sym{n} \times \sym{n'}$. This appendix describes the explicit wreath product structure of this subgroup and gives a closed formula for its order.

The centralizer $C_{\sym{n} \times \sym{n'}}(\tau)$ consists of all pairs $(\nu, \mu) \in \sym{n} \times \sym{n'}$ such that $(\nu \otimes \mu) \tau = \tau(\nu \otimes \mu)$. Equivalently, we have
\begin{align}
    C_{\sym{n} \times \sym{n'}}(\tau) \cong C_{\sym{n+n'}}(\tau) \cap (\sym{n} \times \sym{n'}) \comma
\end{align}
the intersection of the full centralizer of $\tau$ with the Young subgroup $\sym{n} \times \sym{n'} \subset \sym{n+n'}$.

To describe this group explicitly, label the elements $\{1, \dots, n+n'\}$ by a color: $A$ for positions $\{1, \dots, n\}$ and $B$ for $\{n+1, \dots, n+n'\}$. For each cycle $C = (x_0, \dots, x_{L-1})$ of length $L$ in the cycle decomposition of $\tau$, define its \emph{color profile} as the cyclic word $(\text{color}(x_0), \dots, \text{color}(x_{L-1}))$. Two cycles are \emph{equivalent} if their color profiles match up to a cyclic shift. Partition the cycles of $\tau$ into equivalence classes $E$; for each class let $r_E$ be the number of cycles it contains, $L_E$ their common length, and $m_E$ the fundamental period of the color profile word (the smallest non-zero shift preserving the word). Note that $m_E \mid L_E$.

An element $g \in \sym{n+n'}$ commuting with $\tau$ and preserving the $A$/$B$ partition must, for each class $E$: (i) permute the $r_E$ cycles arbitrarily, giving a factor $\sym{r_E}$; and (ii) within each cycle, apply a cyclic shift by a multiple of $m_E$, giving a factor $\Z_{L_E/m_E}$. The full centralizer is therefore a direct product of wreath products~\cite{james1984Representation}:
\begin{align}
    C_{\sym{n} \times \sym{n'}}(\tau) \cong \prod_{E} \left(\Z_{L_E/m_E} \wr \sym{r_E}\right) \pt
\end{align}

The subgroup $G^1_{\tau,n}$ is the projection of this group onto the first factor $\sym{n}$. The projection decomposes over the same cycle classes $E$, and its structure on each class depends on how the cycles intersect $A = \{1, \dots, n\}$:
\begin{itemize}
    \item \textbf{Cycles entirely in $B$.} These contain no elements of $A$; their projection onto $\sym{n}$ is trivial.
    \item \textbf{Cycles meeting $A$.} The valid internal shifts are multiples of $m_E$; such a shift maps every $A$-element to another $A$-element (the color profile is preserved), and distinct shifts produce distinct permutations on $A$. Hence, the internal shift factor $\Z_{L_E/m_E}$ acts faithfully on $A$, as do the block permutations $\sym{r_E}$. The projected component is the full wreath product $\Z_{L_E/m_E} \wr \sym{r_E}$, of order $(L_E/m_E)^{r_E} \, r_E!$.
\end{itemize}

Summing over all classes that contain at least one $A$-element, we obtain
\begin{align}
    G^1_{\tau,n} \cong \prod_{E \not\subseteq B} \bigl( \Z_{L_E/m_E} \wr \sym{r_E} \bigr) \comma
\end{align}
with order
\begin{align}
    \abs{G^1_{\tau,n}} = \prod_{E \not\subseteq B} \bigl(L_E / m_E\bigr)^{r_E} \, r_E! \pt
\end{align}
Analogous formulas hold for the projection $G^2_{\tau,n}$ onto the second factor.

For the permutations $\tau$ that appear in the decomposition of $\tr(X^4)$ (see Section~\ref{sec:computation_matrix_elements}), the number of left cosets $\abs{\sym{n} / G^1_{\tau,n}} = n! / \abs{G^1_{\tau,n}}$ is at most quartic in $n$, and the same bound applies to $G^2_{\tau,n}$. In the algorithm, however, we are free to expand either $P_R$ or $P_S$ first and may pick whichever yields the smaller number of cosets. With this choice, we find experimentally (for $n,n'$ ranging 1 to 30) that
\begin{align}
    \max_\tau \min \left(\abs{\sym{n} / G^1_{\tau,n}}, \abs{\sym{n} / G^2_{\tau,n}}\right)
    = \begin{cases}
        n(n-1) \comma & n = n' \comma \\[2pt]
        \min(n, n') \comma & \abs{n - n'} = 2 \comma \\[2pt]
        1 \comma & \abs{n - n'} = 4 \pt
    \end{cases}
\end{align}
We therefore assume in the complexity analysis that the minimal number of cosets grows quadratically in $n$ when $n=n'$, linearly when $\abs{n-n'}=2$, and is constant when $\abs{n-n'}=4$. Although establishing a formal proof seems feasible, we leave this derivation for future work.

\section{Schur-Weyl duality}
\label{app:schur-weyl-duality}

In this section, we briefly review the Schur-Weyl duality, a fundamental result in representation theory that establishes a deep relationship between representations of the symmetric group $\sym{n}$ and the unitary group $\U{N}$ when they act on a tensor product space. A detailed exposition of the theory for $\mathrm{GL}(N)$ can be found in~\cite[Chapter 9]{goodman2009Symmetry}; the corresponding theory for $\U{N}$ follows via the unitary trick.

Consider the $n$-fold tensor product space $V_N^{\otimes n}$, where $V_N = \C^N$ is the fundamental representation of $\U{N}$. Both $\U{N}$ and $\sym{n}$ act on this space, and these actions commute.

The action of $\U{N}$ is defined by
\begin{align}
    (U^{\otimes n})_I^J \coloneqq U_{i_1}^{j_1}\cdots U_{i_n}^{j_n} \comma
\end{align}
where $I = (i_1,\dots,i_n)$ and $J = (j_1,\dots,j_n)$ are multi-indices. Thus, $\U{N}$ acts identically on each factor of $V_N$ in the tensor product.

The action of $\sym{n}$ permutes the order of factors in the tensor product. For $\sigma \in \sym{n}$, we have
\begin{align}
    \sigma_I^J = \delta_{i_1}^{j_{\sigma_1}}\cdots\delta_{i_n}^{j_{\sigma_n}} \comma
\end{align}
such that, for all $\sigma, \rho \in \sym{n}$, we have $\sigma^K_J \rho^J_I = (\sigma \rho)_I^K$.

Since these two actions commute, they can be simultaneously diagonalized. The \emph{Schur-Weyl duality}~\cite{goodman2009Symmetry} states that
\begin{align}
    \label{eq:schur-weyl-duality}
    V_N^{\otimes n} = \bigoplus_{\substack{R\vdash n \\ \ell(R)\leq N}} V^R_{\U{N}}\otimes V^R_{\sym{n}} \comma
\end{align}
where the sum runs over all Young diagrams $R$ with $n$ boxes and at most $N$ rows (denoted $\ell(R)\leq N$). Here, $V^R_{\U{N}}$ carries the irreducible representation $R$ of $\U{N}$, and $V^R_{\sym{n}}$ carries the irreducible representation $R$ of $\sym{n}$.

The Schur-Weyl decomposition has several important structural implications:
\begin{itemize}
    \item Irreducible tensor representations of both $\U{N}$ and $\sym{n}$ are labeled by Young diagrams. For $\sym{n}$, the diagram must have exactly $n$ boxes; for $\U{N}$, it must additionally satisfy $\ell(R) \leq N$.
    \item In the $n$-fold tensor product $V_N^{\otimes n}$, each irreducible representation $R$ of $\U{N}$ appears with multiplicity equal to $\dim_R$, the dimension of the corresponding $\sym{n}$ representation.
    \item Dually, each irreducible representation $R$ of $\sym{n}$ appears with multiplicity $\mathrm{Dim}_R$, the dimension of the corresponding $\U{N}$ representation.
\end{itemize}

The dimension $\dim_R$ of $V^R_{\sym{n}}$ is given by the hook-length formula~\eqref{eq:hook_length_formula}, which equals the cardinality of $\mathrm{SYT}(R)$ (see appendix \ref{app:young_diagrams} for the definition). The dimension $\mathrm{Dim}_R$ of $V^R_{\U{N}}$ admits a similar combinatorial interpretation: it counts the number of \emph{semistandard Young tableaux} of shape $R$ with entries in $\{1,\dots,N\}$, see~\cite[Corollary 8.1.17]{james1984Representation}. A semistandard Young tableau is a filling of the Young diagram $R$ with entries from $\{1,\dots,N\}$ (repetitions allowed) that are weakly increasing along each row and strictly increasing down each column. In terms of hook lengths, this dimension is given by
\begin{align}
        \mathrm{Dim}_R \coloneqq \dim V_{\U{N}}^R =  \prod_{(i,j)\in R}\frac{N - i + j}{ h_R(i,j)} \pt
\end{align}
As a consequence of the Schur-Weyl decomposition~\eqref{eq:schur-weyl-duality}, the dimension of $V_N^{\otimes n}$ satisfies
\begin{align}
    N^n = \sum_{\substack{R\vdash n \\ \ell(R)\leq N}} \dim_R \mathrm{Dim}_R \pt
\end{align}

In practice, one may realize $V_{\U{N}}^R$ as Weyl modules, obtained by applying an appropriate Young symmetrizer, $Y_{\Theta}$ where $\Theta \in \mathrm{SYT}(R)$, to tensor powers of the fundamental representation. We refer to~\cite[Chapter 6]{fulton2004Representation}, where Weyl's construction is described.

\section{Bosonic Fock space}
\label{app:bosonic_fock_space}

In this appendix, we collect the ingredients needed to describe the action of $\U{N}$ on the bosonic Fock space. We first introduce two standard bases of the Lie algebra $\mathfrak{u}(N)$, then define the associated bosonic Fock space in the $\alpha$-basis and in the matrix-entry basis. Next, we characterize the corresponding gauge-invariant observables, derive the generators of gauge transformations, establish Wick's formula in the matrix-entry basis, and describe the antinormal ordering of the gauge-invariant traces that appear in the Hamiltonian analysis. To keep the notation light, we mostly restrict to a single matrix degree of freedom. In this case, the extension to several matrices is always straightforward.

\subsection{The Lie algebra \texorpdfstring{$\mathfrak{u}(N)$}{u{N}}}
\label{sec:lie_algebra}
The real Lie algebra $\liealgebra{u}{N}$ is the $N^2$-dimensional space of Hermitian $N\times N$ matrices, equipped with the commutator as Lie bracket. We will denote by
\begin{align}
    \enstq{\tau_{\alpha}}{\alpha=0,\dots, N^2-1} \comma
\end{align}
a generic basis of generators. We now introduce two standard bases, both of them orthonormal with respect to the scalar product $(\tau, \nu) \mapsto \tr(\tau^\dagger \nu)=\tr(\tau \nu)$. 

The first is the natural basis. For all $1 \leq k < \ell \leq N$, it is defined as
\begin{align}
    \label{eq:generators_un}
    \quad \tau_{0} &= \tau_{\alpha_1^1} = E_1^1 \comma \\
    \tau_{\ell^2-1} &= \tau_{\alpha_\ell^\ell} = E_{\ell}^\ell \comma\\
    \tau_{\ell^2 + 2(k - \ell) - 1} &= \tau_{\alpha_k^\ell} = \tfrac{1}{\sqrt{2}}(E_k^\ell + E_\ell^k) \comma\\
    \tau_{\ell^2 + 2(k - \ell)} &= \tau_{\alpha_\ell^k} = \tfrac{i}{\sqrt{2}}(E_k^\ell - E_\ell^k) \comma
\end{align}
where $E_k^\ell$ is the matrix with a 1 on the $k$-th row and $\ell$-th column and zeros everywhere else, that is: $(E_k^\ell)_i^j = \delta_i^\ell \delta_k^j$. Notice that the sorting map 
\begin{align}
    \label{eq:index_bijection}
    (\ell, k) \in \{1,\dots,N\}^2 \mapsto \alpha_{k}^\ell \in \{0, \dots, N^2-1\} \comma
\end{align}
is indeed a bijection.

A second convenient choice is obtained by extending the orthonormal basis of $\liealgebra{su}{N}$ described in~\cite{bossion2021General} by adjoining $\tau_{0} = \mathbb{I}_N / \sqrt{N}$ to it, where $\mathbb{I}_N$ is the identity matrix of order $N$. This again yields an orthonormal basis, since $\tr(\tau_\alpha) = 0$ for $\alpha \neq 0$. Compared with~\eqref{eq:generators_un}, this basis keeps the same off-diagonal generators and replaces the diagonal ones by traceless linear combinations, namely, for $\ell=2,\dots, N$
\begin{align}
    \tau_{\ell^2 - 1} = \frac{1}{\sqrt{\ell(\ell-1)}} \left(\sum_{k=1}^{\ell-1} E_k^k + (1 - \ell) E_\ell^\ell\right) \pt
\end{align}

The \emph{totally antisymmetric structure constants} $f_{\alpha\beta\gamma}$ and the \emph{totally symmetric coefficients} $d^{(s)}_{\alpha\beta\gamma}$ are defined by expanding the commutator and anti-commutator in this basis. For all $\alpha,\beta \in \{0,\dots,N^2-1\}$, one has
\begin{align}
    \label{eq:structure_constants}
    \comm{\tau_\alpha}{\tau_\beta} 
    = i \sum_\gamma f_{\alpha\beta\gamma} \tau_{\gamma} \qq{and}
    \acomm{\tau_\alpha}{\tau_\beta} 
    = \sum_\gamma d^{(s)}_{\alpha\beta\gamma} \tau_{\gamma} \pt
\end{align}
Explicit formulas in the second basis are given in~\cite{bossion2021General} for the $\liealgebra{su}{N}$ generators; the extension to the additional generator $\tau_0$ is immediate.

\subsection{Generators for U(N) transformations in different bases}

Consider the operators $A$ and $A^\dagger$ transforming in the adjoint representation of $\U{N}$
\begin{align}
    A \longrightarrow U A U^\dagger \qq{and} A^\dagger \longrightarrow U A^\dagger U^\dagger \pt
\end{align}
When there are more than one matrix, the group acts simultaneously on each matrix degree of freedom $A_I$ and $A_I^\dagger$, for $I=1,\dots, D$.

We consider $\enstq{\tau_\alpha}{\alpha=0,\dots,N^2-1}$ an orthonormal basis of the real Lie algebra $\mathfrak{u}(N)$, as defined in the previous section. Then, we can decompose $A$ and $A^\dagger$ in the \emph{$\alpha$-basis}:
\begin{align}
    \label{eq:decomposition_degree_freedom}
    A = A^\alpha \tau_\alpha \qq{and} A^\dagger = (A^{\dagger})^{\alpha} \tau_\alpha = {A^{\alpha}}^{\dagger} \tau_\alpha \comma
\end{align}
where write $A^{\alpha \dagger} \coloneqq (A^\alpha)^\dagger$ to ease the notation. We impose the canonical commutation relations
\begin{align}
    \label{eq:fock_comm_alpha}
    [ A^\alpha, A^{\beta \dagger} ] = \delta^{\alpha\beta} \comma
\end{align}
on the ladder operators. The \emph{number operator} in mode $\alpha$ is given by
\begin{align}
    N^{\alpha} \coloneqq  A^{\alpha\dagger} A^\alpha = (A^\dagger)^\alpha A^{\alpha} \comma
\end{align}
and satisfies $[N^{\alpha}, (A^{\alpha\dagger})^k] = k (A^{\alpha\dagger})^k$ for all $k\in \mathbb{N}$. The \emph{free (quadratic) Hamiltonian} is
\begin{align}
    \label{eq:free_quadratic_hamiltonian}
    H_0 = \sum_\alpha m \left(N^\alpha + \tfrac{1}{2}\right) \comma
\end{align}
where $m > 0$ denotes the mass of one boson.

The \emph{bosonic Fock space} $\mathcal{F}$ is generated from the unique vacuum $\ket{\Omega_\mathrm{free}}$ defined by
\begin{align}
    A^\alpha \ket{\Omega_\mathrm{free}} = 0,\qquad \forall\alpha\pt
\end{align}
An orthonormal number basis for $\mathcal{F}$ is obtained by acting on the vacuum with creation operators:
\begin{align}
    \ket{\{n_\alpha\}} = \prod_{\alpha} \frac{1}{\sqrt{n_\alpha!}} \left(A^{\alpha \dagger}\right)^{n_\alpha} \ket{\Omega_\mathrm{free}} , \quad n_\alpha\in\mathbb{N}\pt
\end{align}
The ladder operators act in the usual way on number eigenstates:
\begin{align}
    A^\alpha \ket{\cdots, n_\alpha, \cdots} &= \sqrt{n_\alpha}\; \ket{\cdots, n_\alpha-1, \cdots},\\
    A^{\alpha\dagger} \ket{\cdots, n_\alpha, \cdots} &= \sqrt{n_\alpha+1}\; \ket{\cdots, n_\alpha+1, \cdots},\\
    N^\alpha \ket{\cdots, n_\alpha, \cdots} &= n_\alpha \ket{\cdots, n_\alpha, \cdots}\pt
\end{align}
In particular, this shows that $H_0$ counts excitations with energy spacing $m$ per mode.

We can now construct the associated quadrature operators via
\begin{align}
    X^\alpha = \frac{1}{\sqrt{2m}}(A^\alpha + A^{\alpha\dagger}) \qq{and}
    P^\alpha = i \sqrt{\frac{m}{2}}(A^{\alpha\dagger} - A^\alpha) \comma
\end{align}
collect them into a matrix as $X = X^\alpha \tau_\alpha$ and $P = P^\alpha \tau_\alpha$, and use~\eqref{eq:decomposition_degree_freedom} to express them as
\begin{align}
    \label{eq:X_and_P}
    X = \frac{1}{\sqrt{2m}}(A + A^{\dagger}) \qq{and}
    P = i \sqrt{\frac{m}{2}}(A^{\dagger} - A) \pt
\end{align}
They satisfy the commutation relations
\begin{align}
    [X^\alpha,X^{\beta} ] = 0 \comma ~
    [P^\alpha,P^{\beta} ] = 0 \comma ~
    [X^\alpha,P^{\beta} ] = i \delta^{\alpha\beta} \pt
\end{align}

\medskip

We can also decompose the matrices in $\mathfrak{u}(N)$ with respect to canonical coefficient, or \emph{matrix-entry basis}:
\begin{align}
    A = A_k^\ell E_\ell^k \comma
\end{align}
where $E_\ell^k$ is as before the matrix with a 1 on the $\ell$-th row and $k$-th column and zeros everywhere else. This decomposition will allow us to leverage the representation theory for the symmetric group~\cite{koch2024Pedagogical}. We use the same decomposition for all other matricial observables, such as $A^\dagger$, $X$, and $P$.  Expressing the coefficients of $A^\dagger$ with respect to those of $A$, we have:
\begin{align}
    A^\dagger 
    & = (A^\ell_k)^\dagger (E^k_\ell)^\dagger = (A^\ell_k)^\dagger E^\ell_k \pt
\end{align}
In particular: $(A^\dagger)_\ell^k = (A_k^\ell)^\dagger$. Using the sorting map~\eqref{eq:index_bijection}, one can write the operators in the matrix-entry basis in terms of those in the $\alpha$-basis. Taking $X$ and the first basis described in Section~\ref{sec:lie_algebra} as an example, we have
\begin{align}
    X^\ell_k = 
    \begin{cases}
        \tfrac{1}{\sqrt{2}}\left(X^{\alpha^\ell_k} - i X^{\alpha^k_\ell}\right) & \text{if } k < \ell \comma \\
        \tfrac{i}{\sqrt{2}}\left(X^{\alpha^\ell_k} - i X^{\alpha^k_\ell}\right) & \text{if } k > \ell \comma \\
        X^{\alpha_{k}^k} & \text{if } k = \ell \comma
    \end{cases}
\end{align}
and the corresponding identities also hold  for $A$, $A^\dagger$, and $P$. 

The commutation relations for $A_k^\ell$ are given by
\begin{align}
    \label{eq:commutation_relation_2}
    \comm{A^\ell_k}{(A^n_m)^\dagger} = \delta_{k m} \delta^{\ell n} 
\end{align}
which can be shown from \eqref{eq:fock_comm_alpha} taking good care of the indexing. Then, we get
\begin{align}
    \label{eq:commutation_relation_3}
    \comm{A^\ell_k}{(A^\dagger)_m^n} = \delta_{k}^n \delta^{\ell}_m \pt
\end{align}
Unfortunately, there is no compact expression for the (matrix) commutation relation of $A$ and $A^\dagger$. A convenient canonical pairing is $\left(A^\ell_{k}, (A^{\dagger})^k_{\ell}\right)$, for which $\comm{A^\ell_{k}}{(A^{\dagger})^k_{\ell}} = 1$.

The free Hamiltonian~\eqref{eq:free_quadratic_hamiltonian} can be rewritten as 
\begin{align}
    H_0 = \sum_{k,\ell} m (N^\ell_{k} + 1/2) \qq{with} N^\ell_{k} \coloneqq (A^{\dagger})^k_{\ell}  A^\ell_{k} \pt
\end{align}
The vacuum is such that $A^\ell_{k}  \ket{\Omega_\mathrm{free}} = 0$ for all $k\in\N$ and the basis Fock states are
\begin{align}
    \ket{\{n^{\ell}_{k}\}} 
    = \prod_{\ell,k} \frac{1}{\sqrt{n^{\ell}_{k}!}} \left((A^{\dagger})_k^{\ell}\right)^{n^{\ell}_{k}} \ket{\Omega_\mathrm{free}}\comma
\end{align}
where $n^{\ell}_{k}\in\mathbb{N}$.

The $\alpha$-basis and the matrix-entry basis are unitarily related by the change of single-particle modes, hence they produce unitarily equivalent Fock spaces and identical physics.

\subsection{Gauge invariant observables}
\label{app:gauge_invariant_observables}

We now define the observables that are invariant under the $\U{N}$ action. In the one-matrix case, an observable $\mathcal{O}$ acting on the bosonic Fock space is \emph{gauge invariant} if it is fixed by conjugation under the group action
\begin{align}
    \mathcal{G}(U)  \mathcal{O}  \mathcal{G}(U)^{-1} = \mathcal{O} , \quad \forall U \in \U{N} \pt
\end{align}
Equivalently, in infinitesimal form, gauge invariance means that $\mathcal{O}$ commutes with all generators of the symmetry,
\begin{align}
    \comm{\mathcal{O}}{G_\alpha} = 0 , \quad \forall \alpha \comma
\end{align}
in the $\alpha$-basis picture, and
\begin{align}
    \comm{\mathcal{O}}{G^\ell_{k}} = 0 , \quad \forall k,\ell \comma
\end{align}
in the matrix-entry basis picture. The explicit expression for the generators is derived in Section~\ref{app:generator_gauge}. In the several-matrix case, the definition is unchanged, except that the group acts simultaneously by adjoint conjugation on each matrix degree of freedom $A_I$, $A_I^\dagger$, $X_I$, and $P_I$.

For one matrix, an overcomplete set of gauge-invariant observables is provided by traces of words in operators transforming in the adjoint representation. Indeed, the cyclicity of the trace immediately gives
\begin{align}
    \tr(UB_1U^\dagger \cdots UB_nU^\dagger) = \tr(B_1\cdots B_n) \comma
\end{align}
where each $B_r$ may be chosen among $X$ and $P$, or equivalently among $A$ and $A^\dagger$. Therefore, any product of traces of such words is gauge invariant, defining the \emph{gauge invariant traces}. In particular, the free Hamiltonian~\eqref{eq:free_quadratic_hamiltonian} can be rewritten as
\begin{align}
    H_0 = m \left(\tr(A^\dagger A) + N^2/2\right) \comma
\end{align}
and is therefore gauge invariant. The same conclusion holds for observables such as
\begin{align}
    \tr(X^4)\comma \quad \tr(X^2)\tr(P^2) \comma \quad \tr(AA^\dagger A) \pt
\end{align}

In the several-matrix case, an overcomplete basis of gauge-invariant operators can be obtained by enlarging the alphabet to include all matrix species, for instance $\{X_I, P_I\}_{I}$, or equivalently $\{A_I, A_I^\dagger\}_{I}$.

It is convenient to encode the gauge invariant traces in terms of permutations. For example,
\begin{align}
    \tr(XYZ) = X_{\ell_1}^{\ell_2} Y_{\ell_2}^{\ell_3} Z_{\ell_3}^{\ell_1} = \sigma^{\ell_1\ell_2\ell_3}_{k_1k_2k_3} X_{\ell_1}^{k_1} Y_{\ell_2}^{k_2} Z_{\ell_3}^{k_3} \comma
\end{align}
where $\sigma = (123) \in \sym{3}$ and the associated tensor is defined by
\begin{align}
    \sigma^{\ell_1\ell_2\ell_3}_{k_1k_2k_3} \coloneqq \delta_{k_1}^{\ell_{\sigma_1}}\delta_{k_2}^{\ell_{\sigma_2}}\delta_{k_3}^{\ell_{\sigma_3}} \pt
 \end{align}
More generally, following the notation of~\cite{koch2024Pedagogical}, for any permutation $\sigma \in \sym{n}$, we set
\begin{align}
    \ttr{\sigma}{X_1 \otimes\cdots \otimes X_n}
    = \sigma^{\ell_1\cdots \ell_n}_{k_1\cdots k_n} (X_1)_{\ell_1}^{k_1} \cdots (X_n)_{\ell_n}^{k_n}
    = (X_1)_{\ell_1}^{\ell_{\sigma_1}} \cdots (X_n)_{\ell_n}^{\ell_{\sigma_n}} \comma
\end{align}
with $\sigma_K^L = \sigma^{\ell_1\cdots \ell_n}_{k_1\cdots k_n} = \delta_{k_1}^{\ell_{\sigma_1}}\cdots\delta_{k_n}^{\ell_{\sigma_n}}$. This formulation allows us to label each gauge invariant trace of order $n$ by a permutation $\sigma \in \sym{n}$ together with a word of length $n$ drawn from either $\{X_I,P_I\}_I$ or $\{A_I,A_I^\dagger\}_I$.

In particular, it reveals useful invariance properties. When all matrices are equal, $\ttr{\sigma}{X^{\otimes n}}$ depends only on the conjugacy class of $\sigma$; for example,
\begin{align}
    \ttr{(12)(345)}{X^{\otimes 5}} = \tr(X^2)\tr(X^3) \pt
\end{align}
More generally, $\ttr{\sigma}{X_1\otimes \cdots \otimes X_n}$ is invariant under conjugation by every permutation in the Young subgroup $H$ preserving the ordered tuple $(X_1,\dots,X_n)$. Indeed, for every $\tau \in H$,
\begin{align}
    \notag
    \ttr{\sigma}{X_1\otimes \cdots \otimes X_n}
    &= \sigma^L_K (X_1 \otimes\cdots \otimes X_n)_L^K 
    = \sigma^L_K \tau^K_J (X_1 \otimes\cdots \otimes X_n)^J_I (\tau^{-1})_L^I \\
    &= (\tau^{-1}\sigma \tau)_J^I (X_1 \otimes\cdots \otimes X_n)^J_I 
    = \ttr{\tau^{-1}\sigma\tau}{X_1\otimes \cdots \otimes X_n} \pt
\end{align}
For example, if $X_1\otimes \cdots \otimes X_n = Y_1^{\otimes n_1} \otimes \cdots \otimes Y_k^{\otimes n_k}$ with $n_1+\cdots+n_k=n$, then $\ttr{\sigma}{X_1\otimes \cdots \otimes X_n}$ is invariant under the action of $\sym{\underline{\vb*{n}}}=\sym{n_1}\times \cdots \times \sym{n_k} \subset \sym{n}$. When all matrices $Y_i$ are chosen from $\{A_I^\dagger\}_{I}$, the corresponding gauge-invariant operators encode singlet states. The invariant structure described above then provides the starting point for the center-algebra and centralizer-algebra description discussed in respectively Appendix~\ref{app:center_group_algebra} and Appendix~\ref{app:centralizer_group_algebra}.

\subsection{Generators of the gauge transformation}
\label{app:generator_gauge}

In this subsection, we derive the explicit form of the generators of gauge transformations. We show that in the $\alpha$-basis, the generators are given by
\begin{align}
    \label{eq:generator_alpha_basis}
    G_\alpha = i f_{\alpha\beta\gamma} A^{\dagger\beta} A^\gamma \pt
\end{align}
Below, we derive this formula by expanding the infinitesimal transformation $A^\dagger \mapsto U A^\dagger U^\dagger$ with $U \simeq \mathbb{I}_N + i\varepsilon$, with $\epsilon \in \liealgebra{u}{N}$. The generators in the matrix-entry basis can then be obtained either by changing basis from $\{\tau_\alpha\}$ to $\{E_k^\ell\}$, or following the same route as for the $\alpha$-basis case. One finds
\begin{align}
    \label{eq:generator_matrix_basis}
    G^\ell_k = (A^\dagger)^\ell_p A^p_k - (A^\dagger)^p_k A^\ell_p \pt
\end{align}

Remember that the singlet states of the theory are left invariant by the action of the gauge group $\U{N}$, meaning that
\begin{align}
    G_\alpha \ket{\psi} = 0 ,\quad \forall \alpha \pt
\end{align}
In particular, the Hamiltonian~\eqref{eq:hamiltonian_form_1} considered in this work is gauge invariant, so it commutes with all generators.

We now derive the generators explicitly. Consider
\begin{align}
    \ket{\psi} = \prod_{k=1}^{\ell} A^{\dagger\gamma_k} \ket{\Omega_{\mathrm{free}}} \comma
\end{align}
an element of the basis of the bosonic Fock space $\mathcal{F}$. Under a gauge transformation $U \in \U{N}$, this state transforms as
\begin{align}
    \ket{\psi} \longrightarrow \ket{\psi_U} = \mathcal{G}(U) \ket{\psi} \comma
\end{align}
where $U \mapsto \mathcal{G}(U)$ denotes the representation of $\U{N}$ on $\mathcal{F}$. Recall that the matrix of creation operators transforms under the adjoint representation:
\begin{align}
    A^\dagger \longrightarrow UA^\dagger U^\dagger \eqqcolon g(U) A^\dagger \comma
\end{align}
where $g(U)$ acts as a linear map on the $N^2$-dimensional matrix degree of freedom. Consequently,
\begin{align}
    \mathcal{G}(U) \ket{\psi} =  \prod_{k=1}^{\ell} \left(g(U)^{\gamma_k}_\beta A^{\dagger\beta}\right) \ket{\Omega_{\mathrm{free}}} \pt
\end{align}

To extract the infinitesimal generators, consider $U = \mathbb{I}_N + \delta U \in \U{N}$, an infinitesimal perturbation of the identity, with
\begin{align}
    \delta U = i(\delta \lambda)^\alpha \tau_\alpha + \mathcal{O}((\delta\lambda)^2) \comma
\end{align}
where $\{\tau_\alpha\}$ are the generators of the defining representation of $\mathfrak{u}(N)$. Since indices in the $\alpha$-basis transform according to the adjoint representation, the induced transformation reads
\begin{align}
    g(U) = \mathbb{I}_{N^2} + i (\delta \lambda)^\alpha \comm{\tau_\alpha}{\cdot} + \mathcal{O}((\delta\lambda)^2) \comma
\end{align}
Using the orthonormality relation $\tr(\tau_\alpha \tau_\beta) = \delta_{\alpha\beta}$, the matrix elements become
\begin{align}
    g(U)^\gamma_\beta = \delta_\beta^\gamma + i (\delta \lambda)^\alpha \tr(\tau_\gamma \comm{\tau_\alpha}{\tau_\beta}) + \mathcal{O}((\delta\lambda)^2) \comma
\end{align}
Employing the structure constants $f_{\alpha\beta\gamma}$ defined in~\eqref{eq:structure_constants}, this yields
\begin{align}
    g(U)^\gamma_\beta = \delta_\beta^\gamma - (\delta \lambda)^\alpha  f_{\alpha\beta\gamma} + \mathcal{O}((\delta\lambda)^2) \comma
\end{align}
Meanwhile, by definition, the infinitesimal action on Fock space is
\begin{align}
    \mathcal{G}(U) = \mathbb{I}_\mathcal{F} + i(\delta \lambda)^\alpha  G_\alpha + \mathcal{O}((\delta\lambda)^2) \comma
\end{align}
Expanding both sides to first order in $\delta\lambda$ and comparing, we obtain
\begin{align}
   G_\alpha \ket{\psi}
    = \left[\sum_{k=1}^{\ell}
        A^{\dagger\gamma_1} \cdots A^{\dagger\gamma_{k-1}} \left(i f_{\alpha\beta\gamma_k} A^{\dagger\beta} \right)  A^{\dagger\gamma_{k+1}}\cdots A^{\dagger\gamma_k} \right] \ket{\Omega_{\mathrm{free}}} \pt
\end{align}
Then, we use the canonical commutation relations~\eqref{eq:fock_comm_alpha} to rewrite
\begin{align}
   i f_{\alpha\beta\gamma_k} A^{\dagger\beta} 
    =  i f_{\alpha\beta\gamma} A^{\dagger\beta} \delta^{\gamma\gamma_k}
    = i f_{\alpha\beta\gamma} A^{\dagger\beta} \comm{A^{\gamma} }{A^{\dagger\gamma_k}} 
    = \comm{T_\alpha}{A^{\dagger\gamma_k}} \comma 
\end{align}
where $T_\alpha = i f_{\alpha\beta\gamma} A^{\dagger\beta} A^\gamma$. Substituting this into the expression for $G_\alpha \ket{\psi}$ reveals a telescopic sum:
\begin{align}
    \notag
    G_\alpha  \ket{\psi}  
    & = \left[ \left(T_\alpha A^{\dagger\gamma_1} \cdots A^{\dagger\gamma_{\ell}}  - A^{\dagger\gamma_1} T_\alpha A^{\dagger\gamma_{2}}  \cdots A^{\dagger\gamma_{\ell}} \right) \right. \\
    \notag
    & \quad + \left(A^{\dagger\gamma_1} T_\alpha A^{\dagger\gamma_{2}}  \cdots A^{\dagger\gamma_{\ell}} - A^{\dagger\gamma_1}  A^{\dagger\gamma_2} T_\alpha A^{\dagger\gamma_{3}}  \cdots A^{\dagger\gamma_{\ell}} \right)\\
    \notag
    & \quad + \left. \cdots \right] \ket{\Omega_{\mathrm{free}}}  \\
    & = T_\alpha \ket{\psi}  -\left( \prod_{k=1}^{\ell} A^{\dagger\gamma_k}\right)T_\alpha \ket{\Omega_{\mathrm{free}}} \pt
\end{align}
Since $T_\alpha \ket{\Omega_{\mathrm{free}}} = 0$, we conclude that $G_\alpha = T_\alpha$, establishing the claimed result.

\subsection{Wick's formula in the matrix-entry basis}
\label{app:wick_formula}

In this section, we derive the expression for correlators of the form
\begin{align}
    (\mathcal{C}_n)_{I,K}^{J,L} = \mathcal{C}_{i_1 \dots i_n k_1 \dots k_m}^{j_1 \dots j_n \ell_1 \dots \ell_m} 
    \coloneqq \ev{A_{i_1}^{j_1}\cdots A_{i_n}^{j_n} (A^{\dagger})_{k_1}^{\ell_1} \cdots (A^{\dagger})_{k_m}^{\ell_m} } \comma
\end{align}
where we have denoted $\ev{B} \coloneqq \mel{\Omega_{\mathrm{free}}}{B}{\Omega_{\mathrm{free}}}$. Since the vacuum expectation value of a product with an unequal number of creation and annihilation operators vanishes, this correlator is zero if $n\neq m$. For $n=m$, the correlator is given by Wick's formula:
\begin{align}
    \label{eq:wick_formula}
    (\mathcal{C}_n)_{I,K}^{J,L}  
     = \sum_{\sigma \in \sym{n}} (\sigma^{-1})_{I}^L \sigma^J_K
    \coloneqq \sum_{\sigma \in \sym{n}} \left(\prod_{s=1}^{n} \delta^{\ell_{\sigma^{-1}(s)}}_{i_s}\right) \left(\prod_{s=1}^{n} \delta_{k_{s}}^{j_{\sigma(s)}}\right) \pt
\end{align}
We prove this by induction on $n$.

\noindent\textbf{Base case ($n=1$):}
The correlator is $\mathcal{C}_1 = \ev{A_{i_1}^{j_1} (A^\dagger)_{k_1}^{\ell_1}}$. Using the commutation relation~\eqref{eq:commutation_relation_3}, we have:
\begin{align}
    \ev{A_{i_1}^{j_1} (A^\dagger)_{k_1}^{\ell_1}} = \ev{(A^\dagger)_{k_1}^{\ell_1} A_{i_1}^{j_1}} + \delta_{i_1}^{\ell_1}\delta_{k_1}^{j_1} = \delta_{i_1}^{\ell_1}\delta_{k_1}^{j_1} \comma
\end{align}
since the operators annihilate the vacuum on the right and left. This matches~\eqref{eq:wick_formula} for $n=1$, as the only permutation in $\sym{1}$ is the identity.

\noindent\textbf{Inductive step:}
Assume the formula holds for $n-1$. To compute $(\mathcal{C}_n)_{I,K}^{J,L}$, we use the commutation relation~\eqref{eq:commutation_relation_3} to move the operator $(A^{\dagger})_{k_n}^{\ell_n}$ to the left through all the $A$ operators:
\begin{align}
    \label{eq:wick_formula_step_1}
    (\mathcal{C}_n)_{I,K}^{J,L} 
    =  \sum_{r=1}^{n}(\mathcal{C}_{n-1})_{\hat{I}_r,\hat{K}_n}^{\hat{J}_r,\hat{L}_n} \delta_{i_r}^{\ell_n}\delta_{k_n}^{j_r} \comma
\end{align}
where $\hat{I}_r = (i_1, \dots, \hat{i}_r, \dots, i_n)$ denotes the multi-index $I$ with the $r$-th element removed, and similarly for the other multi-indices.

By the induction hypothesis for $n-1$, each term $\mathcal{C}_{n-1}$ can be written as:
\begin{align}
    \label{eq:wick_formula_step_2}
    (\mathcal{C}_{n-1})_{\hat{I}_r,\hat{K}_n}^{\hat{J}_r,\hat{L}_n} = \sum_{\sigma \in \sym{n-1}} \left(\prod_{\substack{s = 1 \\ s\neq r}}^{n} \delta^{\ell_{\hat{\sigma}^{-1}_r(s)}}_{i_s}\right) \left(\prod_{s=1}^{n-1} \delta_{k_{s}}^{j_{\sigma(s)}}\right) \comma
\end{align}
where $\hat{\sigma}_r$ is the permutation of $\sym{n}$ defined by setting $\hat{\sigma}_r(n) = r$ and $\hat{\sigma}_r(s) = \sigma(s)$ for $s < n$. Combining~\eqref{eq:wick_formula_step_2} with~\eqref{eq:wick_formula_step_1}, we get:
\begin{align}
    (\mathcal{C}_n)_{I,K}^{J,L} 
    =  \sum_{r=1}^{n} \sum_{\sigma \in \sym{n-1}} \left(\prod_{\substack{s = 1 \\ s\neq r}}^{n} \delta^{\ell_{\hat{\sigma}^{-1}_r(s)}}_{i_s}\right) \left(\prod_{s=1}^{n-1} \delta_{k_{s}}^{j_{\sigma(s)}}\right) \delta_{i_r}^{\ell_n}\delta_{k_n}^{j_r} \pt
\end{align}
 Summing over all $r$ and all $\sigma \in \sym{n-1}$ is equivalent to summing over all $\hat{\sigma} \in \sym{n}$. This allows us to rewrite the sum as:
\begin{align}
    (\mathcal{C}_n)_{I,K}^{J,L} 
    =  \sum_{\hat{\sigma} \in \sym{n}} \left(\prod_{s=1}^{n} \delta^{\ell_{\hat{\sigma}^{-1}(s)}}_{i_s}\right) \left(\prod_{s=1}^{n} \delta_{k_{s}}^{j_{\tilde{\sigma}(s)}}\right) \comma
\end{align}
which concludes the proof.

For multiple matrices, since the operators of different matrix species commute, the contractions factorize over species, and the Wick sum runs over the Young subgroup $\sym{\underline{\vb*{n}}}=\sym{n_1}\times\cdots\times\sym{n_D}$ rather than the full symmetric group $\sym{n}$:
\begin{align}
    \expval{\left(A_1^{\otimes n_1}\otimes\cdots\otimes A_D^{\otimes n_D}\right)^L_K \left((A_1^\dagger)^{\otimes n_1}\otimes\cdots\otimes (A_D^\dagger)^{\otimes n_D}\right)^N_M}
    = \sum_{\sigma\in\sym{\underline{\vb*{n}}}}(\sigma^{-1})^L_N\,\sigma^K_M \pt
\end{align}

\subsection{Orthogonality of states constructed from restricted Schur polynomials}\label{app:ResSchurOrth}

We use the same notation as in Section~\ref{sec:singlet_states_multi_matrices}. In particular, we consider $R,S\vdash n$ and $\vb*{r}=(r_1,\dots,r_D)$, $\vb*{s}=(s_1,\dots,s_D)$ with $r_I,s_I\vdash n_I$ and $n=n_1+\cdots+n_D$. If $S\not\vdash n$ or $s_I\not\vdash n_I$ for some $I$, the bracket $\braket{R,\vb*{r},a,b}{S,\vb*{s},c,d}$ vanishes trivially, since the two states then belong to different excitation sectors. We therefore assume throughout that the two states carry the same multidegree $\underline{\vb*{n}}=(n_1,\dots,n_D)$.

Using above's Wick sum, the overlap of two states from restricted Schur polynomials is given by
\begin{align}
    \braket{R,\vb*{r},a,b}{S,\vb*{s},c,d}
    &= \mel*{\Omega_{\mathrm{free}}}{\chi^{R,\vb*{r}}_{ba}(A_1,\dots,A_D)\,\chi^{S,\vb*{s}}_{cd}(A_1^\dagger,\dots,A_D^\dagger)}{\Omega_{\mathrm{free}}} \\
    &= \frac{1}{\dim_R\dim_S}\frac{(n!)^2}{(\underline{\vb*{n}}!)^2}
       \sum_{\sigma\in\sym{\underline{\vb*{n}}}}
       \left(P^{R,\vb*{r}}_{ba}\right)^K_L
       \left(P^{S,\vb*{s}}_{cd}\right)^M_N
       (\sigma^{-1})^L_N\,\sigma^K_M \pt
\end{align}

The centralizer basis elements $P^{R,\vb*{r}}_{ab}$ commute with every element of the Young subgroup, $\nu P^{R,\vb*{r}}_{ab}=P^{R,\vb*{r}}_{ab}\nu$ for all $\nu\in\sym{\underline{\vb*{n}}}$. Consequently, each term in the sum over $\sigma$ is independent of $\sigma$, and the sum reduces to the factor $|\sym{\underline{\vb*{n}}}|=\underline{\vb*{n}}!$:
\begin{align}
    \braket{R,\vb*{r},a,b}{S,\vb*{s},c,d}
    &= \frac{1}{\dim_R\dim_S}\frac{(n!)^2}{\underline{\vb*{n}}!}
       \left(P^{R,\vb*{r}}_{ba}\right)^K_L
       \left(P^{S,\vb*{s}}_{cd}\right)^L_K \\
    &= \frac{1}{\dim_R\dim_S}\frac{(n!)^2}{\underline{\vb*{n}}!}
       \left(P^{R,\vb*{r}}_{ba}P^{S,\vb*{s}}_{cd}\right)^K_K \pt
\end{align}

Using the multiplication law of the centralizer basis, $P^{R,\vb*{r}}_{ab}P^{S,\vb*{s}}_{cd}=\delta_{RS}\delta_{\vb*{r}\vb*{s}}\delta_{bc}P^{R,\vb*{r}}_{ad}$, together with the trace identity $\left(P^{R,\vb*{r}}_{bd}\right)^K_K=\dim_{\vb*{r}}\mathrm{Dim}_R\,\delta_{bd}$, we obtain
\begin{align}
    \braket{R,\vb*{r},a,b}{S,\vb*{s},c,d}
    &= \frac{\dim_{\vb*{r}}\mathrm{Dim}_R}{(\dim_R)^2}\frac{(n!)^2}{\underline{\vb*{n}}!}
       \delta_{RS}\delta_{\vb*{r}\vb*{s}}\delta_{ac}\delta_{bd} \\
    &= \frac{n!}{\dim_R}\frac{\dim_{\vb*{r}}}{\underline{\vb*{n}}!}\alpha_R
       \,\delta_{RS}\delta_{\vb*{r}\vb*{s}}\delta_{ac}\delta_{bd} \comma
\end{align}
where in the last step we used $\alpha_R=n!\mathrm{Dim}_R/\dim_R$. Recalling the definition of the normalization constant $\alpha^{R}_{\vb*{r}}=\frac{n!}{\dim_R}\frac{\dim_{\vb*{r}}}{\underline{\vb*{n}}!}\alpha_R$, this reproduces exactly~\eqref{eq:orthogonality_restricted_schur}.

\subsection{Antinormal ordering of gauge invariant observables}
\label{sec:decomposition_power_trace}

We now use the permutation-trace notation introduced in Appendix~\ref{app:gauge_invariant_observables} to express gauge-invariant observables in antinormal order, in a form well suited to our purposes.

We first illustrate this decomposition for the quartic interaction of the single-matrix model, $\tr(X^4)$, and then for the interaction term of the two-matrix model, $\tr(\comm{X_1}{X_2}^2)$. The extension to a general gauge-invariant trace is straightforward.

\begin{table}[ht]
\centering
\caption{Contributions to the decomposition of $4m^2\tr(X^4)$ in antinormal order}
\label{tab:contributions-interactions}
\begin{minipage}[t]{0.49\textwidth}
\vspace{0pt}
\centering
\begin{tabular}{lll}
\toprule
\centering
\textbf{Type} & $\mathbf{\sigma}$ & \textbf{Coefficient} \\
\midrule
Constant & $\id$ & $2N^3+N$ \\
\addlinespace
\multirow[t]{2}{*}{$A^{\otimes 2}$} & $\id$ & $-2$ \\
 & $(12)$ & $-4N$ \\
\addlinespace
\multirow[t]{2}{*}{$A \otimes A^\dagger$} & $\id$ & $-4$ \\
 & $(12)$ & $-8N$ \\
\addlinespace
\multirow[t]{2}{*}{$(A^\dagger)^{\otimes 2}$} & $\id$ & $-2$ \\
 & $(12)$ & $-4N$ \\
\bottomrule
\end{tabular}
\end{minipage}
\hfill
\begin{minipage}[t]{0.49\textwidth}
\vspace{0pt}
\centering
\begin{tabular}{lll}
\toprule
\textbf{Type} & $\mathbf{\sigma}$ & \textbf{Coefficient} \\
\midrule
$A^{\otimes 4}$ & $(1432)$ & $1$ \\
\addlinespace
$A^{\otimes 3} \otimes A^\dagger$ & $(1342)$ & $4$ \\
\addlinespace
\multirow[t]{5}{*}{$A^{\otimes 2} \otimes (A^\dagger)^{\otimes 2}$} & $(1243)$ & $4$ \\
 & $(1324)$ & $2$ \\
\addlinespace
$A \otimes (A^\dagger)^{\otimes 3}$ & $(1243)$ & $4$ \\
\addlinespace
$(A^\dagger)^{\otimes 4}$ & $(1432)$ & $1$ \\[.54em]
\addlinespace
\bottomrule
\end{tabular}
\end{minipage}
\end{table}

Using~\eqref{eq:X_and_P} together with the permutation-trace notation, the interaction term of the single-matrix model expands into 16 terms:
\begin{align}
    \notag
    4m^2\tr(X^4)
    &= \sum_{B_1,B_2,B_3,B_4 \in \{A, A^\dagger\}} \ttr{(1234)}{B_1\otimes B_2 \otimes B_3 \otimes B_4} \\
    &= \sum_{B_1,B_2,B_3,B_4 \in \{A, A^\dagger\}} (B_1)_{k_1}^{k_2} (B_2)_{k_2}^{k_3} (B_3)_{k_3}^{k_4} (B_4)_{k_4}^{k_1} \pt
\end{align}
Using the commutation relation~\eqref{eq:commutation_relation_3}, we can then reorder these contributions antinormally, obtaining 9 distinct types of terms:
\begin{align}
    \notag
    4m^2\tr(X^4) 
    =&~ a_1 + \sum_{\sigma \in \sym{2}} a_2(\sigma) \ttr{\sigma}{A^{\otimes 2}} 
     + \sum_{\sigma \in \sym{2}}a_3(\sigma) \ttr{\sigma}{A \otimes A^\dagger} \\
     \notag
    & + \sum_{\sigma \in \sym{2}}a_4(\sigma) \ttr{\sigma}{(A^\dagger)^{\otimes 2}}
    + \sum_{\sigma \in \sym{4}} a_5(\sigma) \ttr{\sigma}{A^{\otimes 4}} \\
    \notag
    &+ \sum_{\sigma \in \sym{4}} a_6(\sigma) \ttr{\sigma}{A^{\otimes 3}\otimes A^\dagger} 
    + \sum_{\sigma \in \sym{4}} a_7(\sigma) \ttr{\sigma}{A^{\otimes 2}\otimes(A^\dagger)^{\otimes 2}} \\
    &+ \sum_{\sigma \in \sym{4}} a_8(\sigma) \ttr{\sigma}{A\otimes(A^\dagger)^{\otimes 3}}
    + \sum_{\sigma \in \sym{4}} a_9(\sigma) \ttr{\sigma}{(A^\dagger)^{\otimes 4}} \pt
\end{align}
The non-zero contributions, obtained using symbolic computation in Mathematica~\cite{mathematica14.3}, are listed in Table~\ref{tab:contributions-interactions}; in total, 13 terms survive.

\begin{table}[ht]
\centering
\caption{Contributions to the decomposition of $4m^2\tr(\comm{X_1}{X_2}^2)$ in antinormal order}
\label{tab:contributions-interactions-multi}
\begin{minipage}[t]{0.47\textwidth}
\vspace{0pt}
\centering
\begin{tabular}{lll}
\toprule
\textbf{Type} & $\sigma$ & \textbf{Coefficient} \\
\midrule
Constant & $\id$ & $2(N-N^3)$ \\
\addlinespace
\multirow[t]{2}{*}{$A_1^{\otimes 2}$} & $\id$ & $-2$ \\
 & $(12)$ & $2N$ \\

 \addlinespace
 \multirow[t]{2}{*}{$A_1 \otimes A_1^\dagger$} & $\id$ & $-4$ \\
 & $(12)$ & $4N$ \\

 \addlinespace
 \multirow[t]{2}{*}{$(A_1^\dagger)^{\otimes 2}$} & $\id$ & $-2$ \\
  & $(12)$ & $2N$ \\

  \addlinespace
  \multirow[t]{2}{*}{$A_2^{\otimes 2}$} & $\id$ & $-2$ \\
  & $(12)$ & $2N$ \\

  \addlinespace
  \multirow[t]{2}{*}{$A_2 \otimes A_2^\dagger$} & $\id$ & $-4$ \\
  & $(12)$ & $4N$ \\

  \addlinespace
  \multirow[t]{2}{*}{$(A_2^\dagger)^{\otimes 2}$} & $\id$ & $-2$ \\
  & $(12)$ & $2N$ \\
  
  \addlinespace
  \multirow[t]{2}{*}{$A_1^{\otimes 2}\otimes A_2^{\otimes 2}$} & $(1423)$ & $2$ \\
  & $(1432)$ & $-2$ \\
  
  \addlinespace
\multirow[t]{3}{*}{$A_1^{\otimes 2}\otimes A_2 \otimes A_2^\dagger$} & $(1324)$ & $4$ \\
  & $(1342)$ & $-2$ \\
  & $(1432)$ & $-2$ \\
  
  \addlinespace
  \multirow[t]{2}{*}{$A_1^{\otimes 2}\otimes (A_2^\dagger)^{\otimes 2}$} & $(1423)$ & $2$ \\
  & $(1432)$ & $-2$ \\

\bottomrule
\end{tabular}
\end{minipage}
\hfill
\begin{minipage}[t]{0.52\textwidth}
\vspace{0pt}
\centering
\begin{tabular}{lll}
\toprule
\textbf{Type} & $\sigma$ & \textbf{Coefficient} \\
\midrule
  \multirow[t]{6}{*}{$A_1\otimes A_2\otimes A_1^\dagger \otimes A_2^\dagger$} & $(1234)$ & $4$ \\
  & $(1432)$ & $4$ \\
  & $(1243)$ & $-2$ \\
  & $(1342)$ & $-2$ \\
  & $(1324)$ & $-2$ \\
  & $(1423)$ & $-2$ \\
  
  \addlinespace
  \multirow[t]{3}{*}{$A_1\otimes A_1^\dagger \otimes (A_2^\dagger)^{\otimes 2}$} & $(1324)$ & $4$ \\
  & $(1243)$ & $-2$ \\
  & $(1432)$ & $-2$ \\

  \addlinespace
  \multirow[t]{2}{*}{$A_2^{\otimes 2} \otimes (A_1^\dagger)^{\otimes 2}$} & $(1324)$ & $2$ \\
  & $(1432)$ & $-2$ \\
  
  \addlinespace
  \multirow[t]{3}{*}{$A_2 \otimes (A_1^\dagger)^{\otimes 2} \otimes A_2^\dagger$} & $(1243)$ & $4$ \\
  & $(1324)$ & $-2$ \\
  & $(1432)$ & $-2$ \\
  
    \addlinespace
    \multirow[t]{2}{*}{$(A_1^\dagger)^{\otimes 2} \otimes (A_2^\dagger)^{\otimes 2}$} & $(1423)$ & $2$ \\
    & $(1432)$ & $-2$ \\

    \addlinespace
  \multirow[t]{3}{*}{$A_1\otimes A_2\otimes A_2 \otimes A_1^\dagger$} & $(1243)$ & $4$ \\
  & $(1324)$ & $-2$ \\
  & $(1432)$ & $-2$ \\
  [3.37em]

\bottomrule
\end{tabular}
\end{minipage}
\end{table}

For $\tr(\comm{X_1}{X_2}^2)$, we begin by expanding the square of the commutator. Using the cyclicity of the trace together with the fact that operators inside a trace commute pairwise, we obtain:
\begin{align}
    \tr(\comm{X_1}{X_2}^2) = 2 \left(\tr(X_1X_2X_1X_2) - \tr(X_1^2X_2^2)\right) \pt
\end{align}
We then fix an ordering convention on the indices, with $1$ always preceding $2$ in both the creation and annihilation sectors, and reorder every contribution antinormally, with all creation operators placed on the right. This yields 16 possible term types:
\begin{align}
    \notag
     4m^2 & \tr(\comm{X_1}{X_2}^2) \\
     \notag
     = a_1 & + \sum_{j=1}^{2} \left[
        \sum_{\sigma \in \sym{2}} a_2(j;\sigma) \ttr{\sigma}{A_j^{\otimes 2}} \right.
    + \sum_{\sigma \in \sym{2}} a_3(j;\sigma) \ttr{\sigma}{A_j \otimes A_j^\dagger} \\
    \notag
    & + \left. \sum_{\sigma \in \sym{2}} a_4(j;\sigma) \ttr{\sigma}{(A_j^\dagger)^{\otimes 2}} \right]
    + \sum_{\sigma \in \sym{4}} a_5(\sigma) \ttr{\sigma}{A_1^{\otimes 2} \otimes A_2^{\otimes 2}} \\
    & + \sum_{\sigma \in \sym{4}} a_6(\sigma) \ttr{\sigma}{A_1^{\otimes 2} \otimes A_2 \otimes A_2^\dagger}
    + \sum_{\sigma \in \sym{4}} a_7(\sigma) \ttr{\sigma}{A_1 \otimes A_2^{\otimes 2} \otimes A_1^\dagger} \\
    \notag
    & + \sum_{\sigma \in \sym{4}} a_8(\sigma) \ttr{\sigma}{A_1^{\otimes 2} \otimes (A_2^\dagger)^{\otimes 2}}
    + \sum_{\sigma \in \sym{4}} a_9(\sigma) \ttr{\sigma}{A_2^{\otimes 2} \otimes (A_1^\dagger)^{\otimes 2}} \\
    \notag
    & + \sum_{\sigma \in \sym{4}} a_{10}(\sigma) \ttr{\sigma}{A_1 \otimes A_2 \otimes A_1^\dagger \otimes A_2^\dagger}
    + \sum_{\sigma \in \sym{4}} a_{11}(\sigma) \ttr{\sigma}{A_1 \otimes A_1^\dagger \otimes (A_2^\dagger)^{\otimes 2}} \\
    \notag
    & + \sum_{\sigma \in \sym{4}} a_{12}(\sigma) \ttr{\sigma}{A_2 \otimes (A_1^\dagger)^{\otimes 2} \otimes A_2^\dagger}
    + \sum_{\sigma \in \sym{4}} a_{13}(\sigma) \ttr{\sigma}{(A_1^\dagger)^{\otimes 2} \otimes (A_2^\dagger)^{\otimes 2}} \pt
\end{align}
The 39 non-zero contributions are listed in Table~\ref{tab:contributions-interactions-multi}.

\section{From one matrix to \texorpdfstring{$N$}{ N} (non-)interacting fermions}
\label{app:fermion_mapping}

For one matrix, the singlet sector admits an exact reformulation in terms of $N$ one-dimensional (non-)interacting fermions~\cite{brezin1978Planar}. The key point is that, after diagonalization $X=U\Lambda U^\dagger$ with $\Lambda=\mathrm{diag}(\lambda_1,\dots,\lambda_N)$, singlet wavefunctions depend only on eigenvalues and the measure becomes
\begin{align}
    dX = [dU]\prod_{i=1}^N d\lambda_i\Delta(\lambda)^2\comma
    \qquad
    \Delta(\lambda)=\prod_{i<j}(\lambda_i-\lambda_j) \pt
\end{align}
The radial Laplacian identity implies, on singlets,
\begin{align}
    \tr\!\left(\frac{\partial^2}{\partial X^2}\right)\Psi(\lambda)
    = \frac{1}{\Delta(\lambda)}\sum_{i=1}^N \partial_{\lambda_i}^2\!\left[\Delta(\lambda)\Psi(\lambda)\right]\pt
\end{align}
Defining $\widetilde\Psi(\lambda) \coloneqq \Delta(\lambda)\Psi(\lambda)$ the inner product becomes flat, $\langle\Phi|\Psi\rangle=\int \prod_i d\lambda_i\widetilde\Phi^*\widetilde\Psi$, and $\widetilde\Psi$ is totally antisymmetric. Hence, the singlet problem is equivalent to $N$ fermions.

For instance, the singlet Hamiltonian associated with
\begin{align}
    H = \sum_{\ell,k=1}^N m(\hat N_k^\ell + \tfrac12) + g^2\tr(X^4)
\end{align}
acts on $\widetilde\Psi$ as a sum of one-body operators,
\begin{align}
    H_{\mathrm{fermion}}
    = \sum_{i=1}^N\left[-\frac12\partial_{\lambda_i}^2 + \frac{m^2}{2}\lambda_i^2 + g^2\lambda_i^4\right]\pt
\end{align}
Then, by Pauli principle, the energies are obtained by filling distinct one-particle levels.

The same transformation applies to any singlet operator built from products of traces of words in $X$ and $P$, e.g.
\begin{align}
    \mathcal{O}(X,P)=\prod_r \tr(W_r(X,P))\pt
\end{align}
Restricted to singlets, one may first write $\mathcal{O}$ as a symmetric differential operator in $(\lambda_i,\partial_{\lambda_i})$ and then map it by $\widetilde{\mathcal O}=\Delta\mathcal O\Delta^{-1}$.

For single-trace operators such as $\tr(X^4)$ the resulting Hamiltonian consists of only one-body terms and the $N$ fermions are non-interacting. Multi-trace products, by contrast, generate two-body (and higher) interactions; for instance $\tr(X^3)^2 = \sum_{i,j}\lambda_i^3\lambda_j^3$. For finite $N$, such a fermionic many-body system can be challenging to solve numerically.

When $N$ is small to moderate (up to $N\approx 10$), the many-body wavefunction can be expanded in a truncated basis of Slater determinants built from the eigenstates of an appropriate one-body Hamiltonian. Then, the Hamiltonian matrix is assembled using the Slater--Condon rules, and low-lying eigenvalues are estimated via the Lanczos or Arnoldi algorithm. For sextic-like potential as $\tr(X^3)^2$, one typically requires the single-particle cutoff to scale roughly as $3N$, leading to a basis of $\binom{3N}{N}$ Slater determinants, which grows rapidly and becomes prohibitive beyond $N\approx 10$.

The diagonalization $X=U\Lambda U^\dagger$ fixes the unitary gauge freedom; hence this mapping applies only to models with a single matrix. Multi-matrix models cannot be treated by this approach, since one cannot simultaneously diagonalize non-commuting observables.

\acknowledgments
This work is supported by the European Union’s Horizon Europe Framework Programme (HORIZON) under the ERA Chair scheme with grant agreement no. 101087126. This work is supported with funds from the Ministry of Science, Research and Culture of the State of Brandenburg within the Center for Quantum Technologies and Applications (CQTA).
\begin{figure}[!h]
    \centering
    \includegraphics[width=0.1\textwidth]{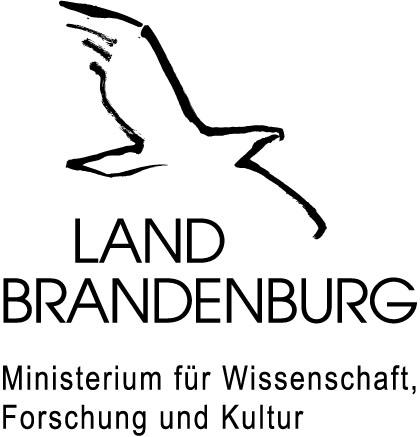}
\end{figure}

\paragraph{Code availability.}
The implementation is publicly available at 
\begin{center}
    \href{https://github.com/jcazalis/matrix-models}{github.com/jcazalis/matrix-models}
\end{center}

\paragraph{Author contributions.}
Jean Cazalis (J.C.) was responsible for the conceptualization, development of the methodology, creation of the software, execution of the investigation, performing the formal analysis, and writing, reviewing, and editing the original draft. Enrico Brehm (E.B.) contributed significantly to the conceptualization, development of the methodology, key aspects of the formal analysis, and writing, reviewing, and editing the original draft. AI tools, in particular Gemini and GitHub Copilot, were used to assist with the implementation and to refine the clarity of the exposition and the presentation of mathematical arguments; all scientific results and interpretations are those of the authors.

\providecommand{\href}[2]{#2}\begingroup\raggedright\endgroup

\end{document}